\documentclass[fleqn,usenatbib]{mnras}

\usepackage{newtxtext,newtxmath}

\usepackage[T1]{fontenc}

\DeclareRobustCommand{\VAN}[3]{#2}
\let\VANthebibliography\thebibliography
\def\thebibliography{\DeclareRobustCommand{\VAN}[3]{##3}\VANthebibliography}

\usepackage{graphicx}	
\usepackage{amsmath}	
\usepackage{times} 
\usepackage{latexsym}
\usepackage{amsfonts}
\usepackage{rotfloat} 
\usepackage{rotating} 
\usepackage{float}
\usepackage{natbib}
\usepackage{supertabular}
\usepackage{longtable}
\usepackage{url}
\usepackage{dcolumn}
\usepackage{placeins}
\usepackage{multirow} 
\usepackage{textcomp}
\usepackage{multicol}
\usepackage{bm}
\usepackage{pdflscape}
\usepackage{verbatim}
\usepackage{ae,aecompl}
\usepackage[utf8]{inputenc}
\usepackage[caption=false]{subfig}

\newcommand{\hi}{\mbox{H\,{\sc i}}}
\newcommand{\hii}{\mbox{H\,{\sc ii}}}
\newcommand{\mgii}{\mbox{Mg\,{\sc ii}}}

\newcommand{\feii}{\mbox{Fe\,{\sc ii}}}

\newcommand{\civ}{\mbox{C\,{\sc iv}}}

\newcommand{\ha}{H\,$\alpha$}
\newcommand{\hb}{H\,$\beta$}
\newcommand{\oii}{[O\,{\sc ii}]}
\newcommand{\oiii}{[O\,{\sc iii}]}
\newcommand{\nii}{[N\,{\sc ii}]}
\newcommand{\sii}{[S\,{\sc ii}]}
\newcommand{\lya}{\ensuremath{{\rm Ly}\alpha}}

\newcommand{\mstar}{$\rm M_*$}
\newcommand{\mhalo}{$\rm M_{\rm h}$}

\newcommand{\kms}{km\,s$^{-1}$}

\newcommand{\cc}{cm$^{-3}$}
\newcommand{\ergs}{$\rm erg\,s^{-1}$}
\newcommand{\ergscm}{$\rm erg\,s^{-1}\,cm^{-2}$}
\newcommand{\ergscmarc}{$\rm erg\,s^{-1}\,cm^{-2}\,arcsec^{-2}$}
\newcommand{\msun}{$\rm M_\odot$}
\newcommand{\msunyr}{$\rm M_\odot yr^{-1}$}

\title[Metal emission from galaxy haloes]{Metal line emission from galaxy haloes at $z\approx1$}

\author[R. Dutta et al.]{Rajeshwari Dutta,$^{1,2}$\thanks{E-mail: rajeshwari.dutta@unimib.it}
Matteo Fossati,$^{1,2}$
Michele Fumagalli,$^{1,3}$
Mitchell Revalski,$^{4}$
Emma K. Lofthouse,$^{1,2}$
\newauthor
Dylan Nelson,$^{5}$
Giulia Papini,$^{1}$
Marc Rafelski,$^{4,6}$
Sebastiano Cantalupo,$^{1}$
Fabrizio Arrigoni Battaia,$^{7}$
\newauthor
Pratika Dayal,$^{8}$
Alessia Longobardi,$^{1,2}$
Celine P\'eroux,$^{9,10}$
Laura J. Prichard,$^{4}$
J. Xavier Prochaska$^{11}$
\\
$^{1}$Dipartimento di Fisica G. Occhialini, Universit\`a degli Studi di Milano Bicocca, Piazza della Scienza 3, 20126 Milano, Italy \\
$^{2}$INAF - Osservatorio Astronomico di Brera, via Bianchi 46, 23087 Merate (LC), Italy \\
$^{3}$INAF – Osservatorio Astronomico di Trieste, via G. B. Tiepolo 11, I-34143 Trieste, Italy \\
$^{4}$Space Telescope Science Institute, 3700 San Martin Drive, Baltimore, MD 21218, USA \\
$^{5}$Universit\"at Heidelberg, Zentrum für Astronomie, Institut f\"ur theoretische Astrophysik, Albert-Ueberle-Str. 2, D-69120 Heidelberg, Germany \\
$^{6}$Department of Physics and Astronomy, Johns Hopkins University, Baltimore, MD 21218, USA \\
$^{7}$Max-Planck-Institut f\"ur Astrophysik, Karl-Schwarzschild-Str. 1,
D-85748 Garching bei M\"unchen, Germany\\
$^{8}$Kapteyn Astronomical Institute, University of Groningen, P.O. Box 800, 9700 AV Groningen, The Netherlands \\
$^{9}$European Southern Observatory, Karl-Schwarzschildstrasse 2, D-85748 Garching bei M\"unchen, Germany \\
$^{10}$Aix Marseille Universit\'e, CNRS, LAM (Laboratoire d'Astrophysique de Marseille) UMR 7326, F-13388, Marseille, France \\
$^{11}$Department of Astronomy \& Astrophysics, UCO/Lick Observatory, University of California, 1156 High Street, Santa Cruz, CA 95064, USA
}

\date{Accepted XXX. Received YYY; in original form ZZZ}

\pubyear{2022}

\begin{document}
\label{firstpage}
\pagerange{\pageref{firstpage}--\pageref{lastpage}}
\maketitle

\begin{abstract}
{We present a study of the metal-enriched halo gas, traced using \mgii\ and \oii\ emission lines, in two large, blind galaxy surveys --- the MUSE (Multi Unit Spectroscopic Explorer) Analysis of Gas around Galaxies (MAGG) and the MUSE Ultra Deep Field (MUDF). By stacking a sample of $\approx$600 galaxies (stellar masses \mstar\ $\approx10^{6-12}$\,\msun), we characterize for the first time the average metal line emission from a general population of galaxy haloes at $0.7 \le z \le 1.5$. The \mgii\ and \oii\ line emission extends farther out than the stellar continuum emission, on average out to $\approx$25\,kpc and $\approx$45\,kpc, respectively, at a surface brightness (SB) level of $10^{-20}$\,\ergscmarc. The radial profile of the \mgii\ SB is shallower than that of the \oii, suggesting that the resonant \mgii\ emission is affected by dust and radiative transfer effects. The \oii\ to \mgii\ SB ratio is $\approx 3$ over $\approx 20-40$\,kpc, also indicating a significant in situ origin of the extended metal emission. The average SB profiles are intrinsically brighter by a factor $\approx 2-3$ and more radially extended by a factor of $\approx 1.3$ at $1.0 < z \le 1.5$ than at $0.7 \le z \le 1.0$. The average extent of the metal emission also increases independently with increasing stellar mass and in overdense group environments. When considering individual detections, we find extended \oii\ emission up to $\approx 50$\,kpc around $\approx30-40$ percent of the group galaxies, and extended ($\approx 30-40$\,kpc) \mgii\ emission around two $z\approx1$ quasars in groups, which could arise from outflows or environmental processes.}
\end{abstract}




\section{Introduction}
\label{sec:introduction}

The circumgalactic medium \citep[CGM;][]{tumlinson2017,peroux2020a}, which acts as the interface between galaxies and the wider environment, is a key aspect of galaxy evolution. Not only does the CGM mediate the accretion and ejection of baryons to and from galaxies, it also modulates larger-scale interactions between galaxies and the environment \citep[e.g.][]{burchett2018,fossati2019b,dutta2020,dutta2021}. The most viable method of studying the diffuse halo gas has been through absorption against a bright background source such as a quasar \citep{sargent1980}. Significant progress has been made over the last few decades in statistically characterizing the distribution and physical conditions of the multiphase halo gas by cross-correlating absorption lines and galaxy surveys at $z<2$ \citep[e.g.][]{prochaska2011a,tumlinson2013,chen2020,dutta2020,dutta2021,wilde2021,berg2022,peroux2022,weng2022}, and at $z>2$ \citep[e.g.][]{rudie2012,turner2014,bielby2020,lofthouse2020,lofthouse2023,muzahid2021,galbiati2023}. 

However, absorption line measurements along pencil-beam sightlines are generally unable to provide a complete mapping of the halo gas distribution. Systems in which it is possible to conduct spatially-resolved studies of the CGM in absorption either via a higher spatial density of background sources such as lensed \citep[e.g.][]{rauch2001,chen2014,rubin2015,zahedy2016,augustin2021} and multiple quasars \citep[e.g.][]{martin2010,bowen2016,lehner2020,zheng2020,beckett2021,mintz2022}, or via spatially-resolved background sources \citep[e.g. lensed or extended galaxies;][]{lopez2018,peroux2018,tejos2021} are rare. 

In contrast, spatially-resolved emission allows us to directly map the gas, at least at moderate to high densities, in and around galaxies and to place more stringent constraints on the extent and physical properties of gas, particularly in cases where multiple line diagnostics can be obtained. However, it has been challenging to detect the low-surface brightness (SB) halo gas at cosmological distances in emission. Past efforts to detect the CGM in emission have mostly focused on \lya\ emission around high redshift quasars, whose larger photoionizing flux can greatly boost the emission \citep[e.g.][]{cantalupo2005,cantalupo2014,christensen2006,hennawi2015}, stacking analysis of massive Lyman break galaxies \citep[LBGs;][]{steidel2011}, or low redshift ($z\le0.2$) galaxies \citep[e.g.][]{hayes2016,zhang2016}.

In the past few years, optical integral field unit (IFU) spectrographs, such as the Multi Unit Spectroscopic Explorer \citep[MUSE;][]{bacon2010} on the Very Large Telescope (VLT) and the Keck Cosmic Web Imager \citep[KCWI;][]{morrissey2018}, have ushered in a new era of CGM observations in emission. Using these instruments, extended \lya\ nebulae are now routinely detected around $z\gtrsim2-4$ quasars \citep[e.g.][]{borisova2016,arrigoni2019,cai2019,farina2019,fossati2021,mackenzie2021}. Moreover, the sensitive SB limits reached by these instruments have enabled detection of \lya\ haloes around typical star-forming galaxies \citep[e.g.][]{wisotzki2016,leclercq2017,leclercq2020,kusakabe2022} and extended \lya\ emission from larger-scale structures \citep[e.g.][]{wisotski2018,umehata2019,bacon2021} at $z\gtrsim3$.

The metal-enriched halo gas has been more challenging to detect \citep{rickards2019}, as expected from theoretical grounds due to their lower SB \citep{bertone2012,frank2012,piacitelli2022}. This technique has nevertheless seen many recent successes, such as the detection of extended nebulae in \oii\ and \oiii\ emission around quasars \citep{johnson2018,johnson2022,helton2021}, galaxy groups and clusters \citep{epinat2018,boselli2019,chen2019}, and ultraluminous galaxies \citep{rupke2019} at $z<1$. Furthermore, extended \mgii\ emission has been detected around a few galaxies at $z<2$ using long-slit spectroscopy \citep{rubin2011,martin2013} and IFU observations \citep{rupke2019,burchett2021,zabl2021,shaban2022}. Recently, there has been a remarkable detection of extended \mgii\ emission from the intragroup medium of a $z\approx1.3$ group using deep MUSE observations \citep{leclercq2022}.

Despite this rapid acceleration in direct detections of the CGM in emission, observations of metal line emission from galaxy haloes to date have been mostly limited to extreme systems such as active or starbursts galaxies or highly overdense groups. To put robust constraints on galaxy formation models \citep[e.g.][]{corlies2016,nelson2021,piacitelli2022}, it is imperative to spatially resolve and map the metal line emission from gaseous haloes around more normal galaxies, and conduct complete systematic studies in larger samples. While detection of metal-emitting haloes around individual galaxies at $z\gtrsim1$ is still challenging, the average metal emission around a population of galaxies can be studied through stacking the IFU data of a statistical sample of galaxies. Such an investigation is now achievable thanks to the medium-deep data available from recent large, blind galaxy surveys with IFUs. In this work, we utilise two such surveys, which are complementary in survey volume and depth --- the medium-deep ($\approx5-10$\,h) MUSE Analysis of Gas around Galaxies \citep[MAGG;][]{lofthouse2020}, and the highly-sensitive ($\approx150$\,h) MUSE Ultra Deep Field \citep[MUDF;][]{fossati2019b} --- to study the average \mgii\ and \oii\ line emission from galaxy haloes as well as from individual systems at $z\approx0.7-1.5$. 

The structure of this paper is as follows. Firstly, we provide a brief description of the galaxy sample built from the MAGG and MUDF surveys in Section~\ref{sec:data}. Then, in Section~\ref{sec:methods}, we describe the methodology adopted to stack the MUSE data. Next, we present the results in Section~\ref{sec:results}, first from the stacking of metal line emission around galaxies in Section~\ref{sec:results_stack}, and then from the search for extended emission in groups of galaxies in Section~\ref{sec:results_individual}. We discuss our results and compare them with literature results based on observations and simulations in Section~\ref{sec:discussion}. Finally, we summarize our results in Section~\ref{sec:summary}. Throughout this work, we adopt a Planck 15 cosmology with $H_{\rm 0}$ = 67.7\,\kms\,Mpc$^{-1}$ and $\Omega_{\rm M}$ = 0.307 \citep{planck2016}, express the distances in proper units, and the magnitudes in the AB system.

\section{Data and sample selection}
\label{sec:data}

\begin{figure}
 \includegraphics[width=0.5\textwidth]{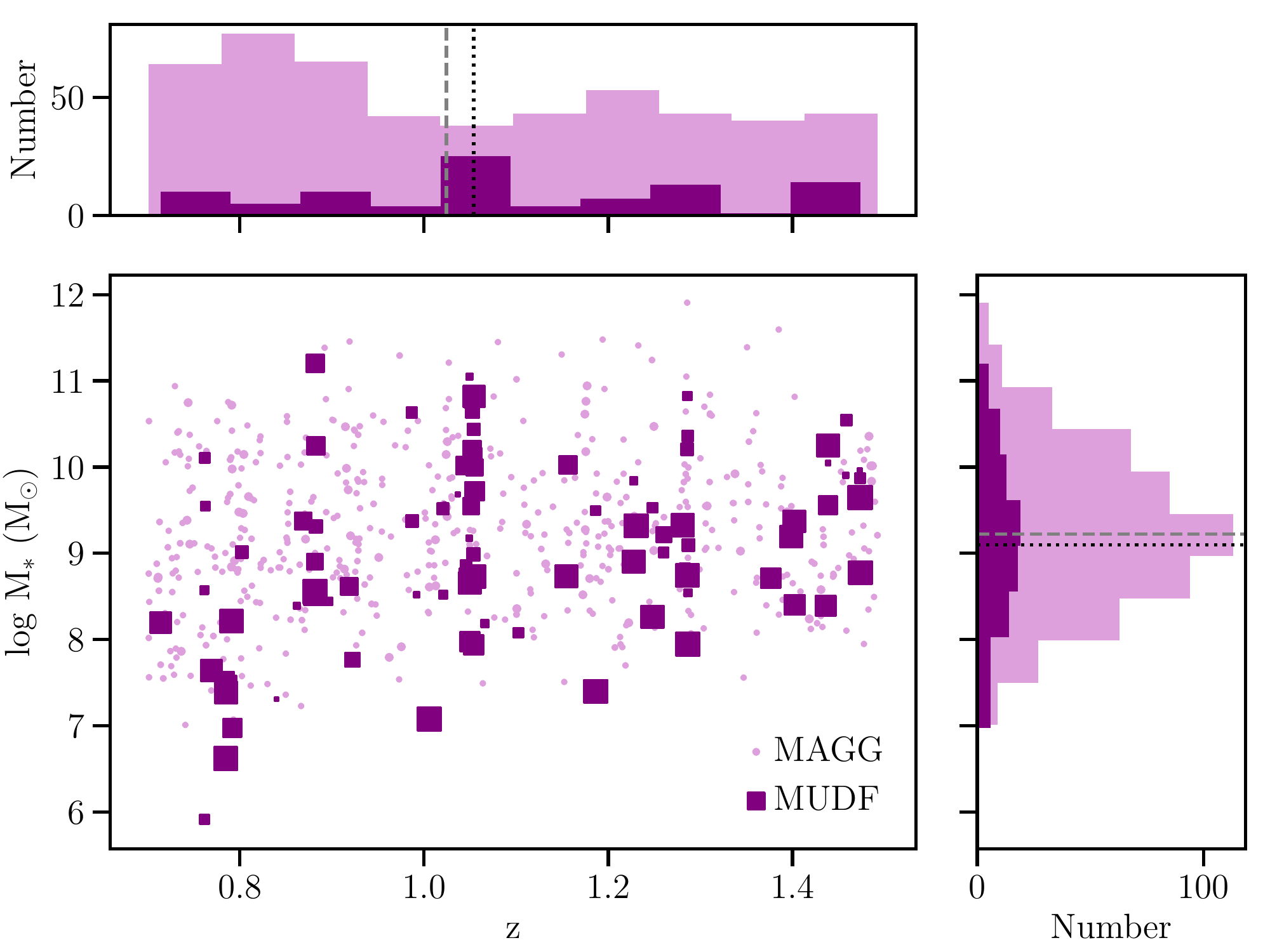}
 \caption{Stellar mass versus redshift of the galaxies used for stacking. Galaxies from the MAGG survey are shown as pink circles, while those from the MUDF survey are shown as purple squares. The size of the symbols represents the MUSE observing time, which ranges from $\approx4$ to $\approx10$\,h in the case of MAGG galaxies, and $\approx2$ to $\approx112$\,h in the case of MUDF galaxies. The plots to the top and to the right show the histograms (pink for MAGG and purple for MUDF) of the redshifts and stellar masses, respectively. The median values are marked by dashed (MAGG) and dotted (MUDF) lines.
 }
 \label{fig:mstar_zgal}
\end{figure}

\subsection{The MAGG survey}
\label{sec:data_magg}

The MAGG survey is built upon a VLT/MUSE large programme (ID: 197.A-0384, PI: M. Fumagalli) that is supplemented by archival data of a guaranteed time observation programme (PI: J. Schaye). In total there are 28 fields that are centred on quasars at $z\approx3.2-4.5$. Each of these fields has been observed for a total on-source time of $\approx4$\,h, except for two fields that have archival data leading to longer exposure times of $\approx10$\,h. The description of the data and of the reduction process is presented in \citet[][see their table 1 and Section 3.2]{lofthouse2020}. In brief, the MUSE data are first reduced using the standard ESO pipeline \citep[version 2.4.1,][]{weilbacher2015}, and then the data are post-processed using tools ({\sc CubeFix} and {\sc CubeSharp}) that are part of the {\sc CubExtractor} package \citep[CubEx, version 1.8,][and Cantalupo in prep.]{cantalupo2019}.

For this work, we use the sample of galaxies that are detected in continuum in the MUSE white-light images. The compilation of the continuum-detected galaxy sample is described in detail in \citet{lofthouse2020} and \citet{dutta2020}. Briefly, the continuum sources are first identified from the white-light images using {\sc SExtractor} \citep{bertin1996}, then the 1D spectra are extracted from the 3D cubes based on the 2D segmentation maps created by {\sc SExtractor}, and finally the source redshifts are estimated using {\sc marz} \citep{hinton2016}. The MAGG sample of continuum-detected sources is 90 per cent complete down to $r\approx26.3$\,mag and the sample of sources with reliable redshifts is 90 per cent complete down to $r\approx24.85$\,mag \citep[see figure 5 of][]{lofthouse2023}. For the stacking analysis presented in this work, we use a sample of 508 galaxies at $0.7 \le z \le 1.5$ that have reliable spectroscopic redshifts \citep[redshift flag 3 and 4, see section 5.1 of][]{lofthouse2020}. The redshift range is chosen such that both the \mgii\ and \oii\ emission lines are covered in the MUSE spectra (wavelength coverage of 4650--9300\,\AA). The physical properties of the galaxies such as stellar mass (\mstar) and star formation rate (SFR) are obtained by jointly fitting the MUSE spectra and photometry \citep[derived in four top-hat pseudo-filters, see table 2 of][]{fossati2019b} with stellar population synthesis models using the Monte Carlo Spectro-Photometric Fitter \citep[{\sc MC-SPF};][]{fossati2018} as described in \citet{fossati2019b} and \citet{dutta2020}. In brief, {\sc MC-SPF} uses \citet{bruzual2003} models at solar metallicity, the \citet{chabrier2003} initial mass function (IMF), nebular emission lines from the models of \citet{byler2018}, and dust attenuation law of \citet{calzetti2000}. The parameters and assumptions that go into the modeling are consistent with the ones typically used in the literature for extragalactic surveys \citep[e.g.][]{fossati2017}.

\subsection{The MUDF survey}
\label{sec:data_mudf}

The MUDF survey is based upon a recent VLT/MUSE large programme (ID: 1100.A-0528; PI: M. Fumagalli) that has obtained very deep MUSE observations ($\approx150$\,h on-source) of a $1.5\times1.2$\,arcmin$^2$ region centred on two quasars at $z\approx3.2$. In this work, we use the MUSE observations that have been acquired till 2021 November 21, consisting of 344 exposures of 1450\,s each or $\approx138$\,h in total. The MUSE observations of the MUDF field have been designed to collect maximum data in the central regions between the two quasars. Therefore, taking into account the detector gaps, the exposure time goes from $\approx$115\,h in the centre to $\approx$2\,h in the outer regions. The MUSE observations are complemented by a deep near-infrared (NIR) spectroscopic survey (HST ID: 15637; PIs: M. Rafelski, M. Fumagalli) consisting of 90 orbits using the Wide Field Camera 3 (WFC3) G141 grism on the Hubble Space Telescope (HST) and eight-orbit near-ultraviolet (NUV) imaging using HST WFC3/UVIS (ID: 15968; PI: M. Fossati).

The MUSE data reduction is described in detail in \citet{fossati2019b} and follows the same methodology as for the MAGG survey. The HST data reduction is described in detail in \cite{revalski2023}. In brief, the individual F140W exposures were aligned and drizzled using the {\sc TweakReg} and {\sc AstroDrizzle} tools, respectively, that are part of the {\sc DrizzlePac} software \citep{hoffmann2021}. The final F140W image is used as a reference for detection of sources and alignment with other images. A segmentation map of sources in the F140W image was constructed by running {\sc SExtractor} with ``deep'' and ``shallow'' thresholds to better characterize the shapes and extents of faint and bright sources, respectively \citep[see table 3 of][for details of the parameters used]{revalski2023}. The photometry in the five different HST filters (F140W, F125W, F702W, F450W, and F336W) was measured using {\sc SExtractor} in dual-image mode, where sources are first detected in the F140W image and then analysed in each of the filters. 

To extract the MUSE photometry, we used {\sc t-phot} \citep{merlin2015}, and the higher resolution F140W image, point spread function (PSF) model, and source catalogue. A matching kernel between the two PSFs (obtained using the ratio of Fourier transforms) is used to convolve the F140W image to have the same PSF as the MUSE one. The MUSE data are first resampled on to the F140W pixel grid. Then, {\sc t-phot} is run two times, where the second run uses the ``multikernel'' option to regenerate templates using spatially varying transfer kernels generated in the first run. The MUSE photometry is extracted in four top-hat pseudo-filters as defined in table 2 of \citet{fossati2019b}. The MUSE 1D spectra are extracted based on the F140W {\sc SExtractor} segmentation map that is convolved with the MUSE PSF mentioned above. The MUSE optical spectra are then visually inspected using a custom-developed tool by different co-authors (GP, MFo, RD) to estimate the source redshifts. The redshifts are further given a flag following the same classification scheme as adopted for the MAGG survey (see Section~\ref{sec:data_magg}). We are able to estimate reliable redshifts for 90 per cent of the sources down to F140W$\approx20.3$\,mag and for 50 per cent of the sources down to F140W$\approx24.4$\,mag. Finally, the MUSE spectra and photometry and HST photometry are jointly fit by {\sc MC-SPF} to derive the galaxy properties similar to the MAGG survey (see Section~\ref{sec:data_magg}). 

For the stacking of \oii\ emission in this work, we use a sample of 93 galaxies from MUDF with reliable redshifts at $0.7 \le z \le 1.5$. For the stacking of the \mgii\ emission, we exclude certain wavelength ranges (see Section~\ref{sec:methods} for details), which leads to a sample of 67 galaxies at $0.7 \le z \le 1.5$ in MUDF. Fig.~\ref{fig:mstar_zgal} shows the distribution of the stellar mass and redshift of all the galaxies from the MAGG and the MUDF surveys that are used in this work. The stellar masses of the galaxies range from $8\times10^5$\,\msun\ to $8\times10^{11}$\,\msun, with a median mass of $\approx2\times10^9$\,\msun. The uncertainty in the base 10 logarithm of the stellar mass is typically between 0.1 and 0.2 dex. The median redshift of the sample is $z\approx1.04$. The total MUSE exposure time of all the galaxies used for \oii\ emission stacking is $\approx7140$\,h, while it is $\approx5674$\,h in the case of \mgii. While the galaxy sample used for \mgii\ stacking is slightly smaller than the one for \oii, we have checked that the distributions of stellar mass and redshift of the two samples are consistent with being drawn from the same parent population ($p$-value from two-sided Kolmogorov–Smirnov (KS) test $\gtrsim0.99$).

\section{Stacking Methodology}
\label{sec:methods} 

\begin{figure*}
 \includegraphics[width=0.48\textwidth]{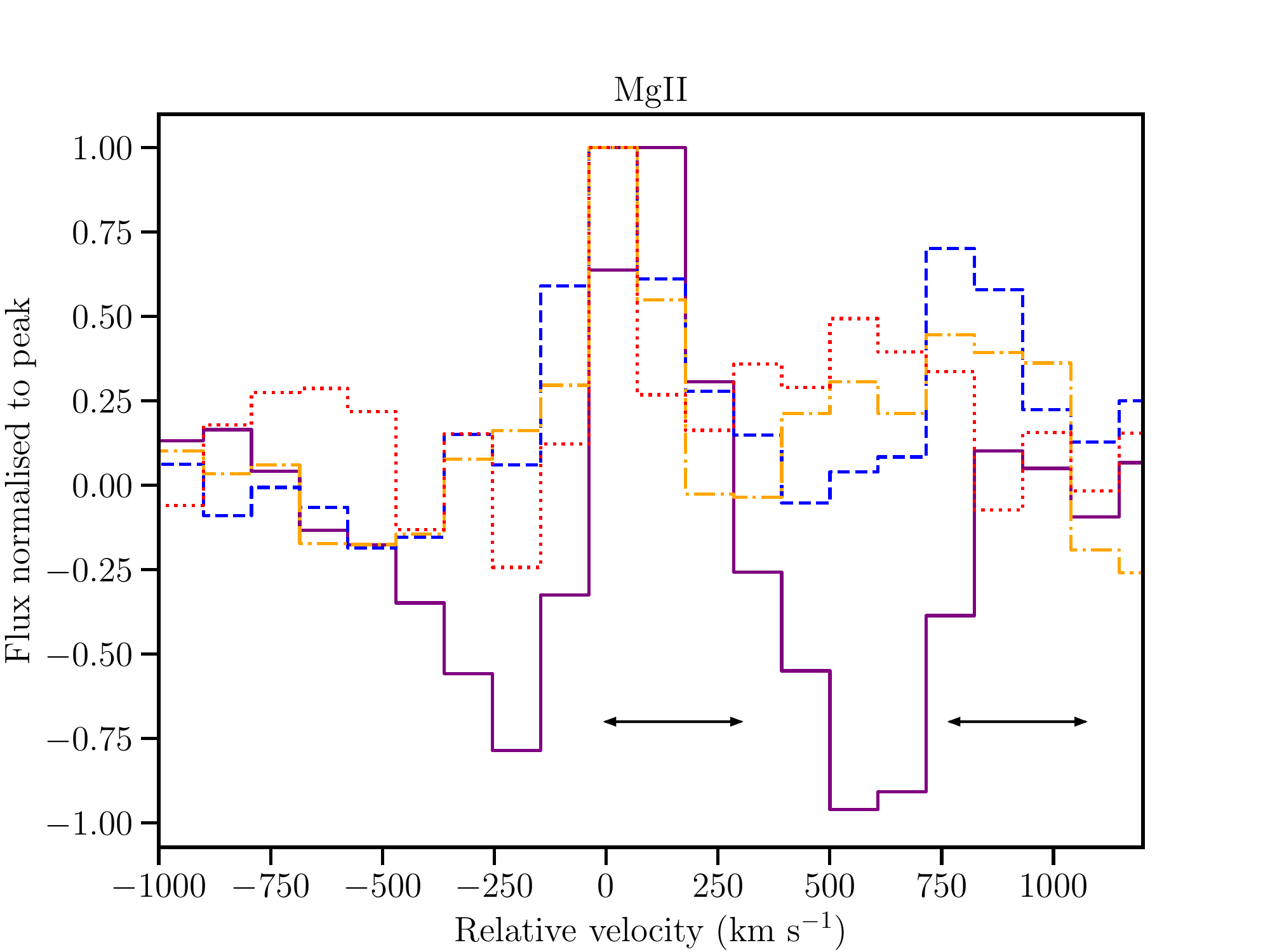}
 \includegraphics[width=0.48\textwidth]{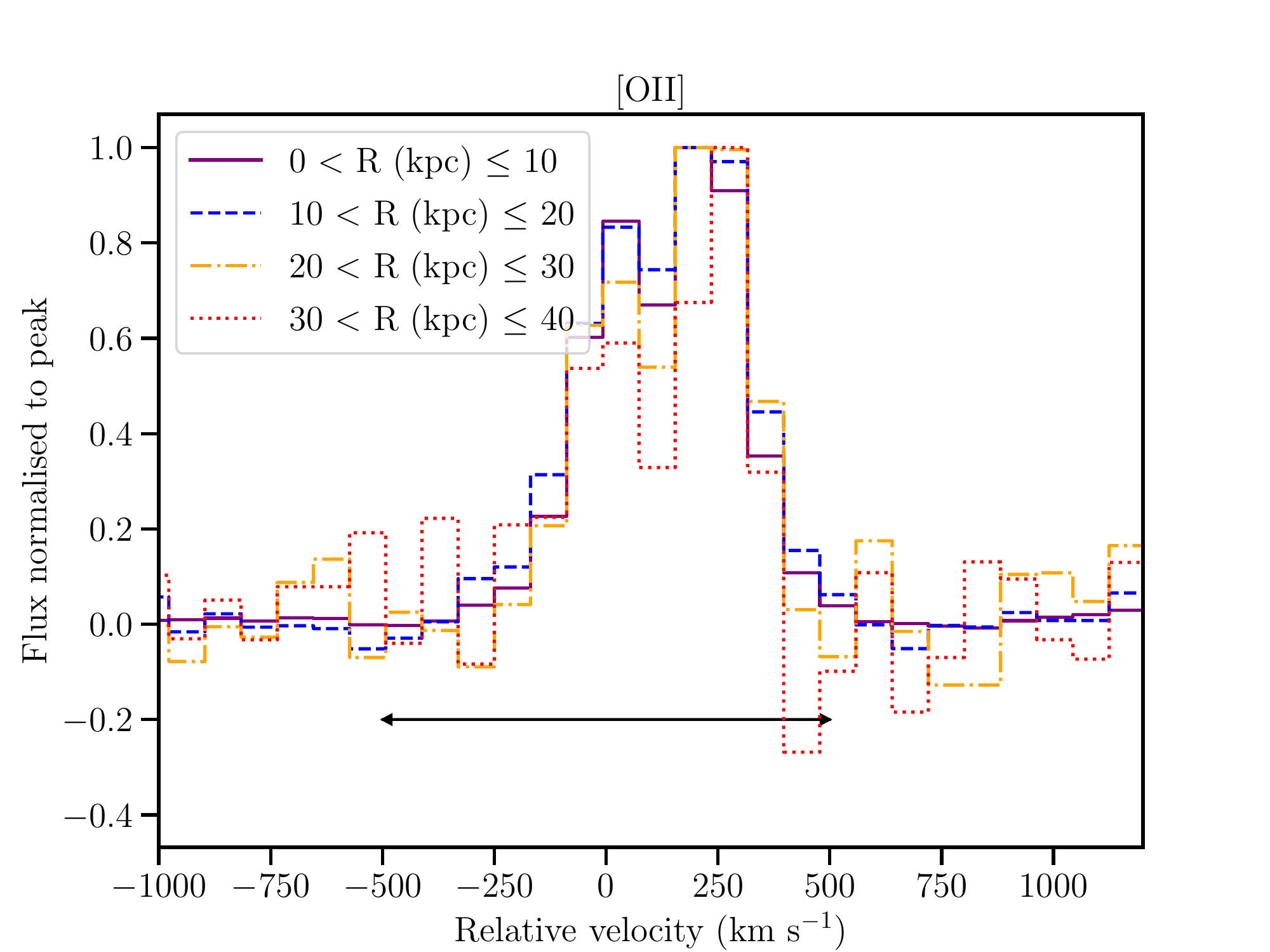}
 \caption{{\it Left:} The median-stacked spectra of \mgii\ emission in circular annuli of 0-10, 10-20, 20-30 and 30-40\,kpc (as indicated in the right panel) from the galaxy centres for the full sample. The flux has been normalised to the peak value. The velocity range used for generating the NB images are marked by black arrows.
 {\it Right:} Same as in the left plot for \oii\ emission.
 }
 \label{fig:stack_spectra}
\end{figure*}
\begin{figure*}
 \includegraphics[width=1.0\textwidth]{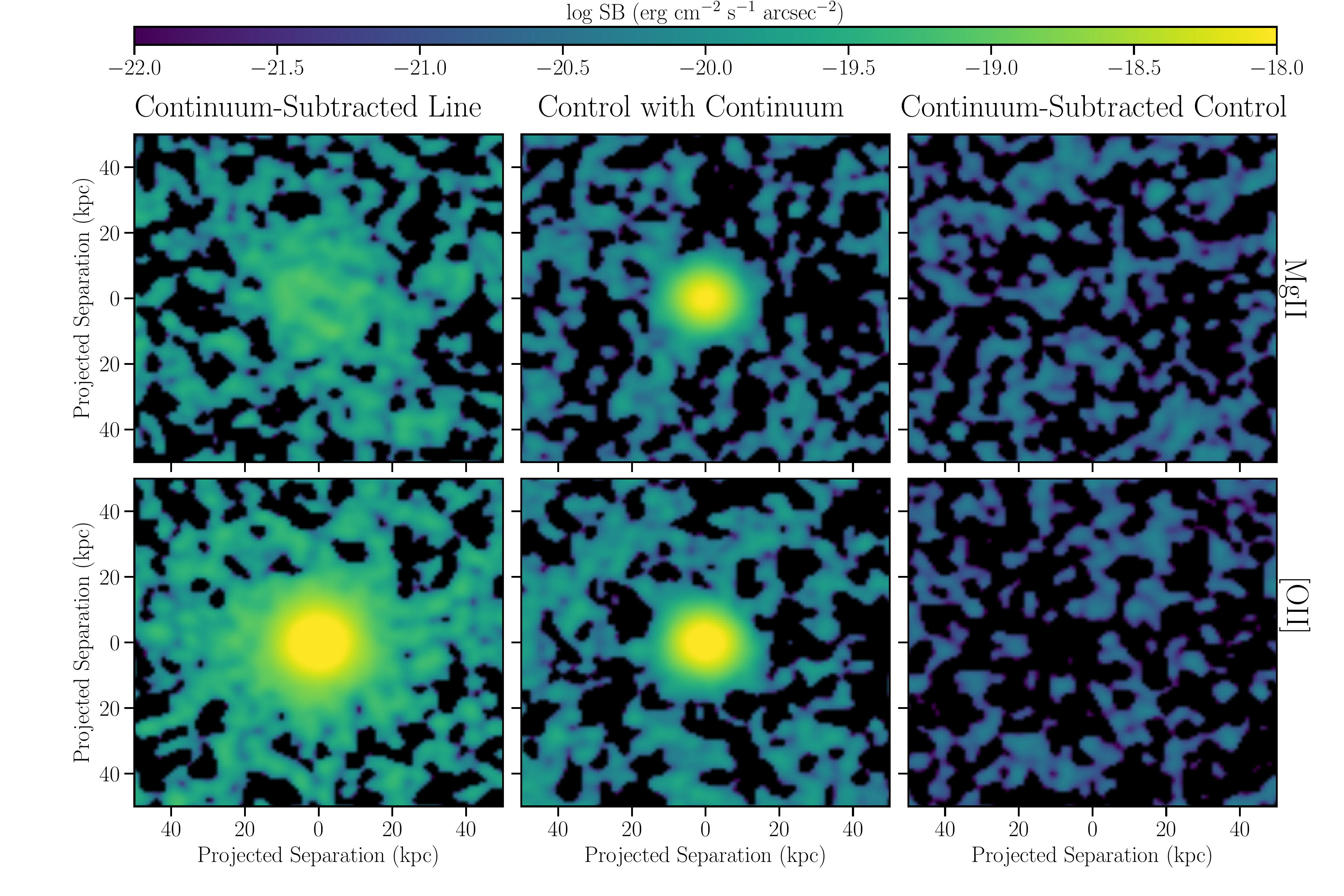}
 \caption{Median-stacked pseudo-NB images for the full sample. The top panels show the results for \mgii\ and the bottom panels show the results for \oii. From left to right, the panels show the stacked line emission after subtracting the continuum, the stacked continuum emission in control windows, and the stacked emission in control windows after subtracting the continuum (see Section~\ref{sec:methods} for details).
 }
 \label{fig:nbimg_all}
\end{figure*}
\begin{figure*}
 \includegraphics[width=0.48\textwidth]{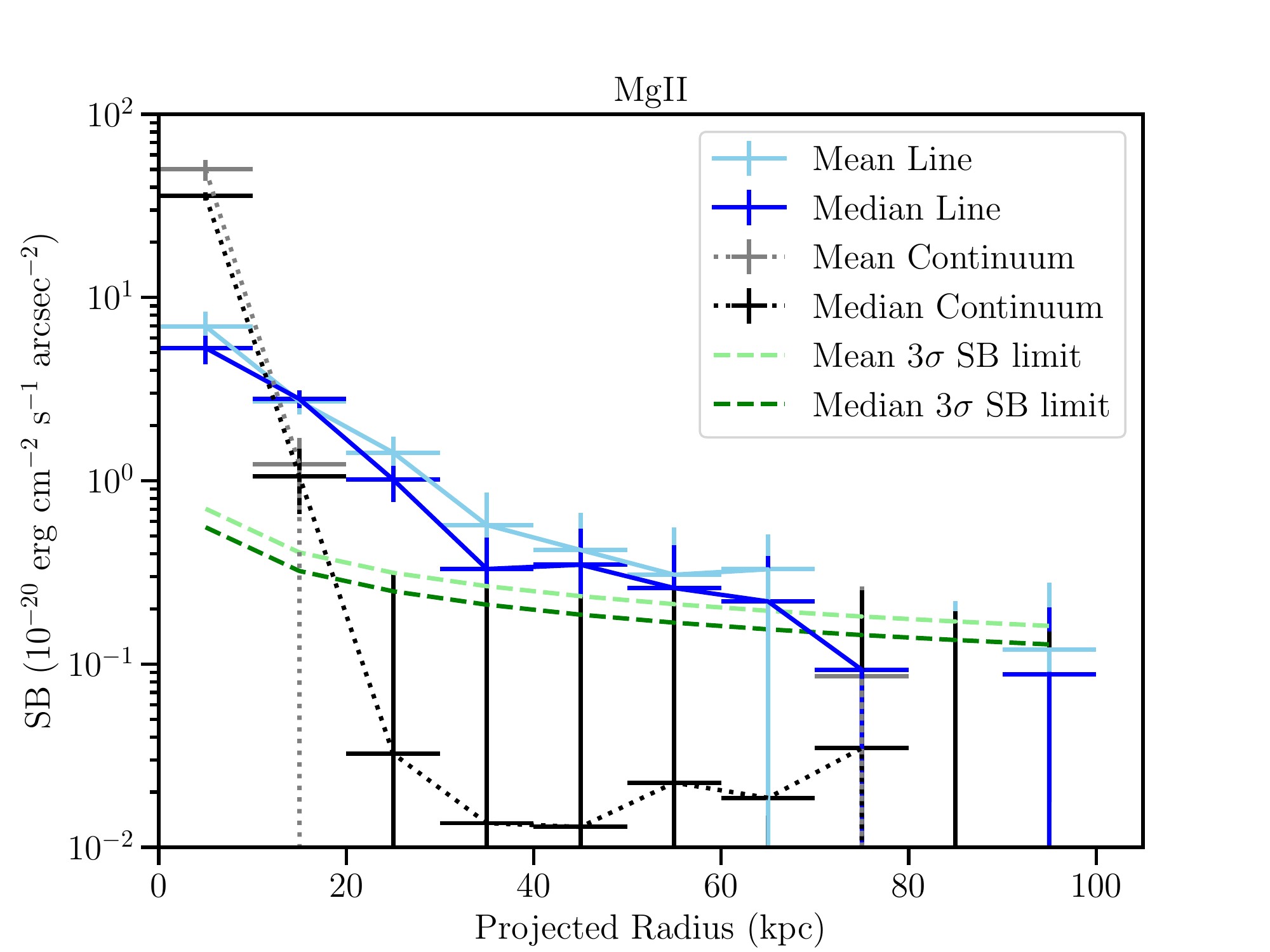}
 \includegraphics[width=0.48\textwidth]{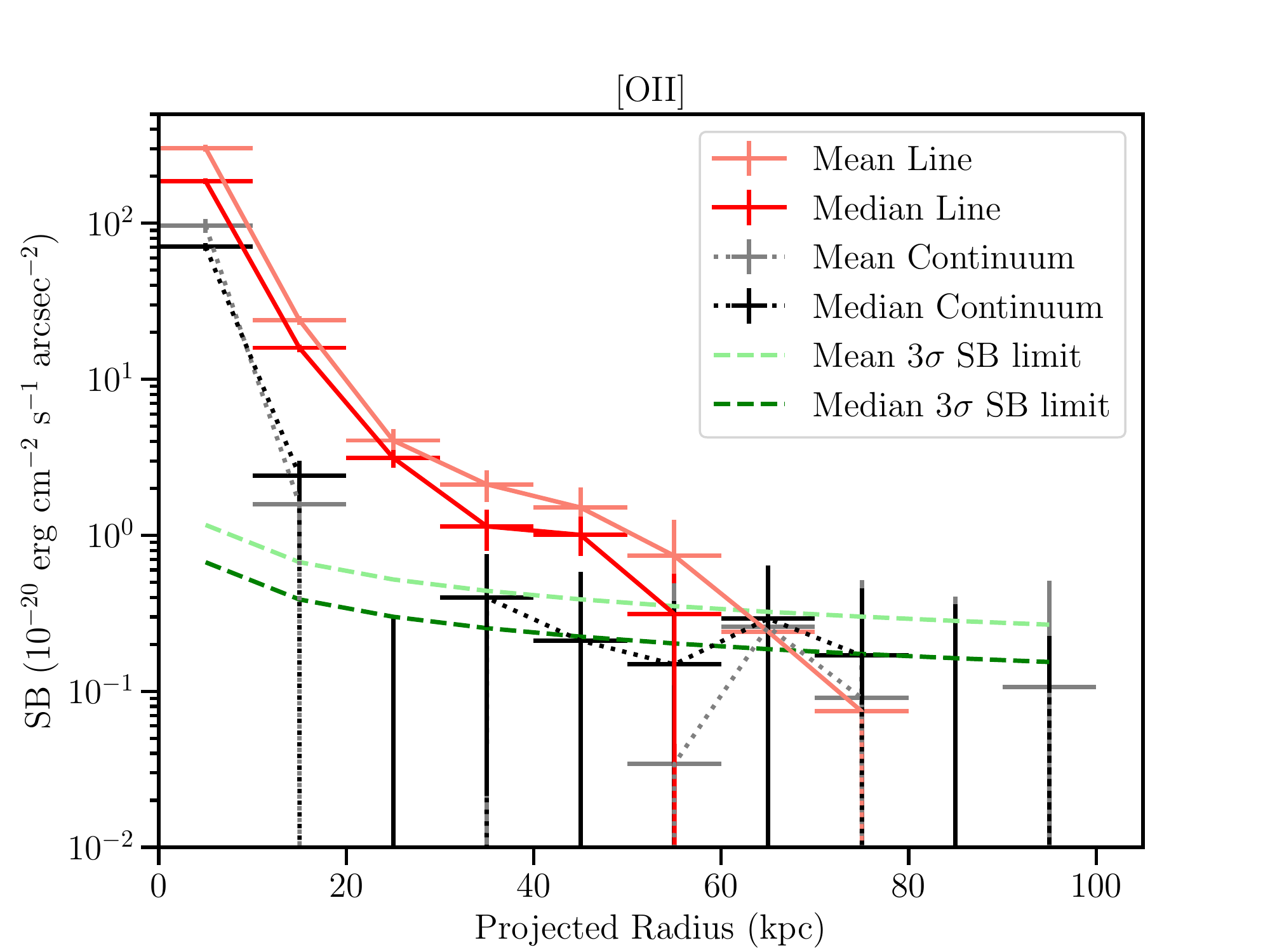}
 \caption{{\it Left:} The mean- and median-stacked azimuthally-averaged radial SB profiles of \mgii\ emission in circular annuli of 10\,kpc from the galaxy centres are shown as light blue and dark blue solid lines, respectively. The mean and median SB profiles of the continuum emission in control windows are shown as grey and black dotted lines, respectively. The $3\sigma$ SB limits for the mean and median profiles are shown as light and dark green dashed lines, respectively.
 {\it Right:} Same as in the left plot for \oii\ emission. The mean and median \oii\ SB profiles are shown as pink and red solid lines, respectively. The error bars represent the $1\sigma$ uncertainties from bootstrapping analysis.
 }
 \label{fig:sbprofile_all}
\end{figure*}

We outline here the method adopted to stack the metal line emission around galaxies in the MUSE 3D cubes. First, we extract sub-cubes centred around each of the galaxies. To do this, we convert the distance of each pixel from the galaxy centre into a physical distance (in kpc) at the redshift of the galaxy. Similarly, we convert the observed spectral wavelength into the rest wavelength. We mask all the continuum sources using the {\sc SExtractor} segmentation map except for the galaxy of interest. Then we interpolate the cube such that the galaxy is located at the centre of a common grid. In the case of MUDF, because it is a single-field observation, we further rotate the cube by a random angle around the galaxy position during the interpolation in order to ensure that residuals of sky subtraction and instrumental artefacts are randomly positioned in the final grid, and thus averaged down during stacking. This rotation is not found to be necessary in the case of MAGG, since these observations are taken in independent regions of the sky and at different times. In this way, we extract sub-cubes of $200\times200$\,kpc$^2$ centred on each galaxy that cover $\pm100$\,\AA\ around the metal line (\mgii, \oii). We note that we do not take into account inclination and orientation of the galaxies for the primary stacking analysis presented in this work. However, we investigate the effect of orientation on the stacked emission in Section~\ref{sec:results_stack_orientation}.

Next, we stack all the cubes using both mean and median statistics. Since we are interested in probing the extended metal line emission around galaxies, we need to subtract the stellar continuum emission first. We perform a local continuum subtraction for this purpose. At each pixel in the stacked cube, we extract the spectra within $\pm2000$\,\kms\ of the line (\mgii, \oii), fit a third-order spline to the continuum emission excluding the region around the line, and subtract this continuum fit from the spectra. We also investigated an alternate method of continuum subtraction using median-filtering, as commonly done in the literature \citep[e.g., when searching for \lya\ emission lines;][]{fossati2021}. However, given the P-Cygni-like profiles of several \mgii\ emitting galaxies and a non-flat continuum in this wavelength range, this method was found to leave stronger negative residuals. Therefore, we prefer the local continuum subtraction method.

We then generate pseudo-narrow-band (NB) images around the lines from the continuum-subtracted cubes. To generate the NB images, we use the rest-frame velocity range of $\pm500$\,\kms\ around the \oii\ 3727\,\AA\ line (which also covers the \oii\ 3729\,\AA\ line of the doublet), whereas for \mgii, we use the velocity range of $0-300$\,\kms\ around both the \mgii\ doublet lines ($\lambda\lambda$2796\,\AA, 2803\,\AA). The choice of velocity width over which to generate the NB images is motivated by the velocity range over which the line emission is found in the 1D spectra extracted from the stacked cube of all the galaxies. The 1D spectra extracted in annuli of 10\,kpc up to 40\,kpc around the galaxy centre and normalised to the peak value are shown in Fig.~\ref{fig:stack_spectra}. In the central 10\,kpc region, the \mgii\ emission line shows a P-Cygni-like profile \citep[as found in the interstellar medium of some galaxies; e.g.][]{weiner2009,rubin2010,prochaska2011b,erb2012,martin2012}, with blue-shifted absorption and red-shifted emission over $0-300$\,\kms. To avoid contamination by absorption in the central region, we select the above velocity range around both the \mgii\ doublet lines for generating NB images. The \oii\ doublet emission profile lies within the velocity range $\pm500$\,\kms\ in all the annuli. Additionally, we performed tests using different velocity windows ($\pm1000$\,\kms\ for \oii\ and $\pm300$\,\kms\ for \mgii) and found that the relative trends found in this work are not affected by this choice.

Finally, we estimated the azimuthally-averaged radial SB profiles based on the NB images in circular annuli of 10\,kpc centred on the galaxy positions. To estimate the uncertainties on the SB, we performed a bootstrap analysis where we repeated the above stacking process 100 times with repetitions. We use the $16^{\rm th}$ and the $84^{\rm th}$ percentiles of the bootstrapped values to obtain an equivalent $1\sigma$ error. We emphasize that the uncertainties determined from the bootstrap method represent the sample variance, which can be significant across a large interval of stellar mass. To estimate and compare the size of the emission in different samples, we performed a curve-of-growth analysis where we calculate the flux in concentric circles with radii increasing by 0.1\,kpc. In our analysis (see Section~\ref{sec:results_stack}), we find that the line emission in the NB images typically extends up to $\approx$30\,kpc in different samples. Therefore, we define here $R_{50}$ and $R_{90}$ as the radii of the circular apertures which contain 50 and 90 per cent of the total flux within a circular aperture of 30\,kpc, respectively. We note that the ratios of the sizes of the line and the continuum emission, and of the \mgii\ and \oii\ line emission, are not sensitive to the choice of the radius at which we normalise the flux. The $1\sigma$ errors on $R_{50}$ and $R_{90}$ are obtained from the $16^{\rm th}$ and the $84^{\rm th}$ percentiles of the bootstrapped values.

Furthermore, in order to check whether the average metal line emission is more spatially extended than the average stellar continuum emission, we generated NB images from the stacked cubes before subtracting the continuum emission. For this we averaged over four control windows that are located at $\pm2000$\,\kms\ and $\pm4000$\,\kms\ around the line, and that have the same velocity width as that used for generating the NB image of the line. We have checked that using control windows at different velocity locations does not affect the results. The azimuthally-averaged radial SB profile of the continuum emission was then estimated in circular annuli of 10\,kpc similar to the line emission. We note that the SB profile of the continuum emission remains at a constant level (typically $\approx2\times10^{-20}$\,\ergscmarc) at large distances ($>50$\,kpc) from the galaxy centres because of low-level sky residuals. We estimated the average emission in the outermost annuli and subtracted this constant background value from the NB images and SB profiles of the continuum emission. Lastly, as a quality check, we generated NB images from the continuum-subtracted cubes, averaged over the four control windows mentioned above.

Because the MUSE observations of the MUDF survey are obtained in the Laser assisted Adaptive Optics (AO) mode, the wavelength range of 5760--6010\,\AA\ is excluded by the Na notch filter that masks the laser emission. Therefore, for the stacking of the \mgii\ emission from MUDF galaxies, we exclude the redshift range, $z=1.05-1.15$. In addition, while inspecting the 1D spectra of the galaxies, we noticed that the \mgii\ wavelength range in a few cases is affected by strong spikes that could be due to sky subtraction residuals, instrumental artefacts or contamination from Raman lines caused by the Na laser during the AO observations. These strong spikes at $<7000$\,\AA\ are also seen in the stacked sky spectra obtained from the full MUDF cube. Therefore, we excluded in total 26 galaxies in the final \mgii\ stack. In the case of the MAGG survey, we checked that removing the corresponding redshift ranges for the \mgii\ stack does not lead to any difference in the final results. The spikes noted in the MUDF stacked sky spectra are not found in the stacked sky spectra of the MAGG survey. This is reasonable since the MAGG survey comprises MUSE observations of 28 different fields taken at different times, and thus any systematics due to sky or instrument should be averaged out by the stacking.

The wavelength range used for stacking the \oii\ emission in this work, particularly at $z>1$, is affected by sky lines. We checked that excluding some of the wavelength ranges that are most affected by sky lines does not change the results for the \oii\ stacked emission. Since we stack in the rest-frame wavelength, that should mitigate any systematic effects of sky subtraction residuals on average. Finally, to check that the stacked line emission is not dominated by individual sources, we performed the 3D stacking after removing at random a few of the galaxies that show strong line (\mgii, \oii) emission in the 1D spectra. This exercise was not found to affect the results. In any case, this would be accounted for in the uncertainties from the bootstrapping analysis, and the results based on median-stacking should be robust against individual outliers. 

In Section~\ref{sec:results_stack}, we present the NB images and SB profiles obtained from the 3D stacking of the MUSE cubes for \mgii\ and \oii\ emission for the full sample and also for different redshift, stellar mass and environment sub-samples. The NB images have been smoothed using a Gaussian kernel of standard deviation 0.3 arcsecond for display purpose, but the SB profiles and size estimates are obtained from the unsmoothed NB images. We present in most cases the stacking results using median statistics. These are found to be overall consistent with the results using mean statistics within the $1\sigma$ uncertainties. The SB profiles shown here are the observed ones, except for when we compare the SB profiles at different redshifts, in which case we correct the profiles for cosmological SB dimming using the median redshift of the underlying sub-sample. We tested that the results obtained using this method are consistent within $1\sigma$ uncertainties with that obtained if we instead correct the individual sub-cubes for cosmological SB dimming before stacking.

\begin{figure}
 \includegraphics[width=0.48\textwidth]{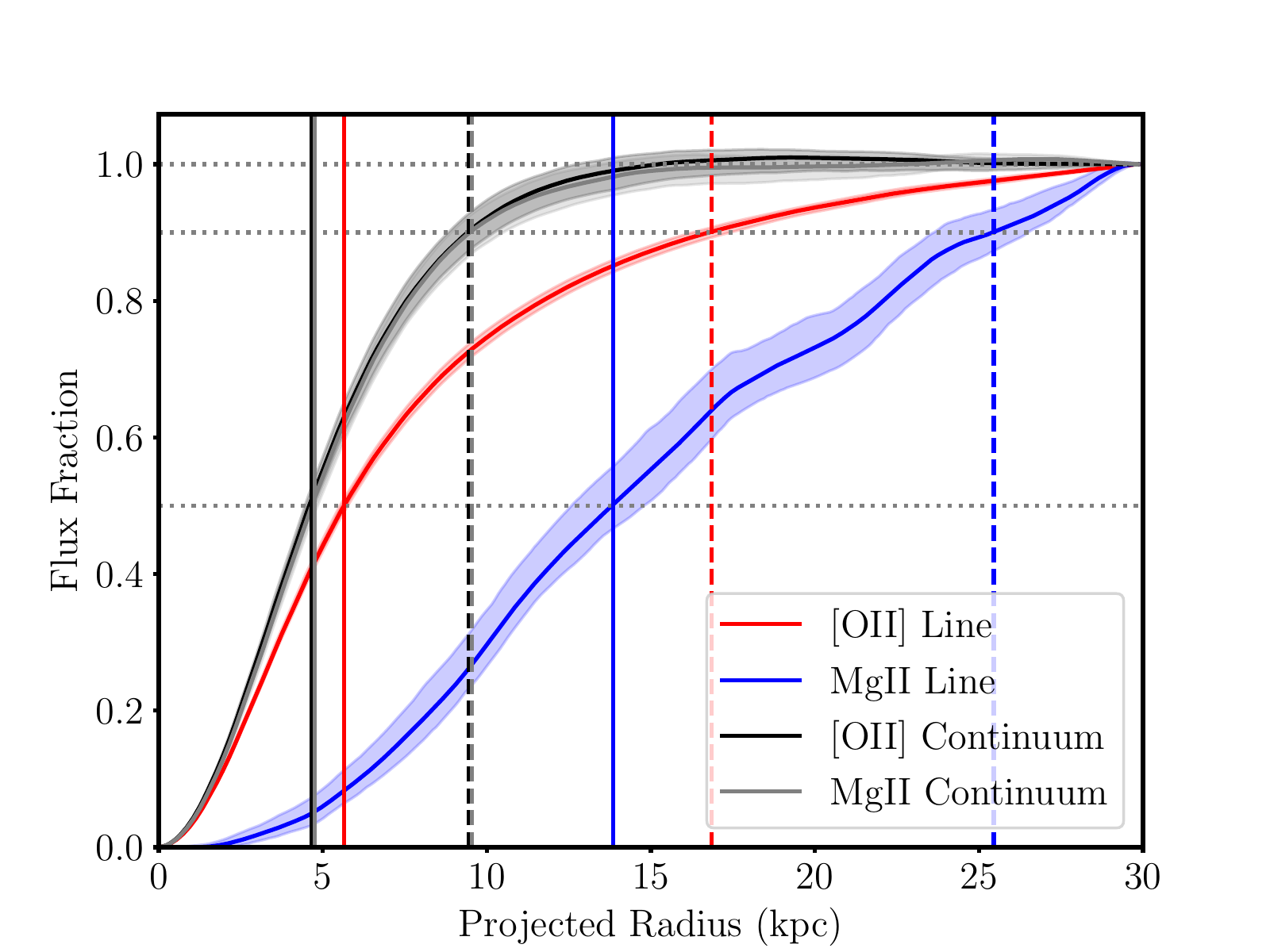}
 \caption{The flux curve-of-growth normalised to the total flux within 30\,kpc --- in blue for \mgii\ line emission, in red for \oii\ line emission, in grey for \mgii\ continuum emission, and in black for \oii\ continuum emission. The shaded regions represent the $1\sigma$ uncertainties from bootstrapping analysis. The vertical solid and dashed lines indicate the radii, $R_{50}$ and $R_{90}$, of the circular apertures which contain 50 and 90 per cent of the total flux within a circular aperture of 30\,kpc, respectively. The color coding for the vertical lines is the same as for the flux fraction curves. The horizontal dotted lines indicate flux fractions of 0.5 and 0.9 and 1.0.
 }
 \label{fig:cumprofile_all}
\end{figure}
\begin{table*}
\caption{Estimates of the size of the emission in different samples. $R_{50}$ and $R_{90}$ are the radii in kpc of the circular apertures which contain 50 and 90 per cent of the total flux within a circular aperture of 30\,kpc. The errors are the 1$\sigma$ uncertainties from bootstrapping analysis.}
\centering
\begin{tabular}{ccccccccc}
\hline
 & \multicolumn{4}{c}{$R_{50}$ (kpc)} & \multicolumn{4}{c}{$R_{90}$ (kpc)} \\
Sample & \mgii\ Line & \mgii\ Continuum & \oii\ Line & \oii\ Continuum & \mgii\ Line & \mgii\ Continuum & \oii\ Line & \oii\ Continuum \\
\hline
Full                             & $13.8^{+1.0}_{-1.2}$ & $4.7^{+0.2}_{-0.2}$ & $5.6^{+0.2}_{-0.2}$ & $4.6^{+0.1}_{-0.1}$ & $25.4^{+1.1}_{-1.8}$ & $9.5^{+1.0}_{-0.6}$  & $16.8^{+0.6}_{-0.6}$ & $9.4^{+0.8}_{-0.6}$  \\
$z\le1$                          & $12.9^{+2.9}_{-1.9}$ & $4.7^{+0.2}_{-0.2}$ & $5.6^{+0.2}_{-0.2}$ & $4.6^{+0.2}_{-0.2}$ & $24.2^{+3.6}_{-1.4}$ & $9.3^{+1.3}_{-0.9}$  & $16.8^{+0.8}_{-0.8}$ & $9.6^{+1.0}_{-0.7}$  \\
$z>1$                            & $13.9^{+1.2}_{-1.2}$ & $4.7^{+0.2}_{-0.2}$ & $5.7^{+0.2}_{-0.2}$ & $4.7^{+0.2}_{-0.2}$ & $25.4^{+1.4}_{-2.2}$ & $10.1^{+1.2}_{-1.1}$ & $16.9^{+0.7}_{-0.7}$ & $9.6^{+1.3}_{-1.2}$  \\
$7\le$log\,\mstar\,(\msun)$<8$   & $2.5^{+7.0}_{-1.0}$  & $4.1^{+1.3}_{-1.0}$ & $3.7^{+0.3}_{-0.3}$ & $2.9^{+1.0}_{-1.1}$ & $4.0^{+22.0}_{-2.0}$ & $8.3^{+2.0}_{-17.0}$ & $8.2^{+1.7}_{-1.6}$  & $8.0^{+2.0}_{-19.0}$  \\
$8\le$log\,\mstar\,(\msun)$<9$   & $8.8^{+1.4}_{-0.9}$  & $4.4^{+0.3}_{-0.3}$ & $4.8^{+0.2}_{-0.2}$ & $3.7^{+0.3}_{-0.4}$ & $22.9^{+3.9}_{-7.2}$ & $9.0^{+11.0}_{-2.0}$ & $13.9^{+1.6}_{-1.7}$ & $9.0^{+2.3}_{-1.3}$  \\
$9\le$log\,\mstar\,(\msun)$<10$  & $14.1^{+0.9}_{-1.4}$ & $4.5^{+0.3}_{-0.3}$ & $5.7^{+0.2}_{-0.2}$ & $4.4^{+0.3}_{-0.2}$ & $25.3^{+1.0}_{-3.2}$ & $9.0^{+1.1}_{-0.9}$  & $16.8^{+0.6}_{-0.7}$ & $9.1^{+2.3}_{-1.3}$  \\
$10\le$log\,\mstar\,(\msun)$<12$ & ---$^a$              & $5.5^{+0.3}_{-0.3}$ & $7.1^{+0.3}_{-0.2}$ & $5.2^{+0.2}_{-0.2}$ & ---$^a$              & $11.0^{+1.1}_{-0.9}$ & $19.2^{+0.8}_{-0.8}$ & $11.1^{+0.6}_{-0.6}$ \\
Group                            & $14.5^{+2.2}_{-2.4}$ & $4.9^{+0.4}_{-0.3}$ & $5.9^{+0.3}_{-0.2}$ & $4.8^{+0.3}_{-0.3}$ & $27.0^{+1.0}_{-1.0}$ & $10.8^{+2.3}_{-1.8}$ & $19.3^{+0.8}_{-0.9}$ & $11.7^{+3.3}_{-2.2}$ \\
Single                           & $13.5^{+1.3}_{-1.4}$ & $4.6^{+0.3}_{-0.2}$ & $5.6^{+0.3}_{-0.2}$ & $4.9^{+0.3}_{-0.5}$ & $25.1^{+2.0}_{-4.0}$ & $8.9^{+1.7}_{-1.0}$  & $15.9^{+1.2}_{-0.7}$ & $10.3^{+2.4}_{-2.4}$ \\
\hline
\end{tabular}
\label{tab:sizes_table}
\begin{flushleft}
$^a$ \mgii\ is detected in absorption in the central region in this stellar mass bin
\end{flushleft}
\end{table*}
\begin{table}
\caption{Median values of the metal line surface brightness in units of $10^{-20}$\,\ergscmarc\ within a circular aperture of radius 30\,kpc for different samples. The errors are the 1$\sigma$ uncertainties from bootstrapping analysis.}
\centering 
\begin{tabular}{ccc}
\hline
Sample & \mgii\ & \oii\ \\
\hline
Full                             &  1.93$_{-0.28}^{+0.18}$ &   6.27$_{ -0.55}^{+ 0.41}$ \\
$z\le1^a$                          & 18.13$_{-3.20}^{+5.27}$ &  77.95$_{ -9.44}^{+ 7.86}$ \\
$z>1^a$                            & 53.34$_{-5.55}^{+6.91}$ & 173.14$_{-21.17}^{+13.21}$ \\
$7\le$log\,\mstar\,(\msun)$<8$   &  0.59$_{-0.55}^{+0.71}$ &   3.64$_{ -1.07}^{+ 1.62}$ \\
$8\le$log\,\mstar\,(\msun)$<9$   &  1.44$_{-0.35}^{+0.28}$ &   4.06$_{ -0.56}^{+ 0.95}$ \\
$9\le$log\,\mstar\,(\msun)$<10$  &  2.85$_{-0.47}^{+0.38}$ &   8.96$_{ -0.75}^{+ 0.94}$ \\
$10\le$log\,\mstar\,(\msun)$<12$ &  1.70$_{-0.57}^{+0.56}$ &  13.53$_{ -1.18}^{+ 1.27}$ \\
Group                            &  2.60$_{-0.50}^{+0.47}$ &   9.47$_{ -0.91}^{+ 0.94}$ \\
Single                           &  1.85$_{-0.43}^{+0.42}$ &   7.71$_{ -1.04}^{+ 0.88}$ \\
\hline
\end{tabular}
\label{tab:sb_table}
\begin{flushleft}
$^a$ corrected for cosmological surface brightness dimming
\end{flushleft}
\end{table}

\section{Results}
\label{sec:results}

\begin{figure*}
 \includegraphics[width=1.0\textwidth]{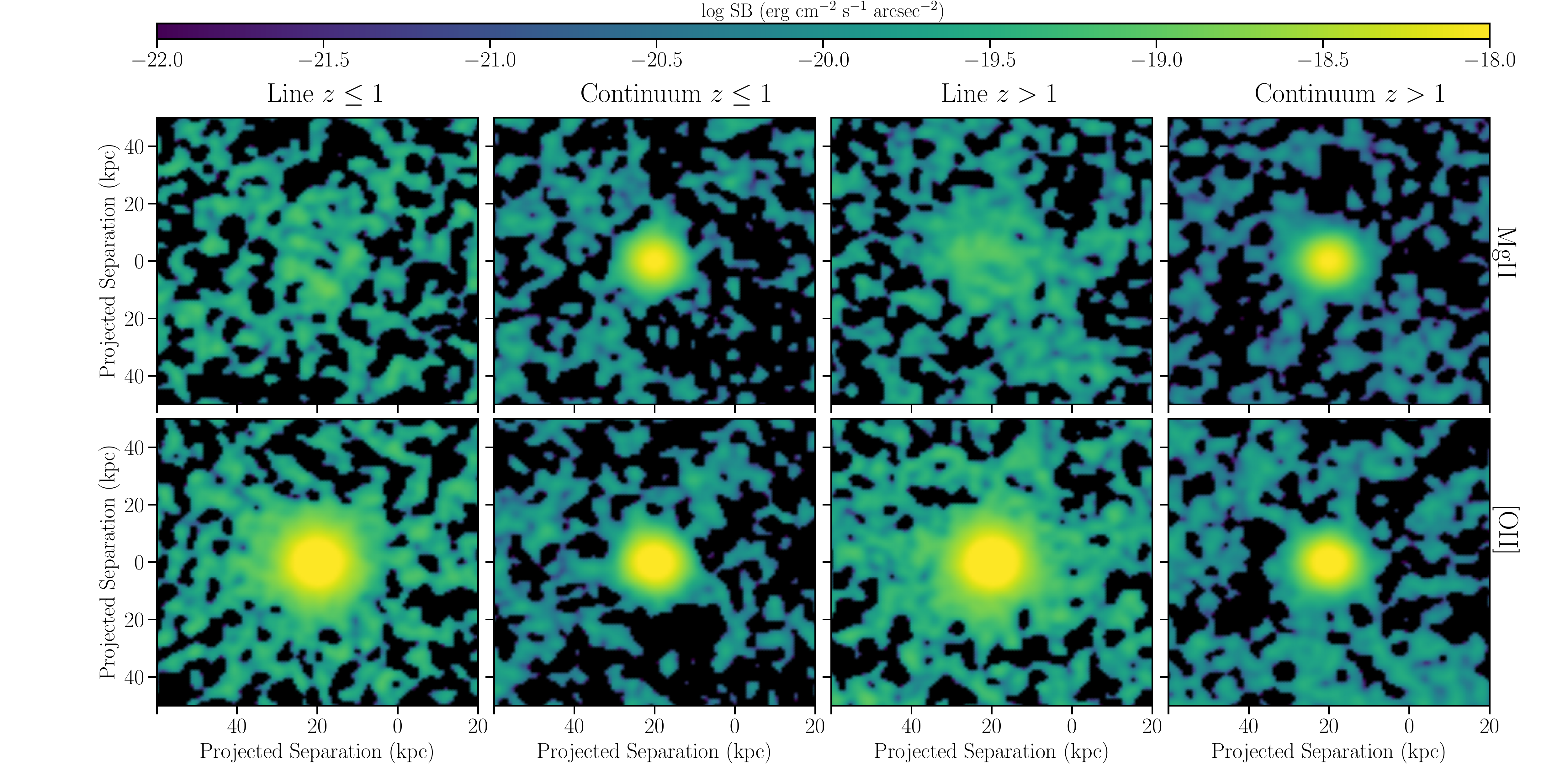}
 \caption{Median-stacked pseudo-NB images at different redshift bins. The top panels show the results for \mgii\ and the bottom panels show the results for \oii. From left to right, the panels show the stacked line emission after subtracting the continuum at $z\le1$, the stacked continuum emission in control windows at $z\le1$, the stacked line emission after subtracting the continuum at $z>1$, and the stacked continuum emission in control windows at $z>1$.
 }
 \label{fig:nbimg_z}
\end{figure*}
\begin{figure*}
 \includegraphics[width=0.48\textwidth]{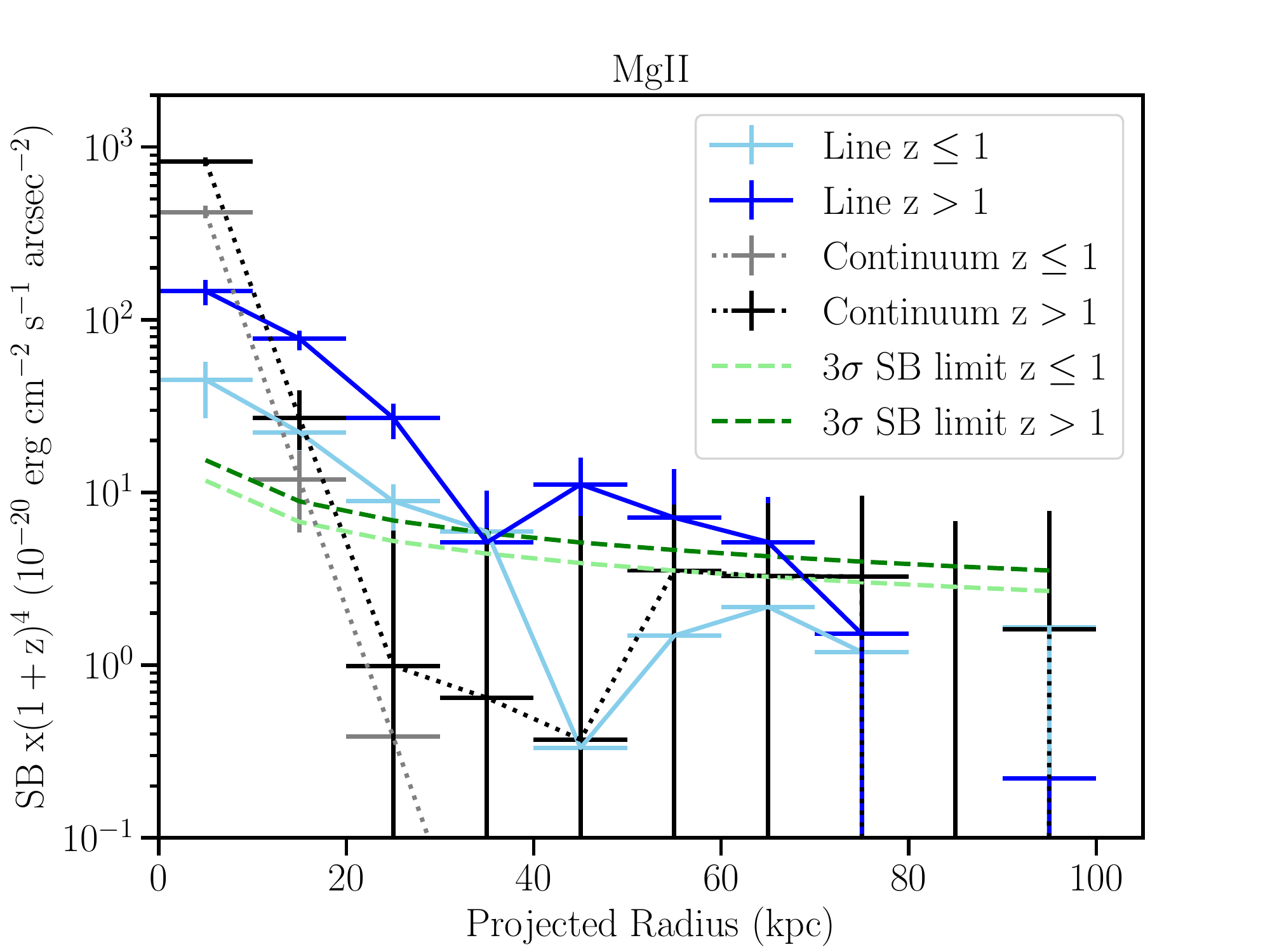}
 \includegraphics[width=0.48\textwidth]{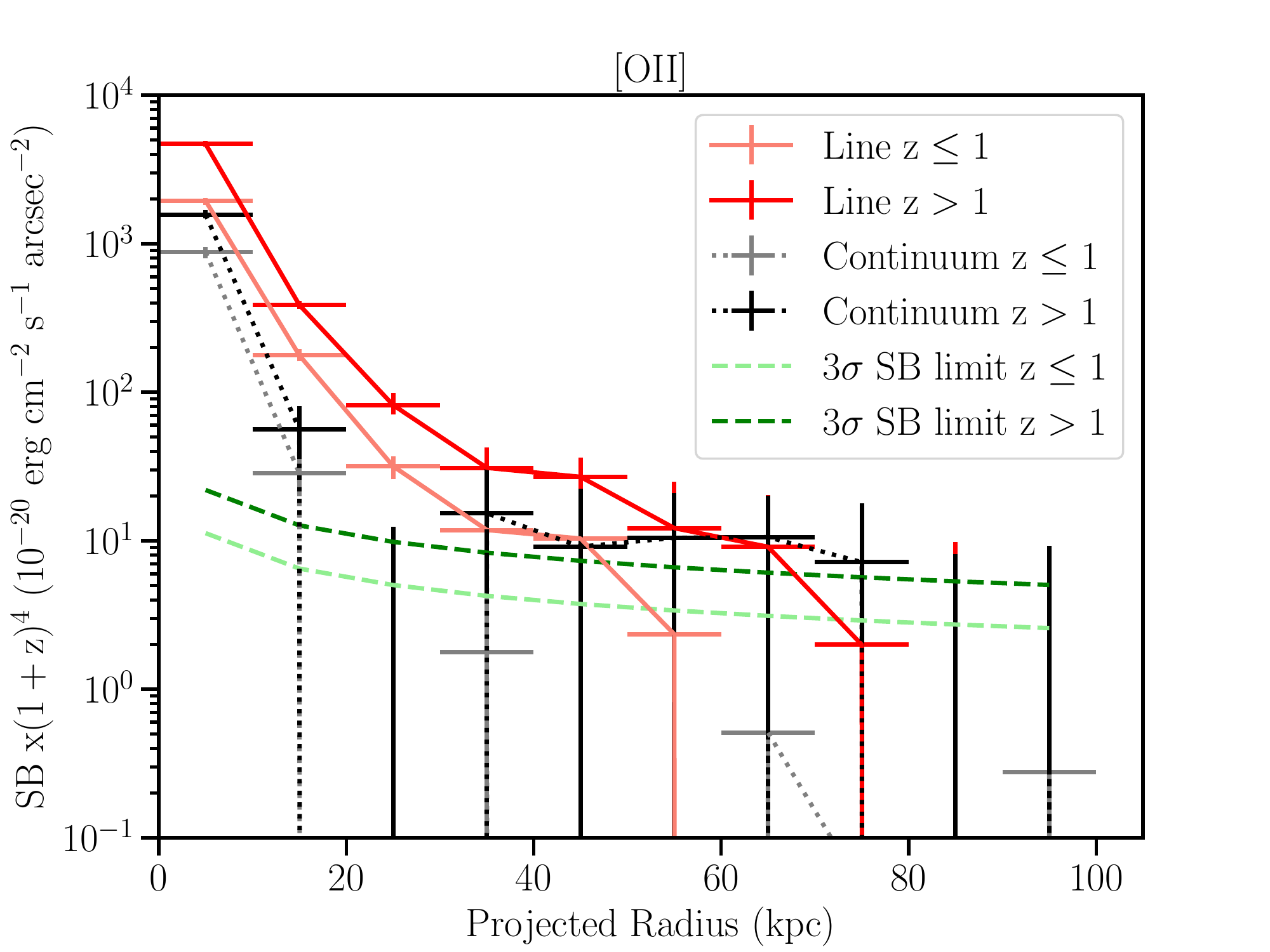}
 \caption{{\it Left:} The median-stacked azimuthally-averaged radial SB profiles of \mgii\ emission (corrected for cosmological SB dimming) in circular annuli of 10\,kpc from the galaxy centres at $z\le1$ and $z>1$ are shown as light blue and dark blue solid lines, respectively. The SB profiles of the control continuum emission at $z\le1$ and $z>1$ are shown as grey and black dotted lines, respectively. The $3\sigma$ SB limits for the $z\le1$ and $z>1$ profiles are shown as light and dark green dashed lines, respectively.
 {\it Right:} Same as in the left plot for \oii\ emission. The \oii\ SB profiles at $z\le1$ and $z>1$ are shown as pink and red solid lines, respectively. The error bars represent the $1\sigma$ uncertainties from bootstrapping analysis.
 }
 \label{fig:sbprofile_z}
\end{figure*}

In Section~\ref{sec:results_stack} we investigate the average metal line emission around galaxies in our sample through stacking. Then, in Section~\ref{sec:results_individual}, we investigate whether there is extended metal line emission in a sub-sample of individual galaxy groups.

\subsection{Extended emission in stacked images}
\label{sec:results_stack}

Fig.~\ref{fig:nbimg_all} shows the reconstructed NB images of \mgii\ (top panel) and \oii\ (bottom panel) line emission obtained from median stacking the MUSE cubes of the full galaxy sample. For comparison, the NB images of the continuum emission and of the continuum-subtracted emission in control windows are also shown. The 3$\sigma$ SB limit for a 1\,arcsec$^2$ aperture and a 1000\,\kms\ velocity window is $10^{-20}$\,\ergscmarc. Both \mgii\ and \oii\ line emission are clearly detected, and they appear to be more extended than the average continuum emission. To quantify the radial profile of the emission, we show in Fig.~\ref{fig:sbprofile_all} the azimuthally-averaged SB profiles in circular annuli of 10\,kpc from the centre for \mgii\ (left) and \oii\ (right). We show the profiles obtained using both mean and median statistics. The mean stacking typically gives slightly larger values of SB than the median stacking, which is expected if the distribution is not symmetric around the median, however the profiles are overall consistent within the $1\sigma$ uncertainty. 

The \mgii\ and \oii\ line emission are detected out to $\approx30-50$\,kpc from the centre, beyond which the SB drops below the $3\sigma$ limit ($\approx2\times10^{-21}$\,\ergscmarc)\footnote{The limits on the average SB profiles (dashed green lines in Fig.~\ref{fig:sbprofile_all}) are estimated based on the noise per pixel in the outer annuli and the number of pixels in each annuli.}. On the other hand, the continuum emission is detected out to $\approx15$\,kpc from the centre. The \mgii\ line emission is more radially extended compared to the continuum emission over $\approx15-30$\,kpc, while the \oii\ line emission is more radially extended over $\approx15-45$\,kpc. At a fixed SB level of $10^{-20}$\,\ergscmarc, the mean-(median-)stacked \mgii\ emission has a radial extent of $\approx30(25)$\,kpc or $\approx2$ times larger than that the radial extent of the continuum emission, while the radial extent of the mean-(median-)stacked \oii\ emission is $\approx50(45)$\,kpc or $\approx3$ times larger than that of the continuum emission. This can be considered as the average extent of the metal-emitting halo gas at the observed SB level. We additionally note that the line emission is more extended than the typical PSF of the MUSE data (full-width at half-maximum $\lesssim$0.6\,arcsec or $\lesssim$5\,kpc at $z=1$).

To compare the sizes of the emission, we estimate the flux curve-of-growth normalised to the total flux within 30\,kpc as shown in Fig.~\ref{fig:cumprofile_all}. The estimates of $R_{50}$ and $R_{90}$ for \mgii, \oii\ and the continuum emission are provided in Table~\ref{tab:sizes_table}. The ratios of $R_{50}$ and $R_{90}$ of \mgii\ line to continuum emission are $\approx$2.9 and $\approx$2.7, respectively, while the corresponding ratios of \oii\ line to continuum emission are $\approx$1.2 and $\approx$1.8, respectively. The ratios of $R_{50}$ and $R_{90}$ of \mgii\ to \oii\ line emission are $\approx$2.5 and $\approx$1.5. From these ratios and Fig.~\ref{fig:cumprofile_all}, it is evident that the \oii\ emission follows the continuum emission in the inner region ($\lesssim$5\,kpc), but it becomes more extended than the continuum with increasing radius. The \mgii\ emission is clearly more extended than the continuum and the \oii\ emission, with a shallower profile starting from the inner region. 

If we consider that the central circular region of radius 15\,kpc traces the stellar disc, and that the circular annulus between radii 15 and 30\,kpc traces the halo, then we find that the total luminosity of the \oii\ emission at $z\approx1$ from the disc ($\approx2\times10^{41}$\,\ergs) is an order of magnitude higher than that from the halo ($\approx2\times10^{40}$\,\ergs). On the other hand, the total \mgii\ luminosity at $z\approx1$ in the central 15\,kpc disc ($\approx7\times10^{39}$\,\ergs) is comparable to that from the halo ($\approx6\times10^{39}$\,\ergs). We discuss and compare the \mgii\ and \oii\ emission further in Section~\ref{sec:discussion_mgii_oii}. In the following sections, we study the evolution of the average metal line emission from galaxy haloes with redshift, stellar mass and environment, and the dependence of the emission on orientation. The average metal line SB within a circular aperture of radius 30\,kpc in different sub-samples are listed in Table~\ref{tab:sb_table}.

\subsubsection{Evolution with redshift}
\label{sec:results_stack_redshift}

In order to check how the average metal line emission evolves with redshift, we divide the full sample into two redshift bins: $0.7 \le z \le 1.0$ (259 galaxies, median $z\approx 0.82$, median \mstar\ $\approx10^{9}$\,\msun) and $1.0 < z \le 1.5$ (342 galaxies, median $z\approx 1.22$, median \mstar\ $\approx2\times10^{9}$\,\msun). In Fig.~\ref{fig:nbimg_z}, we show the median-stacked pseudo-NB images of \mgii\ (top panel) and \oii\ (bottom panel) line emission in these two redshift ranges along with that of the corresponding control continuum emission. \mgii\ and \oii\ line emission are detected in both redshift ranges, with the emission at $z>1$ being more extended. In Fig.~\ref{fig:sbprofile_z}, we show the azimuthally-averaged SB profiles that have been corrected for cosmological dimming using the median redshift of the above two redshift bins for \mgii\ (left) and \oii\ (right). 

Firstly, the cosmological-corrected metal SB profiles are brighter at the higher redshift range. The \mgii\ SB profile at $z>1$ is $\approx3$ times brighter than that at $z\le1$ within a circular aperture of radius 30\,kpc, while the \oii\ SB profile at $z>1$ is $\approx2$ times brighter than that at $z\le1$ within the same aperture (see Table~\ref{tab:sb_table}). Secondly, the metal line SB profiles at both redshift ranges are more radially extended compared to the continuum emission SB profiles, being brighter than the continuum emission over $\approx15-30$\,kpc. Lastly, the metal line SB profiles at $z>1$ are also more radially extended compared to those at $z\le1$. At a (corrected) SB level of $10^{-19}$\,\ergscmarc, the \mgii\ profile at $z\le1$ has an extent of $\approx25$\,kpc, whereas the \mgii\ profile at $z>1$ has an extent of $\approx33$\,kpc. Similarly, at a (corrected) SB level of $5\times10^{-19}$\,\ergscmarc, the extent of the \oii\ profile at $z\le1$ is $\approx24$\,kpc, whereas it is $\approx32$\,kpc at $z>1$. From Table~\ref{tab:sizes_table}, it can be seen that, similar to the full sample, the metal line emission have shallower profiles than the continuum emission, and the \oii\ line emission is more centrally concentrated than the \mgii\ emission in both the redshift bins. Despite the $R_{50}$ and $R_{90}$ estimates of the metal line emission at $z\le1$ and at $z>1$ being consistent within the uncertainties, there is a hint of both \mgii\ and \oii\ emission being more extended at higher redshift. 

As higher redshift samples are naturally more dominated by massive systems, we investigate to what extent the above evolution is driven by redshift and is independent of stellar mass. To do so, we formed two samples at $z\le1$ and at $z>1$ that are matched in the base 10 logarithm of the stellar mass within $\pm0.3$\,dex. This led to \mstar-matched samples consisting of 183 and 190 galaxies for \mgii\ and \oii, respectively, in each of the two redshift bins. Based on a two-sided KS test, the maximum difference between the cumulative distributions of the stellar mass in the two redshift bins is $\approx$0.05 and the $p$-value is $\approx$0.98 and $\approx$0.95 for \mgii\ and \oii, respectively. On comparing the SB profiles (corrected for cosmological dimming) in the two redshift sub-samples that are matched in stellar mass, we find that, although the uncertainties are larger due to the smaller sample sizes in this case, the metal line SB profiles are $\approx2-3\times$ brighter and more extended at $z>1$ than at $z\le1$. This is similar to what we find for the full sample, indicating that there is indeed likely to be an evolution in the average metal line emission with redshift that is independent of stellar mass.

This evolution could arise from the higher SFR of galaxies, larger ionizing radiation or cool gas density at higher redshifts. Indeed, the average SFR, obtained from SPS fitting (Section~\ref{sec:data}), of the $z>1$ galaxy sample is $\approx$2.5 times larger than that of the $z\le1$ sample, even after matching the two samples in stellar mass. To check for the dependence of the redshift evolution on the SFR, we performed a second control analysis where we formed two samples at $z\le1$ and at $z>1$ that are matched in both the stellar mass and SFR within $\pm0.3$\,dex in the base 10 logarithm values. We thus selected 132 and 143 galaxies for \mgii\ and \oii, respectively, in each of the two redshift bins. The maximum differences between the cumulative distributions of the stellar mass and SFR in the two redshift bins are $\lesssim$0.05 and the $p$ values are $\gtrsim$0.96 from two-sided KS tests. We find that the difference between the metal SB in the two redshift sub-samples reduces when SFR is controlled for. The cosmologically-corrected metal SB profiles at $z>1$ are on an average $\approx1.4\times$ brighter than at $z\le1$, however they are consistent within the 1$\sigma$ uncertainties. This indicates that the redshift evolution of the metal line emission could be predominantly driven by the higher SFR of galaxies at $z>1$, although there could be intrinsic evolution or additional factors at play that need to be verified with a larger sample.

\citet{zhang2016} stacked the Sloan Digital Sky Survey (SDSS) 1D spectra of background galaxies to statistically map the \ha\ + \nii\ emission around foreground galaxies at $z<0.2$. They detected significant line emission out to a projected radius of 100\,kpc. The mean value of the emission within 50\,kpc is $(2.19 \pm 0.62) \times 10^{-20}$\,\ergscm\AA$^{-1}$, which translates to $(3.7 \pm 1.0) \times10^{-20}$\,\ergscmarc\ considering their typical emission line width of 12\,\AA\ and SDSS fibre area of 7\,arcsec$^2$. After correcting for the cosmological SB dimming at $z\approx0.1$, and assuming a typical \oii/\ha\ ratio of 0.62 (not corrected for extinction) at these redshifts \citep{mouchine2005}, we obtain the average \oii\ SB of $(3.4 \pm 1.0) \times10^{-20}$\,\ergscmarc. Comparing these to the cosmologically-corrected average \oii\ SB we obtain in this work (Table~\ref{tab:sb_table}), there is a factor of $\approx$20 increase in the SB going from $z<0.2$ to $0.7 \le z \le 1.0$. However, we note that there are certain caveats to the above comparison. The low redshift result is based on stacking of 1D spectra to statistically map the line emission, whereas this work utilises 3D stacking to directly map the emission around galaxies. In addition to the different techniques, the sample selection differs, with the low redshift sample comprising of luminous galaxies ($10^{10-11}$\,L$_\odot$), and our sample comprising of \mstar\ $\approx10^{6-12}$\,\msun\ galaxies. The conversion of the low redshift result from flux density to SB would be wrong if the emission is concentrated in structures smaller than the SDSS fibre size. Finally, although we adopt an average value of the \oii/\ha\ ratio, it shows a large scatter, and correlates with metallicity and ionization state \citep{mouchine2005}.

\begin{figure*}
 \includegraphics[width=1.0\textwidth]{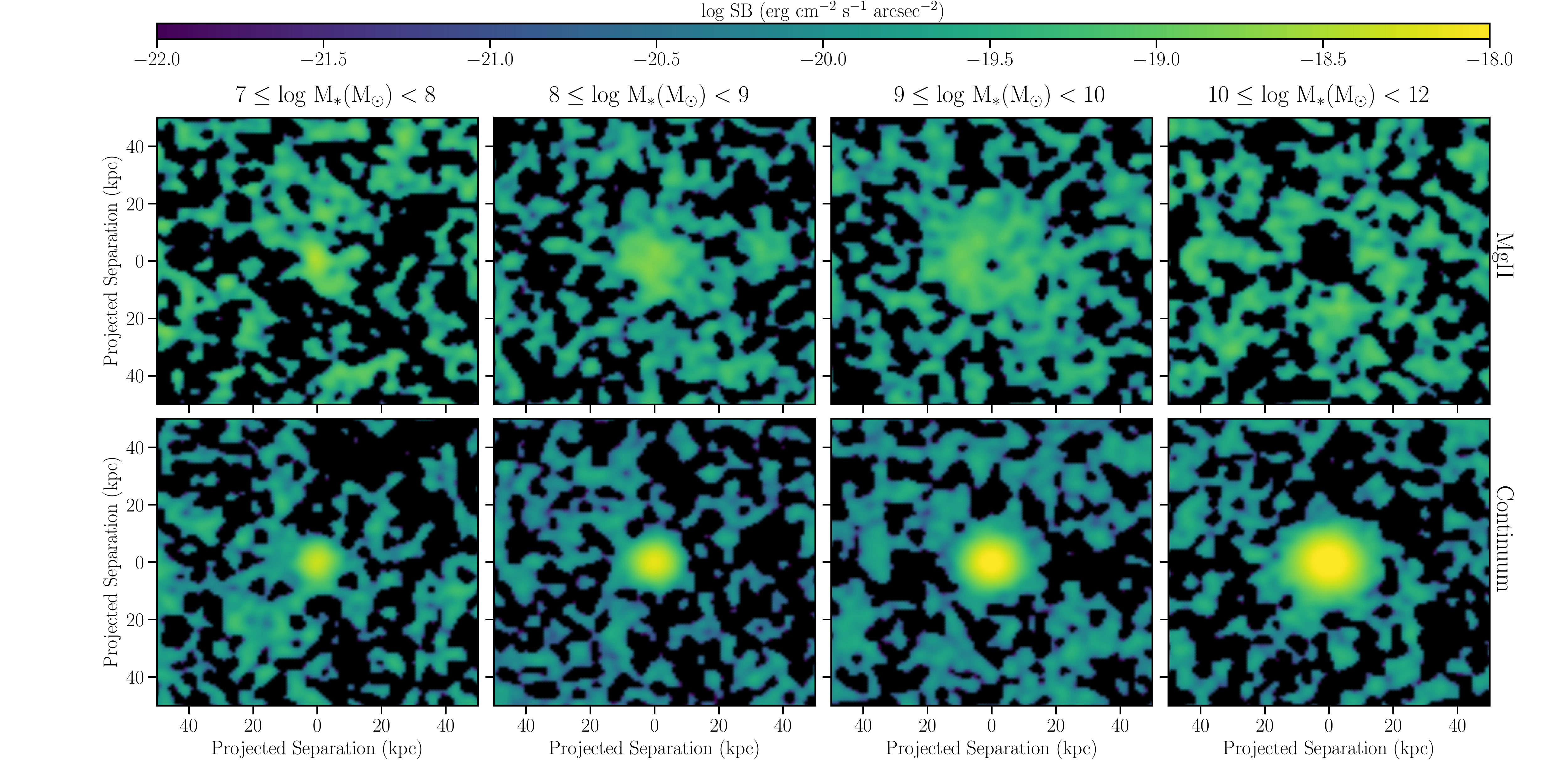}
 \caption{Median-stacked pseudo-NB images at different stellar mass bins. The top panels show the results for stacked \mgii\ line emission after subtracting the continuum for different stellar mass bins as marked in the panels. The bottom panels show the corresponding stacked continuum emission in control windows for the different stellar mass bins. The central region in the \mgii\ NB image of the highest stellar mass bin is dominated by absorption.
 }
 \label{fig:nbimg_mass_mgii}
\end{figure*}
\begin{figure*}
 \includegraphics[width=1.0\textwidth]{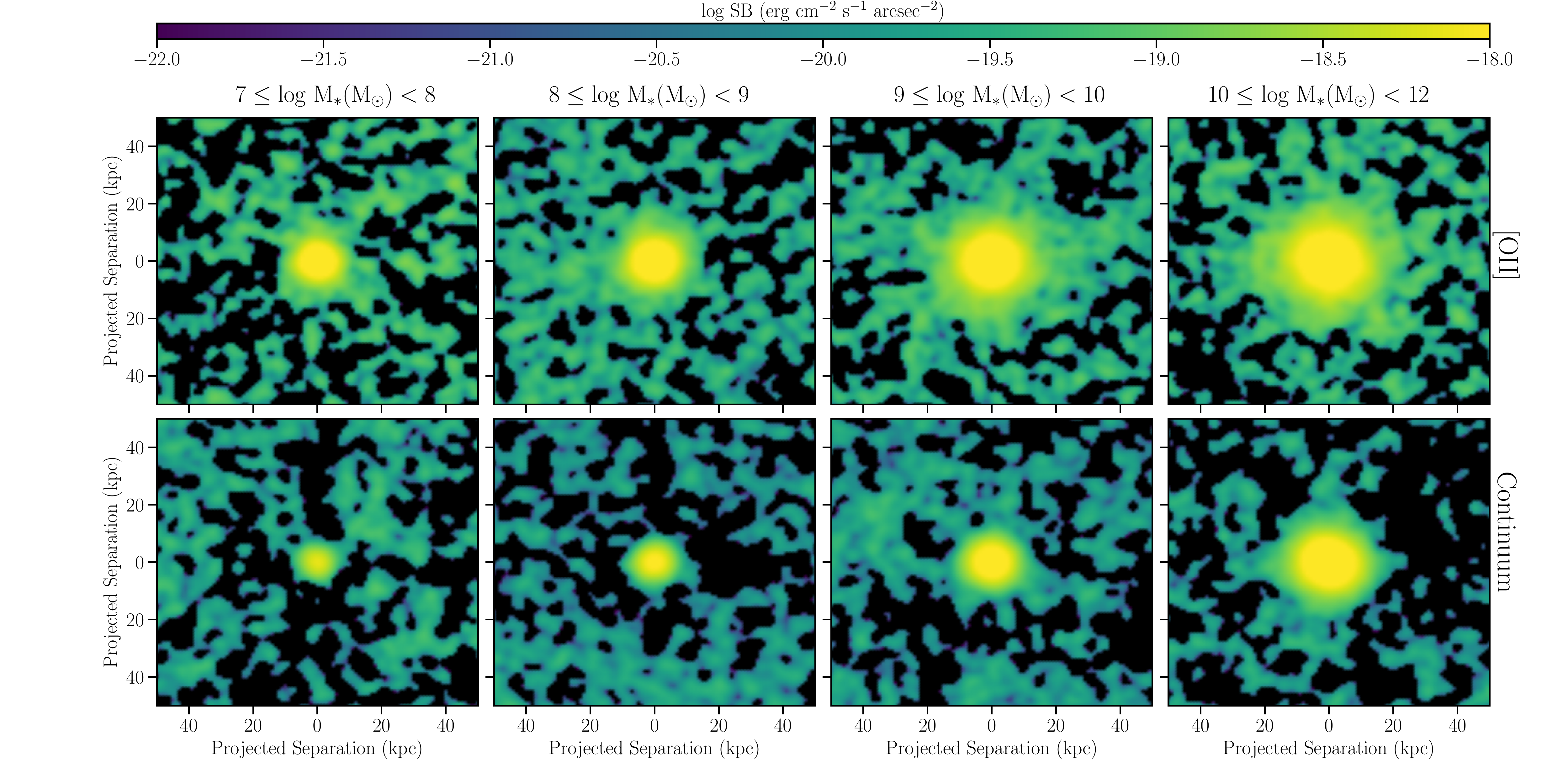}
 \caption{Same as in Fig.~\ref{fig:nbimg_mass_mgii} for \oii\ emission.}
 \label{fig:nbimg_mass_oii}
\end{figure*}
\begin{figure*}
 \includegraphics[width=0.48\textwidth]{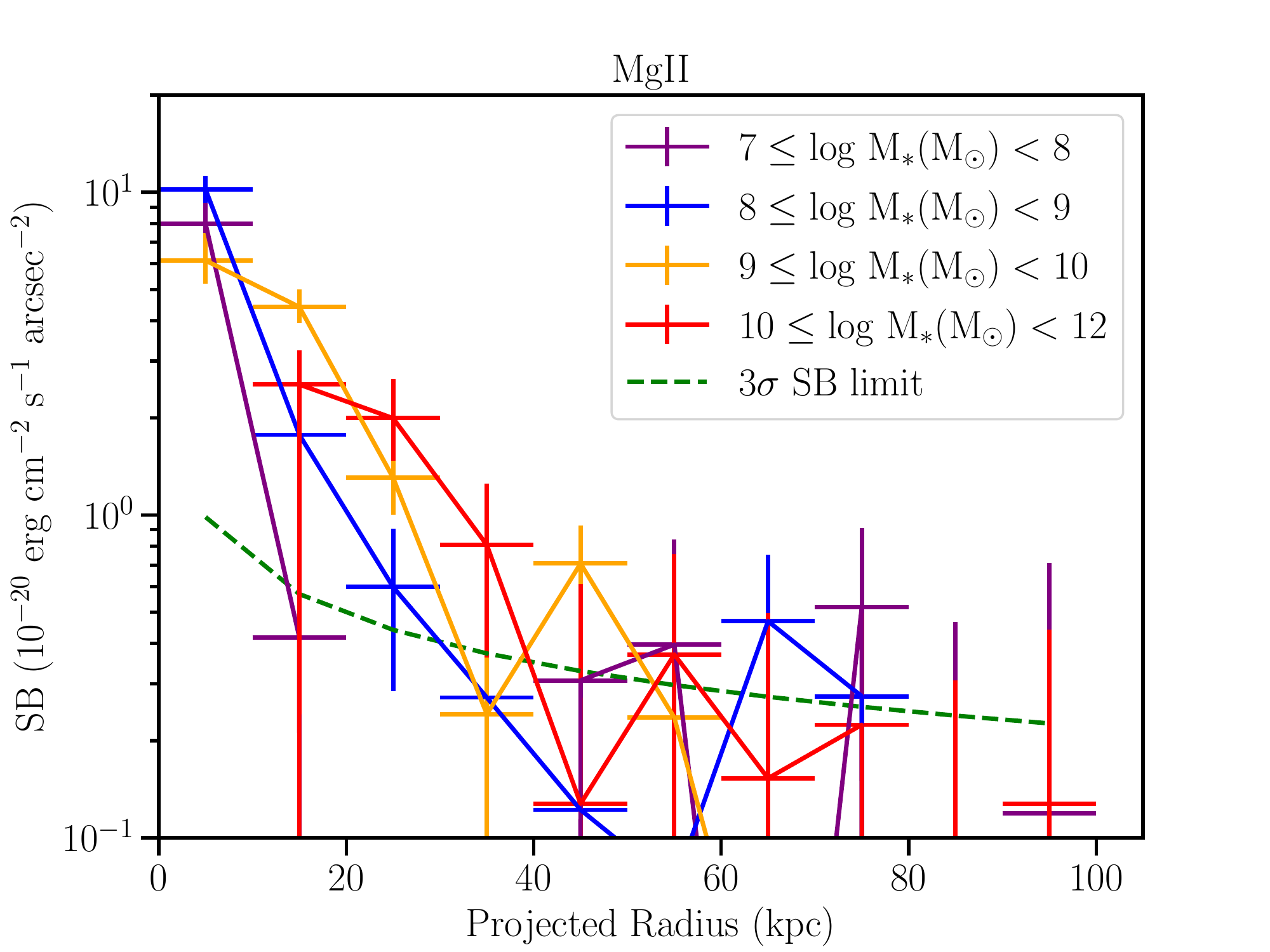}
 \includegraphics[width=0.48\textwidth]{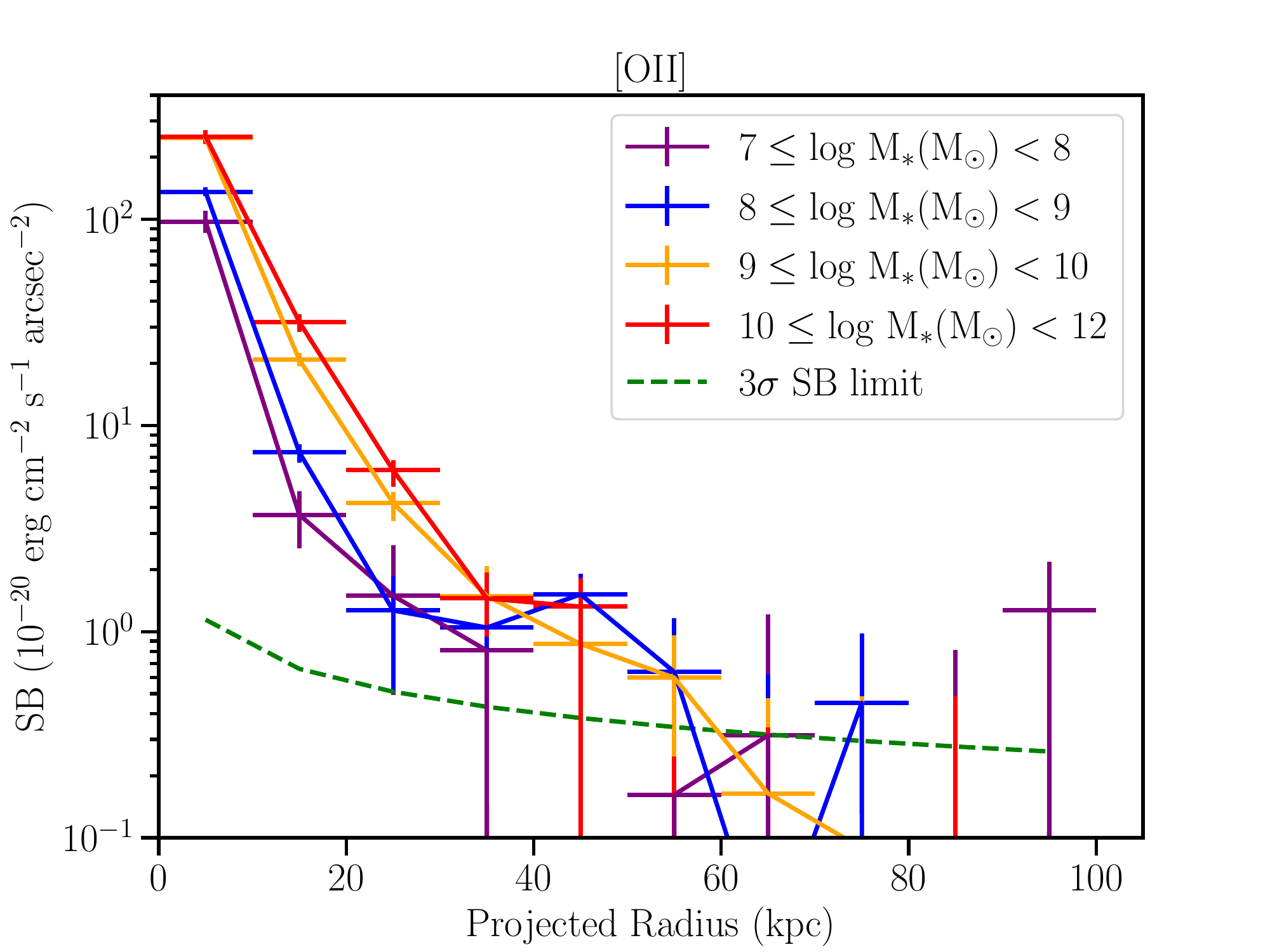}
 \caption{{\it Left:} The median-stacked azimuthally-averaged radial SB profiles of \mgii\ emission in circular annuli of 10\,kpc from the galaxy centres for different stellar mass bins are shown as marked in the figure. The error bars represent the $1\sigma$ uncertainties from bootstrapping analysis. For clarity, only the $3\sigma$ SB limit for the stellar mass bin of log\,\mstar(\msun) = 8--9 is shown as green dashed line. {\it Right:} Same as in the left plot for \oii\ emission. 
 }
 \label{fig:sbprofile_mass}
\end{figure*}

\subsubsection{Evolution with stellar mass}
\label{sec:results_stack_mass}

To study the evolution of the metal line emission as a function of stellar mass, at first we divide the full sample into four different stellar mass bins: \mstar\ = $10^{7-8}$\,\msun\ (48 galaxies, median \mstar\ $\approx5\times10^{7}$\,\msun, median $z\approx0.8$), $10^{8-9}$\,\msun\ (190 galaxies, median \mstar\ $\approx4\times10^{8}$\,\msun, median $z\approx1.0$), $10^{9-10}$\,\msun\ (229 galaxies, median \mstar\ $\approx2\times10^{9}$\,\msun, median $z\approx1.1$) and $10^{10-12}$\,\msun\ (131 galaxies, median \mstar\ $\approx2\times10^{10}$\,\msun, median $z\approx1.0$). The pseudo-NB images of \mgii\ line emission obtained from median stacking the MUSE cubes in the above four stellar mass bins are shown in the top panel of Fig.~\ref{fig:nbimg_mass_mgii}, while those of the control continuum emission are shown in the bottom panel. Fig.~\ref{fig:nbimg_mass_oii} shows the same for \oii\ emission. For both \mgii\ and \oii, the control continuum emission becomes brighter and more extended with increasing stellar mass. The line emission also becomes brighter and more extended as the stellar mass increases, however the central region is dominated by \mgii\ absorption in the highest stellar mass bin.

Fig.~\ref{fig:sbprofile_mass} shows the SB profiles in the four different stellar mass bins for \mgii\ (left) and \oii\ (right) line emission. We do not show the corresponding SB profiles for the continuum emission in these plots for clarity, however we discuss here how the SB profiles of line and continuum emission compare. For the lowest mass bin ($10^{7-8}$\,\msun), the \mgii\ SB profile is not more extended than that of the continuum emission, as can be also seen from Fig.~\ref{fig:nbimg_mass_mgii}. The \mgii\ SB profile is brighter and more extended than the continuum SB profile on average over $\approx15-20$\,kpc and $\approx15-30$\,kpc for the mass bins $10^{8-9}$\,\msun\ and $10^{9-10}$\,\msun, respectively. The central region up to $\approx15$\,kpc shows \mgii\ in absorption for the highest mass bin ($10^{10-12}$\,\msun). Beyond that, the \mgii\ SB profile is brighter than the continuum one, extending up to $\approx40$\,kpc. When it comes to \oii, the line SB profile is more radially extended compared to the continuum SB profile for all the mass bins, extending up to $\approx20-40$\,kpc.

Comparing the \mgii\ SB profiles across the different stellar mass bins, the brightest emission in the central $\approx10$\,kpc region is seen for the $10^{8-9}$\,\msun\ mass bin, although the SB profiles at the centre are consistent within the $1\sigma$ uncertainties for the three lowest stellar mass bins. The radial extent of the \mgii\ SB profiles increases by a factor of two from the lowest to the highest stellar mass bin. At a fixed SB level of $10^{-20}$\,\ergscmarc, the extent of the \mgii\ SB profile is $\approx14$\,kpc, $\approx21$\,kpc, $\approx27$\,kpc, and $\approx33$\,kpc for the stellar mass bins of $10^{7-8}$\,\msun, $10^{8-9}$\,\msun, $10^{9-10}$\,\msun, and $10^{10-12}$\,\msun, respectively. In the case of \oii, the SB profile becomes brighter by a factor of $\approx3$ on average within a circular aperture of radius 30\,kpc when going from the lowest to the highest stellar mass bin (Table~\ref{tab:sb_table}). Similar to \mgii, the \oii\ SB profile also becomes more radially extended with increasing stellar mass. At a fixed SB level of $5\times10^{-20}$\,\ergscmarc, the \oii\ SB profile extends up to $\approx14$\,kpc, $\approx18$\,kpc, $\approx24$\,kpc, and $\approx27$\,kpc for the stellar mass bins defined above, respectively. The increase in size of the metal line emission with increasing stellar mass can also be seen from the $R_{50}$ and $R_{90}$ estimates in Table~\ref{tab:sizes_table}. 

Next, to check that the above evolution is driven by stellar mass, independent of redshift, we conducted a control analysis similar to that in Section~\ref{sec:results_stack_redshift}. We formed samples in three different stellar mass bins ($10^{8-9}$\,\msun, $10^{9-10}$\,\msun\ and $10^{10-12}$\,\msun) that are matched in redshift within $\pm0.3$, such that the $p$-value from a two-sided KS test is $\approx$0.50--95. The stellar mass bin of $10^{7-8}$\,\msun\ does not have sufficient number of galaxies to form a matched-sample. We were able to form redshift-matched samples consisting of 70 and 79 galaxies for \mgii\ and \oii, respectively, in each of the mass bins. Similar to the full sample, we find that the SB profiles become more radially extended as the stellar mass increases in the matched samples as well. The above exercise along with the control analysis of Section~\ref{sec:results_stack_redshift} suggest that the average metal line emission evolves independently with both redshift and stellar mass.

\subsubsection{Evolution with environment}
\label{sec:results_stack_environment}

\begin{figure*}
 \includegraphics[width=1.0\textwidth]{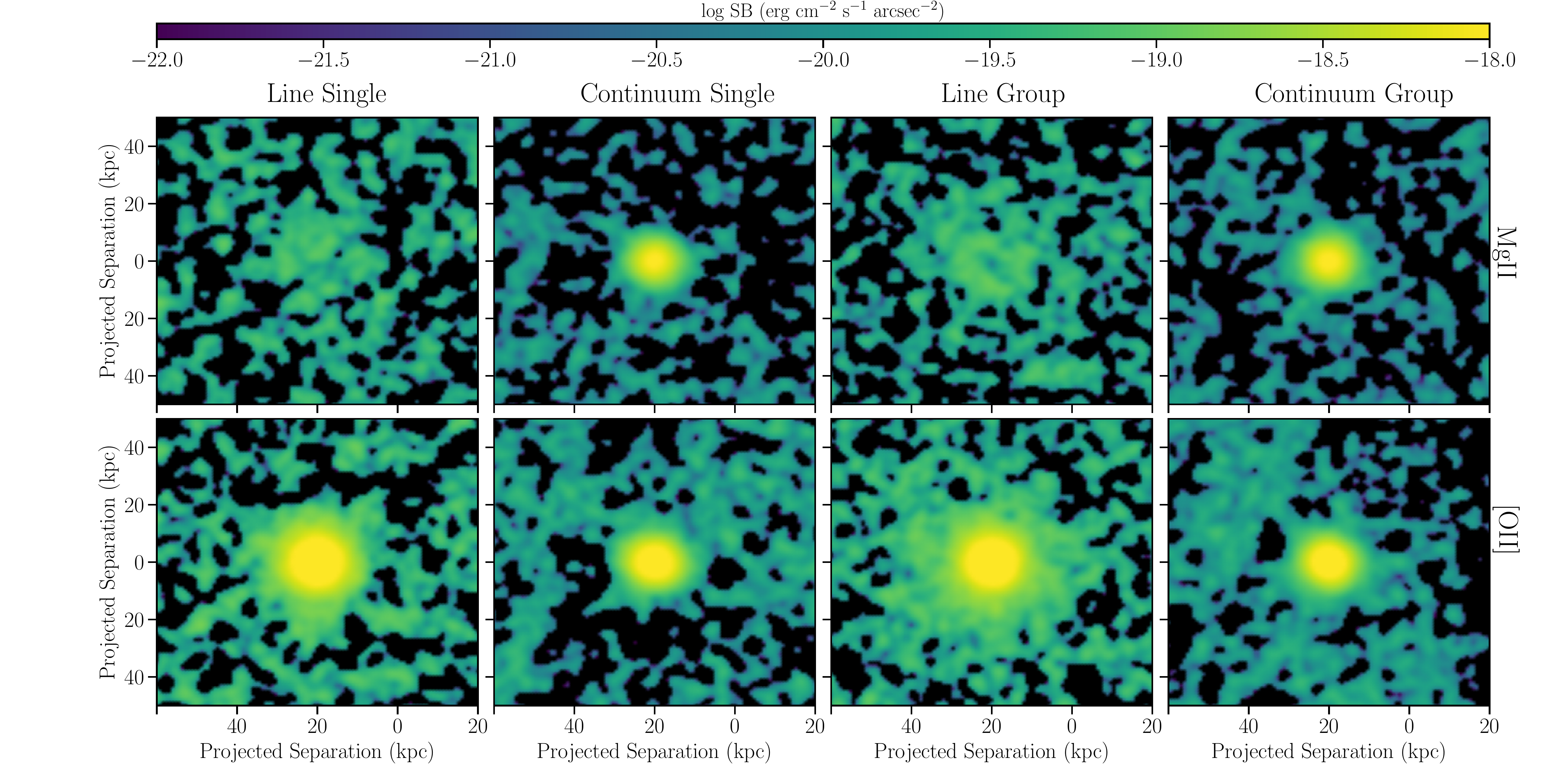}
 \caption{Median-stacked pseudo-NB images for galaxies in different environments. The top panels show the results for \mgii\ and the bottom panels show the results for \oii. From left to right, the panels show the stacked line emission after subtracting the continuum for single galaxies, the stacked continuum emission in control windows for single galaxies, the stacked line emission after subtracting the continuum for group galaxies, and the stacked continuum emission in control windows for group galaxies.
 }
 \label{fig:nbimg_env}
\end{figure*}
\begin{figure*}
 \includegraphics[width=0.48\textwidth]{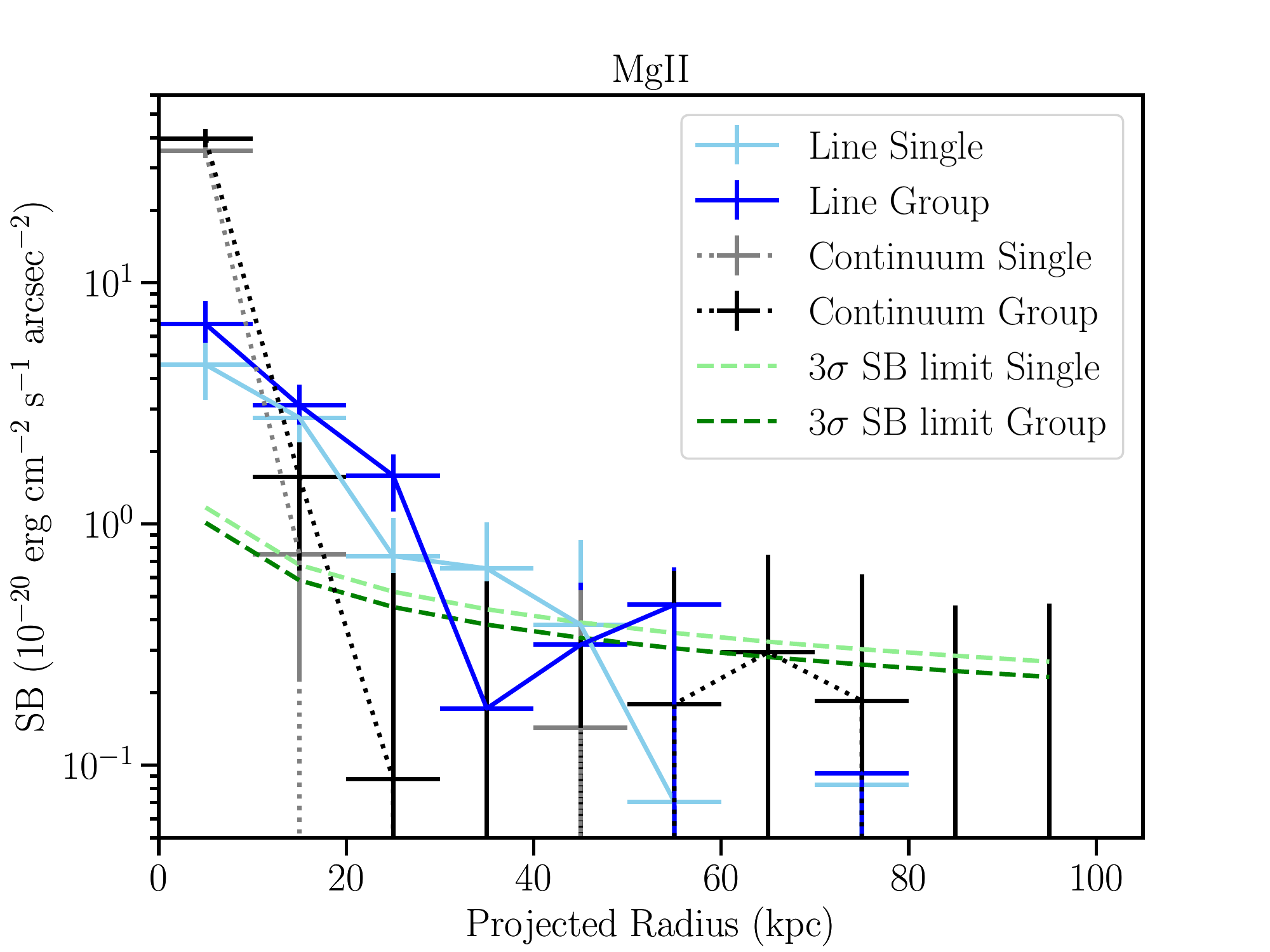}
 \includegraphics[width=0.48\textwidth]{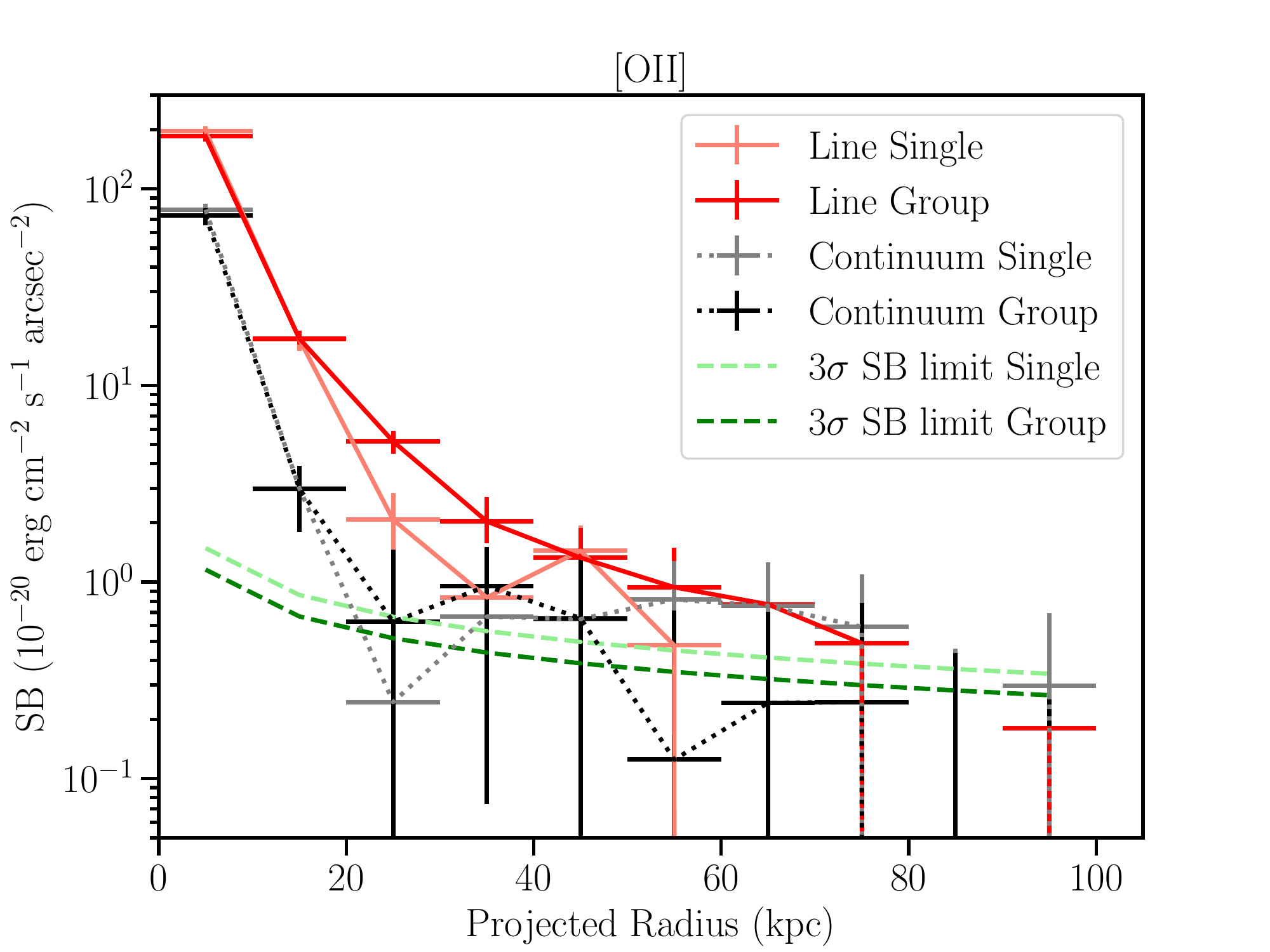}
 \caption{{\it Left:} The median-stacked azimuthally-averaged radial SB profiles of \mgii\ emission in circular annuli of 10\,kpc from the galaxy centres for single and group galaxies are shown as light blue and dark blue solid lines, respectively. The SB profiles of the control continuum emission for single and group galaxies are shown as grey and black dotted lines, respectively. The $3\sigma$ SB limits for the single and group profiles are shown as light and dark green dashed lines, respectively.
 {\it Right:} Same as in the left plot for \oii\ emission. The \oii\ SB profiles for single and group galaxies are shown as pink and red solid lines, respectively. The error bars represent the $1\sigma$ uncertainties from bootstrapping analysis.
 }
 \label{fig:sbprofile_env}
\end{figure*}

Several studies have found dependence of the multiphase CGM gas probed in absorption on the galaxy environment \citep[e.g.][]{chen2010,bordoloi2011,yoon2013,johnson2015,burchett2016,fossati2019b,dutta2020,dutta2021,huang2021}. To characterize the emission properties as a function of galaxy environment, we classify galaxies as either single systems or as belonging to a group. We define a `group' here as an association of two or more galaxies and place no constraints on the halo mass of the structure. In order to find galaxy groups, we adopt a Friends-of-Friends (FoF) algorithm that identifies galaxies that are connected within linking lengths of 400\,kpc along the transverse physical distance, and of 400\,\kms\ along the line-of-sight velocity space. The algorithm and parameters used here are consistent with those generally adopted in the literature \citep[e.g.][]{knobel2009,knobel2012,diener2013}. We find in total 108 groups (90 in MAGG and 18 in MUDF) at $0.7 \le z \le 1.5$ with members ranging from 2 to 20 galaxies. About half of the galaxies in the full sample are found to be part of galaxy groups, similar to other studies with similar sensitivity and group definition \citep[e.g.][]{dutta2021}. With the group catalog in hand, we next formed samples of group and single galaxies that are matched in stellar mass and redshift. For each galaxy in a group, we identified a unique single galaxy within $\pm0.3$\,dex in stellar mass and $\pm0.3$ in redshift. In this way, we were able to form matched-samples consisting of 185 and 190 galaxies, for \mgii\ and \oii, respectively. Based on a two-sided KS test, the maximum difference between the cumulative distributions of the stellar mass in the group and single samples is $\approx$0.05 and the $p$-value is $\approx$0.94-0.99, while for the redshift distributions, the maximum difference and $p$-values are $\approx$0.07 and $\approx$0.75, respectively. 

We stacked the MUSE cubes of the group and single matched-samples separately. Fig.~\ref{fig:nbimg_env} shows the median-stacked pseudo-NB images for the group and single samples for \mgii\ (top panel) and \oii\ (bottom panel). While \mgii\ and \oii\ line emission are detected for both group and single samples, the \mgii\ emission appears to be slightly brighter, and the \oii\ emission more extended in the case of the group sample. This can be also seen from Fig.~\ref{fig:sbprofile_env} that shows the azimuthally-averaged SB profiles in group and single samples for \mgii\ (left) and \oii\ (right). The \mgii\ and \oii\ emission within a circular aperture of radius 30\,kpc around the group sample are brighter by factors of $\approx1.4$ and $\approx1.2$, respectively, compared to the single sample (Table~\ref{tab:sb_table}). The differences in the metal line emission between the group and single samples are more prominent over $\approx20-30$\,kpc, where the \mgii\ SB profile of the group sample is $\approx2\times$ brighter than that of the single sample, and the \oii\ SB profile of the group sample is $\approx2.5\times$ brighter than that of the single sample. The group sample shows relatively larger extent in \oii\ emission compared to in \mgii\ emission. At a SB level of $10^{-20}$\,\ergscmarc, the \mgii\ SB profile of the group sample has a larger radial extent by a factor $\approx1.3$ compared to that of the single sample, while the radial extent of the \oii\ SB profile of the group sample is $\approx1.6\times$ larger compared to that of the single sample. The $R_{50}$ and $R_{90}$ size estimates of \mgii\ emission for the group sample are slightly larger than that of the single sample, but the values are consistent within the uncertainties (see Table~\ref{tab:sizes_table}). On the other hand, the $R_{90}$ size of the \oii\ emission for the group sample is larger than that of the single sample by a factor of $\approx1.2$.

In order to further check the dependence on environment, we performed tests wherein we selected the sample of group galaxies in different ways. In the above analysis, we have considered all the galaxies in groups. We repeated the analysis by considering only the most massive galaxies in each group \citep[defined usually as the central galaxy;][]{yang2008}, groups with three or more members, groups with physical separations between members $<$200\,kpc, and groups with mass ratio between the most and the least massive members within a factor of 10. These group samples are selected in such a way that we have sufficient number of galaxies to form matched single samples for stacking as described above for the full group sample. We find that the dependence of the metal line emission on environment persists, albeit with larger uncertainties due to the smaller sample sizes, when we select the group sample via different ways. The difference is most prominent when we select groups with smaller physical separations, with the metal line emission being up to two times more radially extended compared to the single sample, indicating that environmental interactions or possible overlap from multiple haloes of group members could be a cause of the extended emission.

\subsubsection{Dependence on orientation}
\label{sec:results_stack_orientation}

\begin{figure*}
 \includegraphics[width=1.0\textwidth]{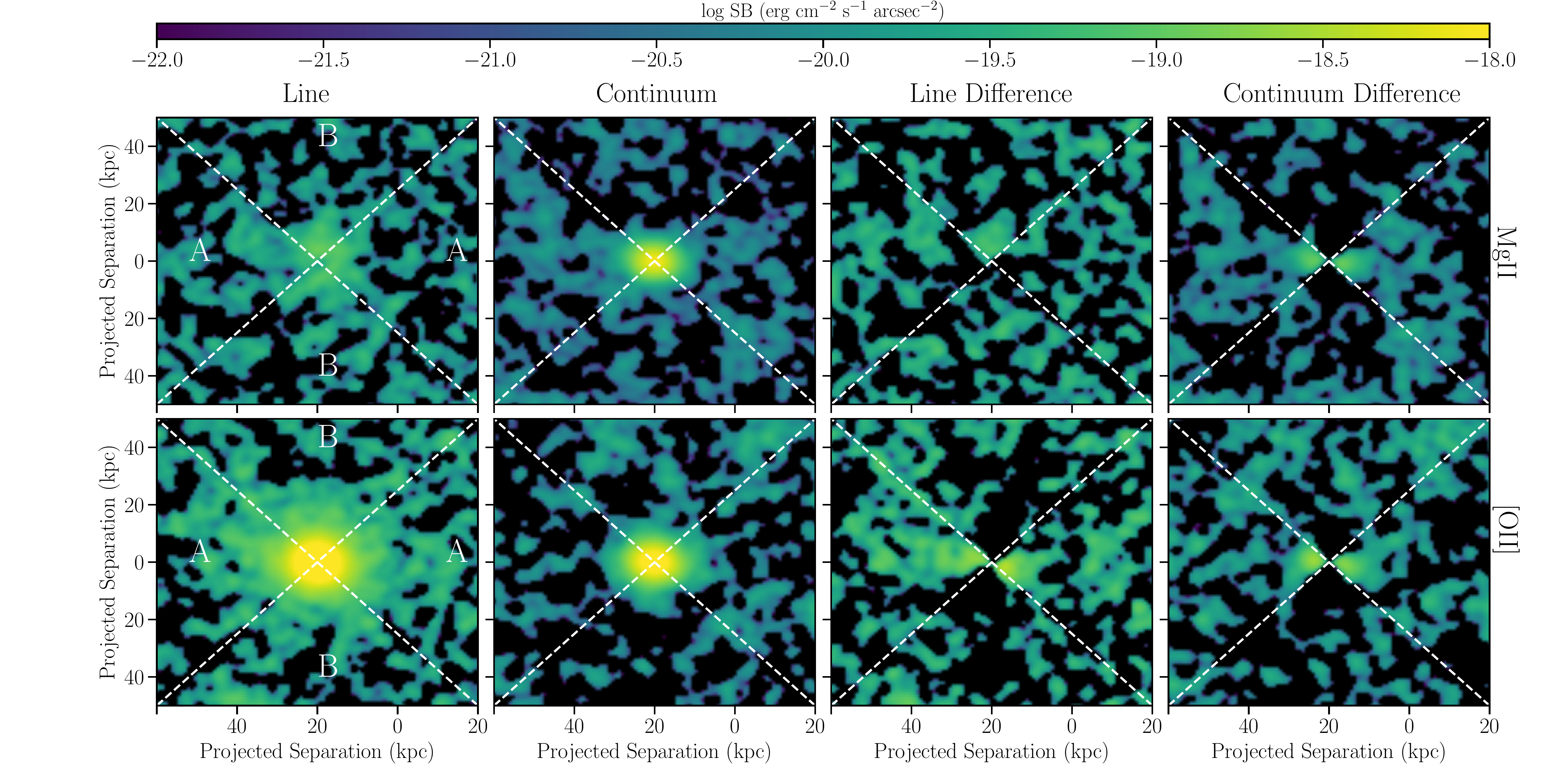}
 \caption{Median-stacked pseudo-NB images for galaxies with inclination greater than 30$^\circ$ in the MUDF that are aligned along the major axis. The top panels show the results for \mgii\ and the bottom panels show the results for \oii. From left to right, the panels show the stacked line emission, the stacked continuum emission, the difference between the stacked line emission and the same rotated by 90$^\circ$, and the difference between the stacked continuum emission and the same rotated by 90$^\circ$. The regions, `A' and `B', in which we obtain the SB profiles are marked by dashed diagonal lines. The two regions marked `A' are along the major axis (aligned with the horizontal axis), while the two regions marked `B' are along the minor axis.
 }
 \label{fig:nbimg_morph}
\end{figure*}
\begin{figure*}
 \includegraphics[width=0.48\textwidth]{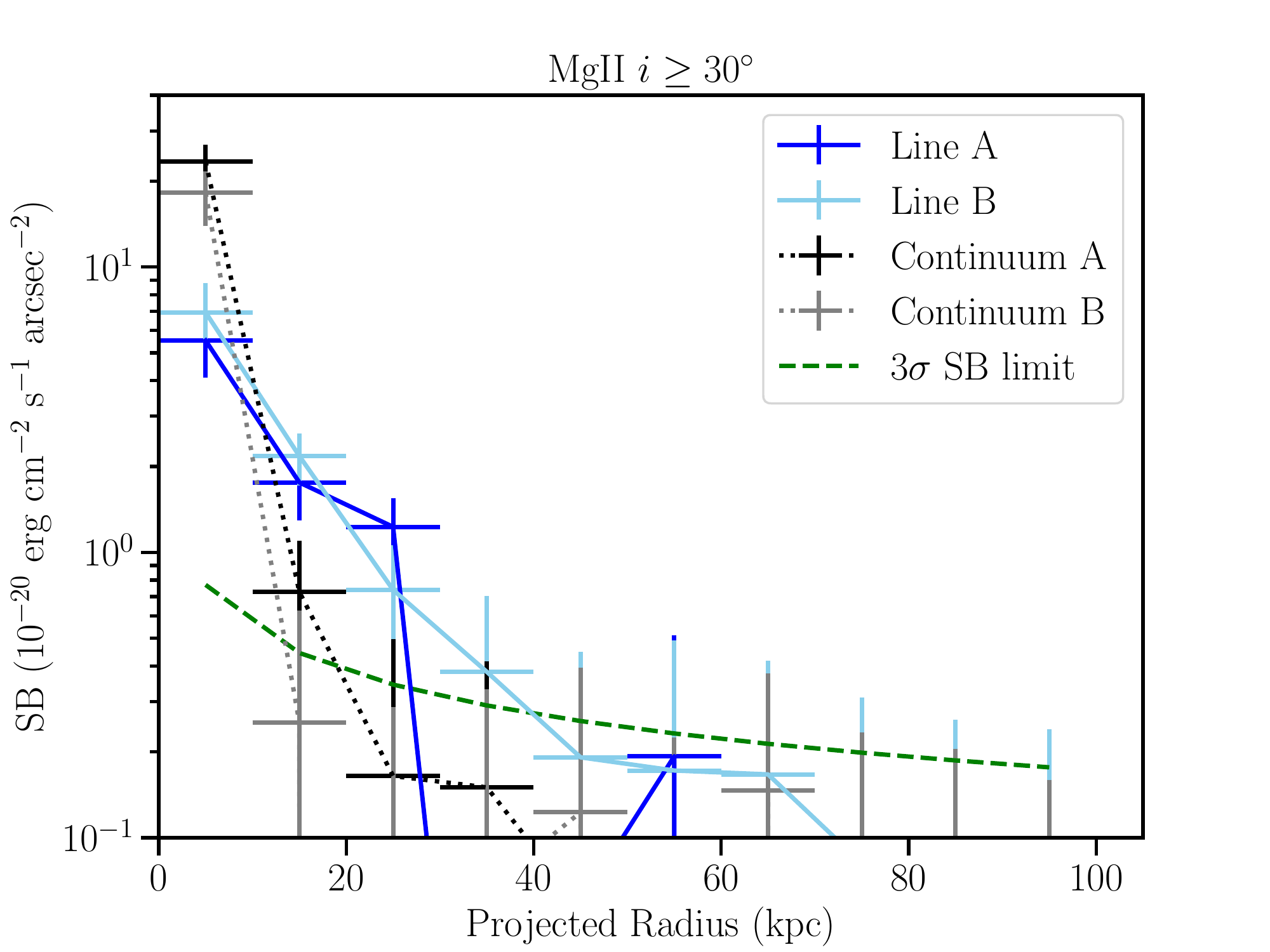}
 \includegraphics[width=0.48\textwidth]{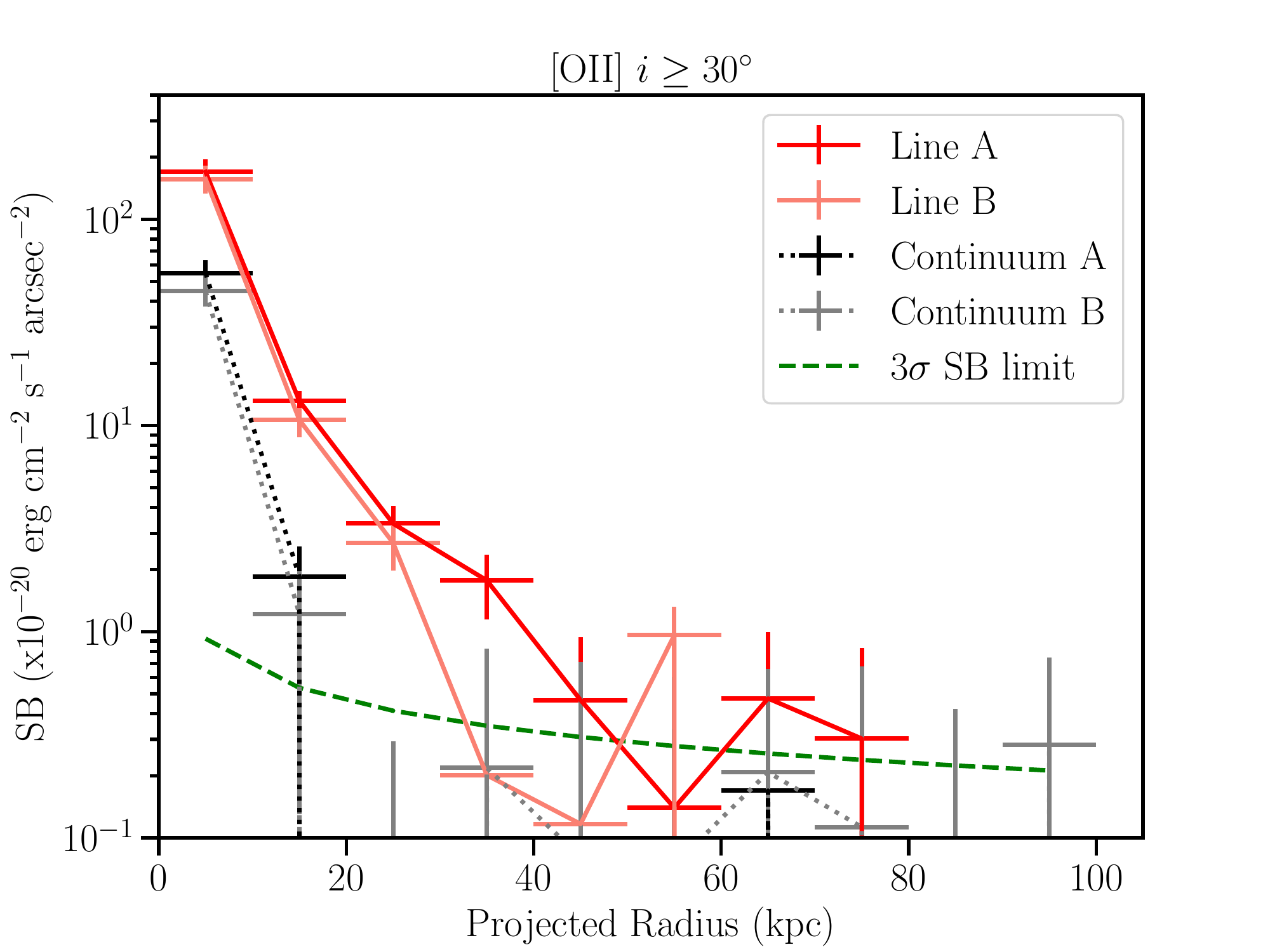}
 \caption{{\it left:} The median-stacked radial SB profiles of \mgii\ emission in circular annuli of 10\,kpc from the galaxy centres in the regions `A' and `B', as marked in Fig.~\ref{fig:nbimg_morph}, are shown as dark blue and light blue solid lines, respectively. The SB profiles of the corresponding control continuum emission in the regions `A' and `B' are shown as black and grey dotted lines, respectively. The $3\sigma$ SB limit is shown as green dashed line.
 {\it Right:} Same as in the left plot for \oii\ emission. The \oii\ SB profiles in the regions `A' and `B' are shown as red and pink solid lines, respectively. The error bars represent the $1\sigma$ uncertainties from bootstrapping analysis.
 }
 \label{fig:sbprofile_morph}
\end{figure*}

The stacking results presented so far have been focused on the average radial extent of the metal emission and do not take into account the orientation of the galaxy discs. Several observational studies have found that the distribution of the metal-enriched halo gas, probed by \mgii\ absorption lines, is anisotropic, suggesting either outflowing gas along the minor axis or accreting gas along the major axis \citep[e.g.][]{bordoloi2011,kacprzak2012,lan2018,schroetter2019}.  Cosmological hydrodynamical simulations also predict that properties of the halo gas such as metallicity and X-ray emission are anisotropic when shaped, e.g., by galactic outflows \citep[][]{nelson2019b,peroux2020b,truong2021}.

We can investigate whether the average metal line emission is isotropically distributed around the disc for a sub-sample of galaxies in the MUDF survey that has deep ($\approx$6.5\,h) HST NIR imaging available in the F140W band ($5\sigma$ limit of 28\,mag). This has been used to derive morphological parameters of the galaxies using {\sc statmorph} \citep{rodriguez2019}. The full description of this procedure is provided in section 6 of \citet{revalski2023}. Using the F140W image, segmentation map and PSF model, {\sc statmorph} has derived non-parametric as well as 2D Sersic fitting-based morphological measurements. There are 39 and 57 galaxies in the \mgii\ and \oii\ samples, respectively, with successful {\sc statmorph} fits and inclination angle greater than 30$^\circ$ (such that the orientation of the major axis can be confidently estimated). The median redshift and stellar mass of these samples are $z\approx1.1$ and $M_*\approx2\times10^9$\,\msun, respectively. We first aligned these galaxies along the direction of the major axis and then performed the stacking following the same procedure as described in Section~\ref{sec:methods}. We present here the results based on the non-parametric measurements derived by {\sc statmorph}, and note that we get similar results if instead we use the measurements derived from Sersic fitting by {\sc statmorph}.

The results from this stacking are shown in Fig.~\ref{fig:nbimg_morph}. The first column shows the median-stacked pseudo-NB images of \mgii\ and \oii\ line emission, where the galaxy major axis is aligned with the horizontal axis. We divide the region around the galaxy centre into four sections defined by cones with a 45$^\circ$ opening angle to the horizontal axis (marked by dashed lines in Fig.~\ref{fig:nbimg_morph}). We estimate the radial SB profiles separately in the left and right quadrants (marked as `A') that lie along the major axis, and in the top and bottom quadrants (marked as `B') that lie along the minor axis. Fig.~\ref{fig:sbprofile_morph} shows the radial SB profiles estimated in these two regions for \mgii\ (left) and \oii\ (right) emission. 

In the case of \mgii, we do not find any significant difference in the average emission along the major and minor axes. The \oii\ emission, on the other hand, is more extended (by a factor of $\approx$1.3 at a SB level of $10^{-20}$\,\ergscmarc) and brighter over $\approx30-40$\,kpc along the major axis compared to along the minor axis. This extension in \oii\ emission along the major axis is further confirmed when we rotate the stacked NB image by 90$^\circ$ and subtract it from the original. As can be seen from Fig.~\ref{fig:nbimg_morph}, there is excess \oii\ emission in the `A' quadrants along the major axis in the difference NB image. On the other hand, when we repeat the same exercise for \mgii, there is no significant residual emission, indicating a more isotropic distribution. We caution, however, that these results are based on stacking a relatively small number of galaxies, and need to be verified with a larger and deeper sample in the future. The present sample also has too few ($\lesssim$10) face-on galaxies (inclination angle $\le30^\circ$) to check for dependence of the line emission on inclination.

The \oii\ emission along the major axis could be tracing the extended diffuse ionized gas disc or the disc-halo interface. In the local Universe, observations of \hi\ 21-cm emission have revealed that the neutral gas disc is typically more extended than the stellar disc of galaxies, and that it could be arising from a combination of gas accretion and recycling \citep[e.g.][]{oosterloo2007,sancisi2008,chemin2009,kamphuis2013,zschaechner2015,marasco2019}. The extended \oii\ emission could be probing the higher redshift ionized disc counterparts of these local thick \hi\ discs.

\subsection{Emission in individual groups}
\label{sec:results_individual}

Group environments, where the gaseous component of galaxies is susceptible to stripping due to gravitational and hydrodynamic interactions, are regions of interest to search for extended emission around and between galaxies. Motivated by this, we have carried out a search for extended metal line emission in the richer groups in our sample (see Section~\ref{sec:results_stack_environment} for group identification), defined here as those consisting of three or more galaxies. Search for extended emission in the full galaxy sample and detailed discussion of extended emission in individual systems will be the subjects of future work. The goal of the present search is to complement the stacking analysis by checking whether we detect instances of extended intragroup medium between galaxies or extended emission around group members at the observed SB level. 

For this part of the analysis, we use tools from the {\sc CubEx} package \citep[version 1.8,][and Cantalupo in prep.]{cantalupo2019}. At first, we subtract the PSF of the quasar and of the stars in the field from the MUSE cubes using {\sc CubePSFSub}. This code takes into account the wavelength dependence of the full-width at half-maximum of the seeing by generating PSF images from NB images obtained in bins of 250 MUSE spectral pixels \citep[see][for further details]{fossati2021}. Then we subtract the remaining continuum sources from the cubes using {\sc CubeBKGSub}, which is based on a fast median-filtering approach as described in \citet{borisova2016}. Afterwards, we select a sub-cube that covers $\pm25$\,\AA\ around the metal line at the group redshift from the PSF- and background-subtracted cube and run {\sc CubEx} on it. Before running the detection algorithm, {\sc CubEx} applies a two-pixel boxcar spatial filter on the cube. 

We apply the following parameters in order to detect extended emission: (i) signal-to-noise ratio (SNR) threshold of individual voxel of 2 or 2.5 (depending on the noise in individual fields); (ii) integrated SNR threshold over the 3D mask of 5; (iii) minimum number of voxels of 5000; (iv) minimum number of spatial pixels of 100; (v) minimum number of spectral pixels of 3; (vi) maximum number of spectral pixels of 250 (to exclude residuals from continuum sources). We experimented with different values for the detection parameters before settling on the above values that were found to lead to the most reliable extraction of extended emission with minimum number of contaminants. 

{\sc CubEx} produces 3D segmentation cubes, pseudo-NB images and 1D spectra of all the extended emission candidates. In order to verify whether the extended emission is real, we visually inspected all of the above {\sc CubEx} products to rule out cases of emission from sources at different redshifts, residual continuum emission from bright sources, noise at the edge of the field of view, and low-level widespread systematic noise. Below we discuss the results from running the above procedure on the groups in MAGG and MUDF. 

\subsubsection{Emission in MAGG groups}
\label{sec:results_individual_magg}

\begin{figure*}
 \includegraphics[width=0.33\textwidth]{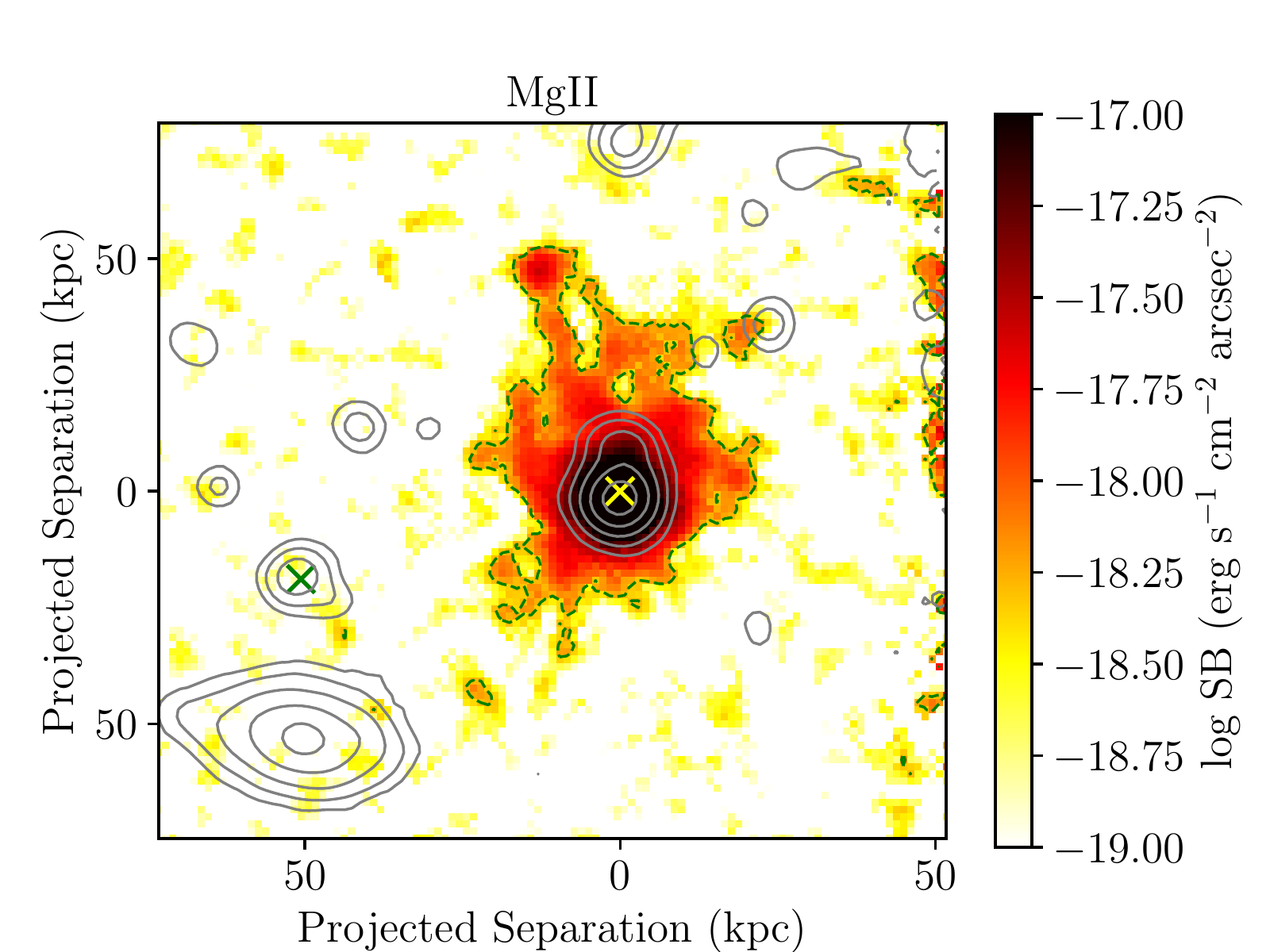}
 \includegraphics[width=0.33\textwidth]{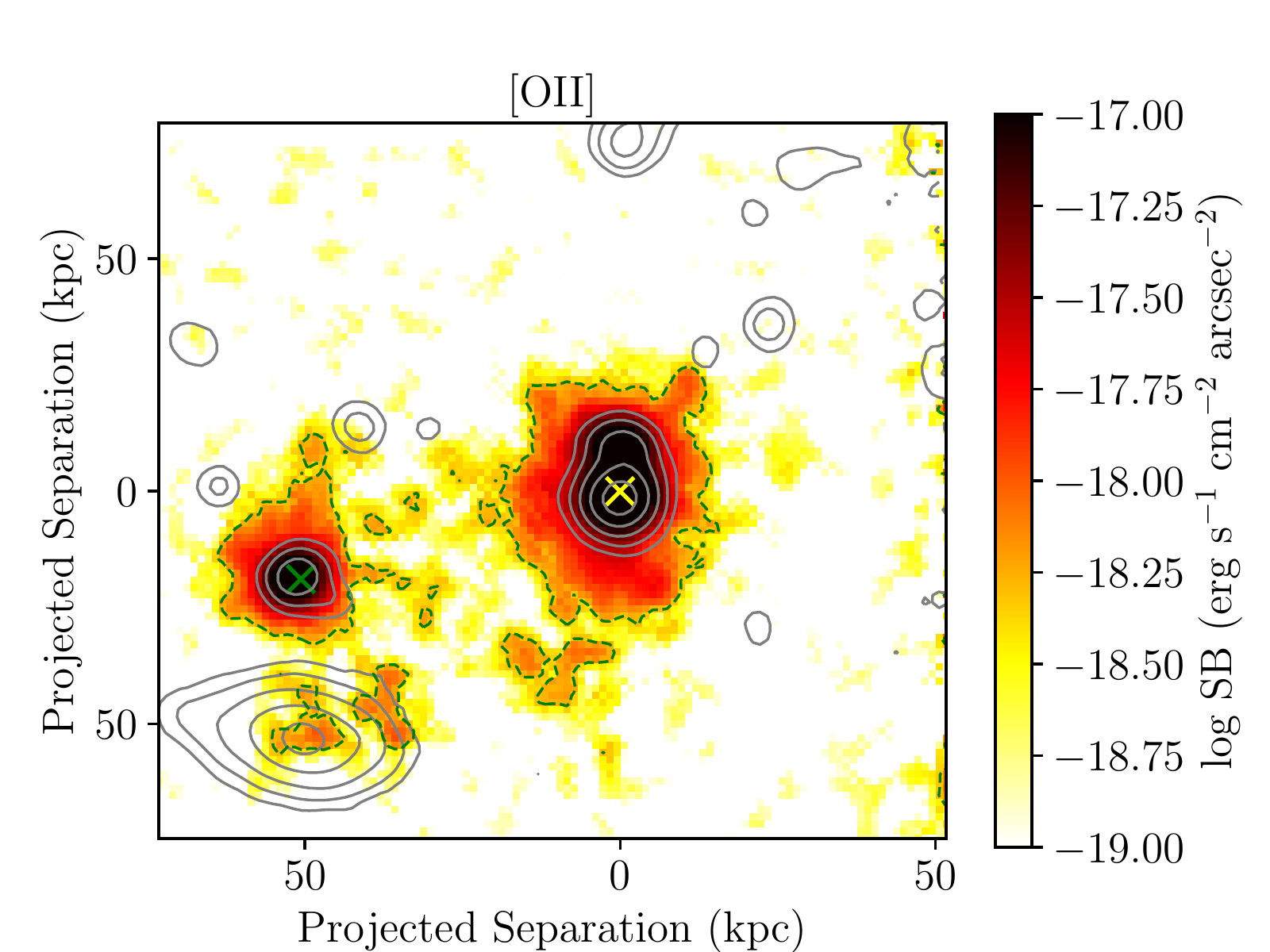}
 \includegraphics[width=0.33\textwidth]{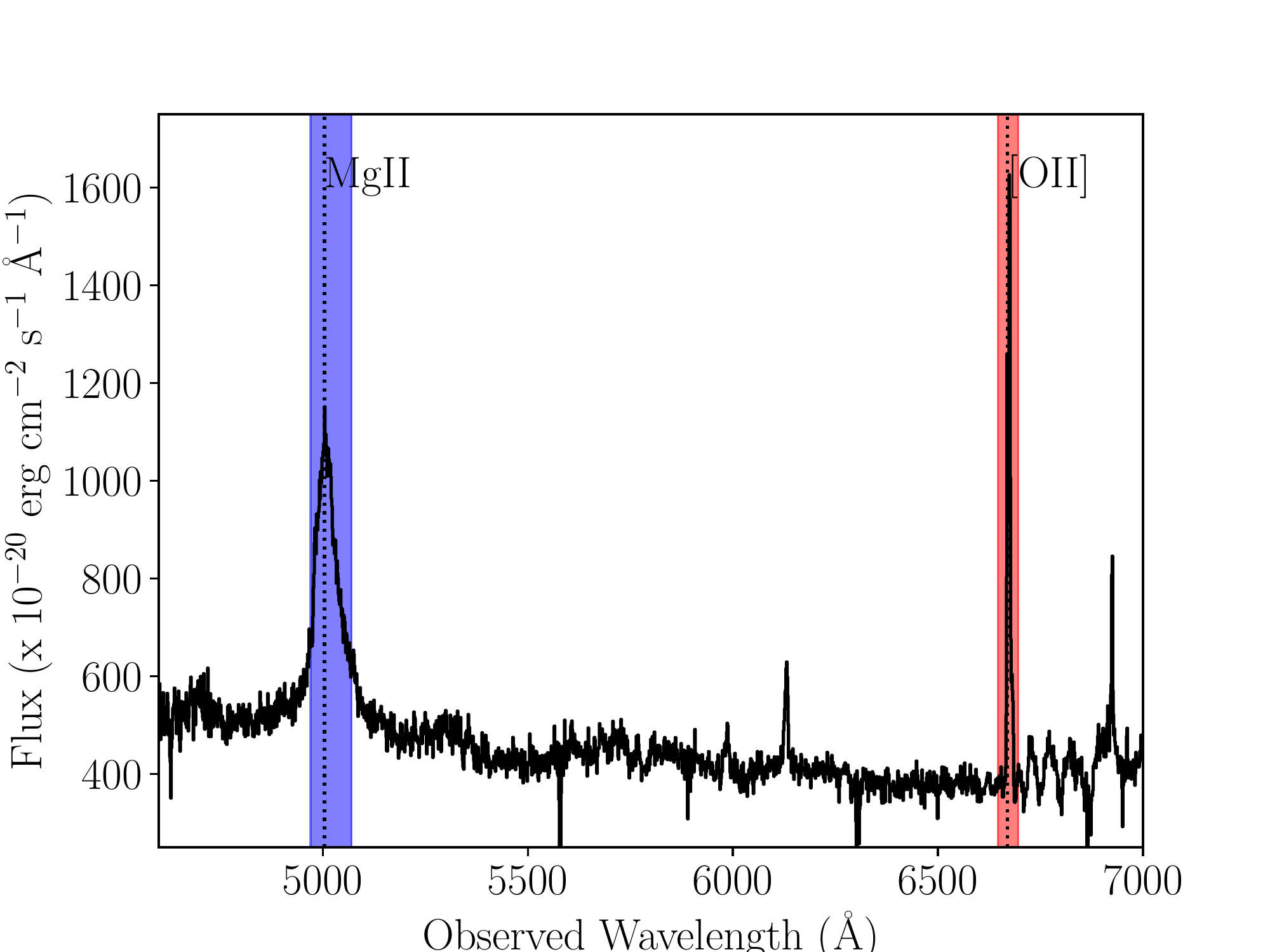}
 \caption{{\it Left:} The observed SB map of the optimally extracted \mgii\ emission from a galaxy at $z\approx0.8$ in MAGG that shows quasar-like broad \mgii\ emission in its spectrum. The emission has been smoothed using a top-hat kernel of width 0.4\,arcsec. The position of the galaxy is marked by an yellow ``X'', while that of a nearby galaxy belonging to the same group is marked by a green ``X''. The green contour marks the SB level of $5\times10^{-19}$\,\ergscmarc. The grey contours show the continuum emission at levels of 22, 23, 24, 25, 26 and 27\,mag\,arcsec$^{-2}$.
 {\it Centre:} Same as in the left panel for \oii\ emission.
 {\it Right:} The MUSE spectrum of the galaxy. The observed wavelengths of the \mgii\ and \oii\ emission lines are marked by dotted vertical lines. The wavelength ranges around the \mgii\ and \oii\ emission lines that are used to search for extended emission in {\sc CubEx} are marked by blue and red shaded regions, respectively.
 }
 \label{fig:magg_quasar}
\end{figure*}
\begin{figure*}
 \includegraphics[width=1.0\textwidth]{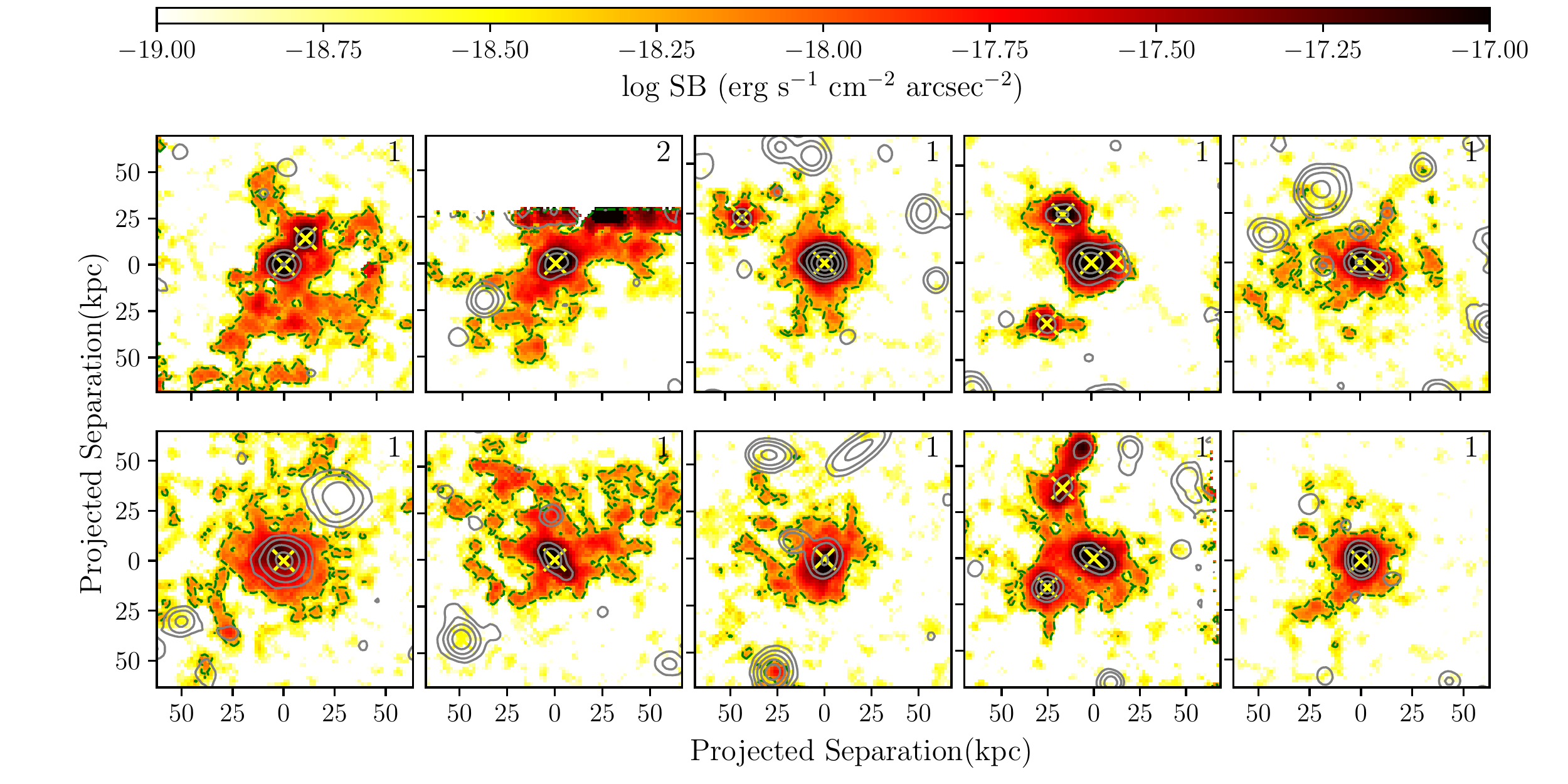}
 \caption{Observed SB maps of the optimally extracted \oii\ emission around selected group galaxies in MAGG. The emission has been smoothed using a top-hat kernel of width 0.4\,arcsec. The galaxy position is marked by an X. The green contour marks the SB level of $5\times10^{-19}$\,\ergscmarc. The grey contours show the continuum emission at levels of 22, 23, 24, 25, 26 and 27\,mag\,arcsec$^{-2}$. The number in each panel indicates the confidence flag of the detection (1 = higher; 2 = lower). In one of the top panels, we show an example of a system that is given a flag 2 because it is at the edge of the field of view.
 }
 \label{fig:magg_oii_individual}
\end{figure*}

We search for extended \mgii\ and \oii\ line emission in 34 galaxy groups in MAGG that are composed of three or more galaxies at $0.7 \le z \le 1.5$. The maximum number of group members in this sample is six. The typical $3\sigma$ noise in the pseudo-NB images of width $\pm500$\,\kms\ around the group redshifts is $\approx7\times10^{-18}$\,\ergscmarc. We detect extended \mgii\ line emission from only one of the group galaxies, i.e. $\approx1$ per cent of the sample. This galaxy belongs to a group of three galaxies and exhibits a broad \mgii\ emission line at $z\approx0.8$ in its MUSE spectrum (see right panel of Fig.~\ref{fig:magg_quasar}), indicating that it is likely to be a quasar. Fig~\ref{fig:magg_quasar} shows the SB maps of the optimally extracted \mgii\ and \oii\ emission from this galaxy. Consistently with previous studies \citep{borisova2016,arrigoni2019,fossati2021}, these optimally extracted maps are obtained by collapsing the voxels in the continuum-subtracted MUSE cube within the {\sc CubEx} segmentation cube along the spectral direction and filling in the pixels outside of the segmentation cube with data from the central wavelength layer of the line emission. Both the \mgii\ and \oii\ emission extends beyond the continuum emission from this source up to $\approx$30--40\,kpc. Extended \oii\ emission is also detected from another galaxy (\mstar\ $\approx5\times10^{9}$\,\msun) at $\approx$50\,kpc separation that is part of the same group. The total luminosity of the \mgii\ emission is $\approx10^{42}$\,\ergs, and that of the \oii\ emission is $\approx7\times10^{41}$\,\ergs.

For $\approx$32 per cent of the galaxies, we detect \oii\ emission around them that extend beyond their continuum emission. Fig.~\ref{fig:magg_oii_individual} shows the SB maps of the optimally extracted \oii\ emission from some of these galaxies that are obtained as described above. We give a flag to each case to signify how reliable the extended emission is, using 1 for higher confidence and 2 for lower confidence. In $\approx$15 per cent of the cases, due to the presence of contamination from residual continuum emission from a nearby star or quasar, or due to the source being close to the edge of the field of view, all of the emission detected by {\sc CubEx} may not be associated with the galaxy, and hence these are given a flag 2. The \oii\ emission can be seen extending beyond the continuum emission from the galaxies and arising from filamentary-like structures up to 50\,kpc in some cases. In $\approx$30 per cent of the cases, the emission forms a common envelope or bridge-like structure around two or more nearby galaxies. The extended \oii\ emission could be arising as a result of tidal interactions between galaxies in the group environment or ram-pressure stripping by the intra-group medium. Based on the average \oii\ emission around group galaxies being more extended than the average continuum emission in the stacking analysis (Section~\ref{sec:results_stack_environment}), such emission is likely to be present also around other group galaxies, but at a level that is below the observed SB limit.

\subsubsection{Emission in MUDF groups}
\label{sec:results_individual_mudf}

\begin{figure*}
 \includegraphics[width=1.0\textwidth, trim={2cm 0 2cm 0}, clip]{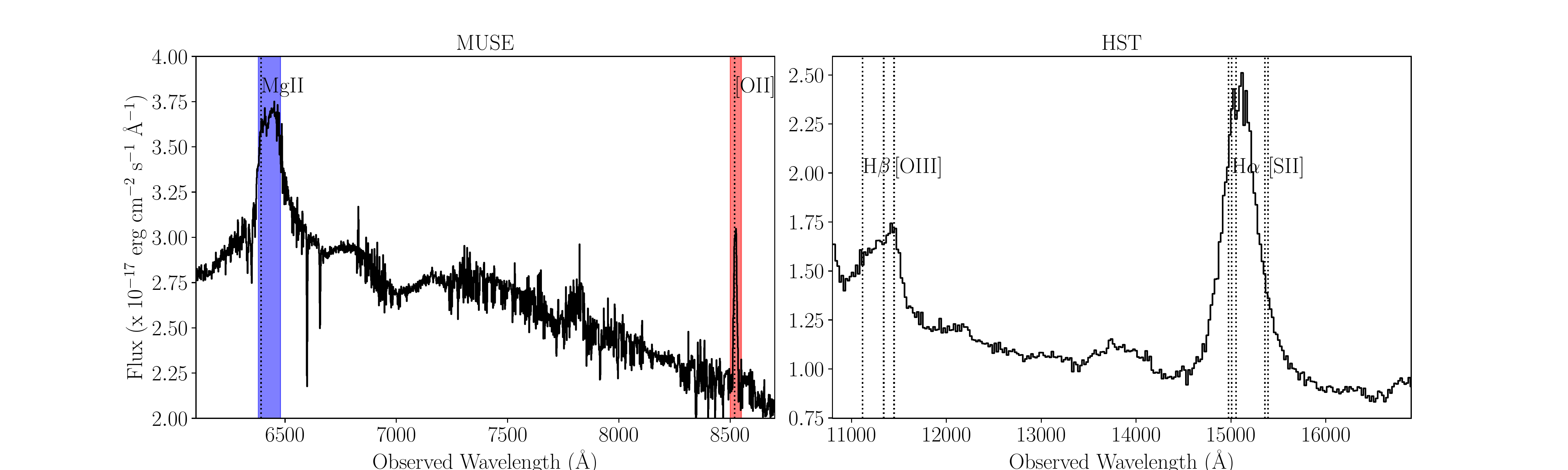}
 \includegraphics[width=0.48\textwidth]{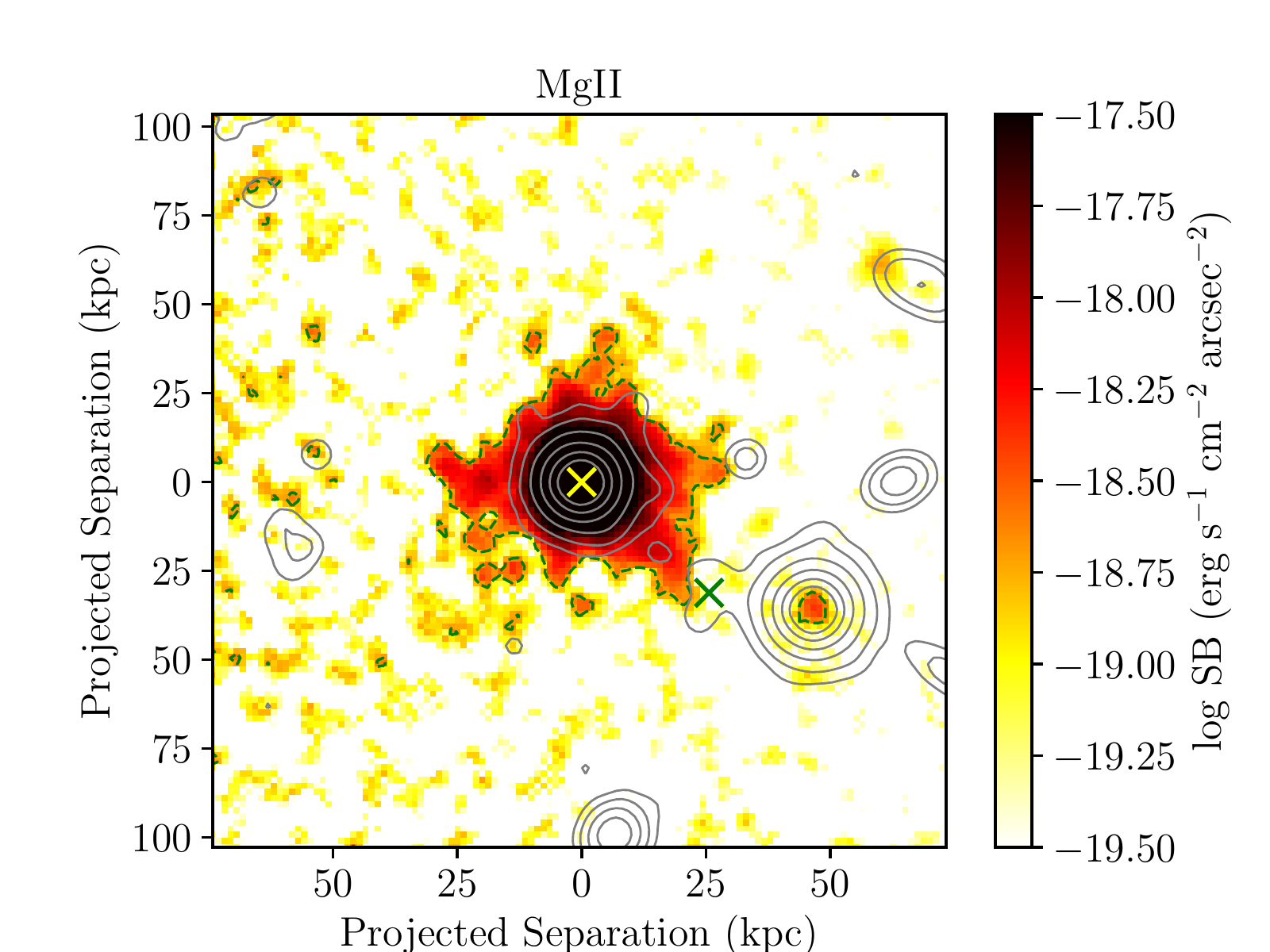}
 \includegraphics[width=0.48\textwidth]{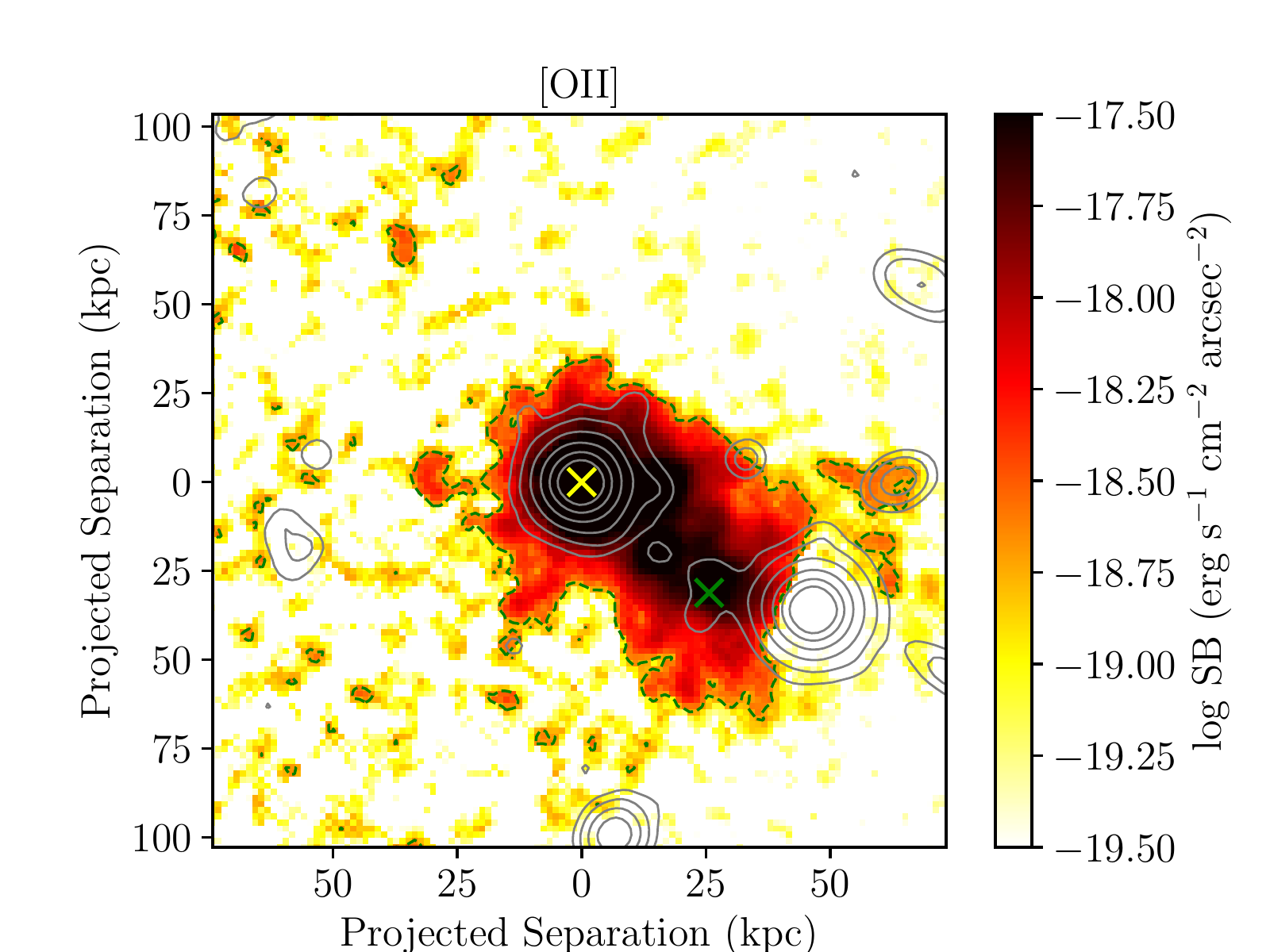}
 \caption{{\it Top:} The MUSE optical (left) and HST NIR (right) spectra of a galaxy at $z\approx1.3$ in MUDF that shows quasar-like broad emission features. The observed wavelengths of the \mgii\, \oii, \hb, \oiii, \nii, \ha\ and \sii\ emission lines are marked by dotted vertical lines. The wavelength ranges around the \mgii\ and \oii\ emission lines that are used to search for extended emission in {\sc CubEx} are marked by blue and red shaded regions, respectively.
 {\it Bottom:} The observed SB maps of the optimally extracted \mgii\ (left) and \oii\ (right) emission from the galaxy. The emission has been smoothed using a top-hat kernel of width 0.4\,arcsec. The position of the galaxy is marked by an yellow ``X'', while that of a nearby galaxy belonging to the same group is marked by a green ``X''. The green contour marks the SB level of $2\times10^{-19}$\,\ergscmarc. The grey contours show the continuum emission at levels of 22, 23, 24, 25, 26 and 27\,mag\,arcsec$^{-2}$.
 }
 \label{fig:mudf_quasar}
\end{figure*}
\begin{figure*}
 \includegraphics[width=1.0\textwidth]{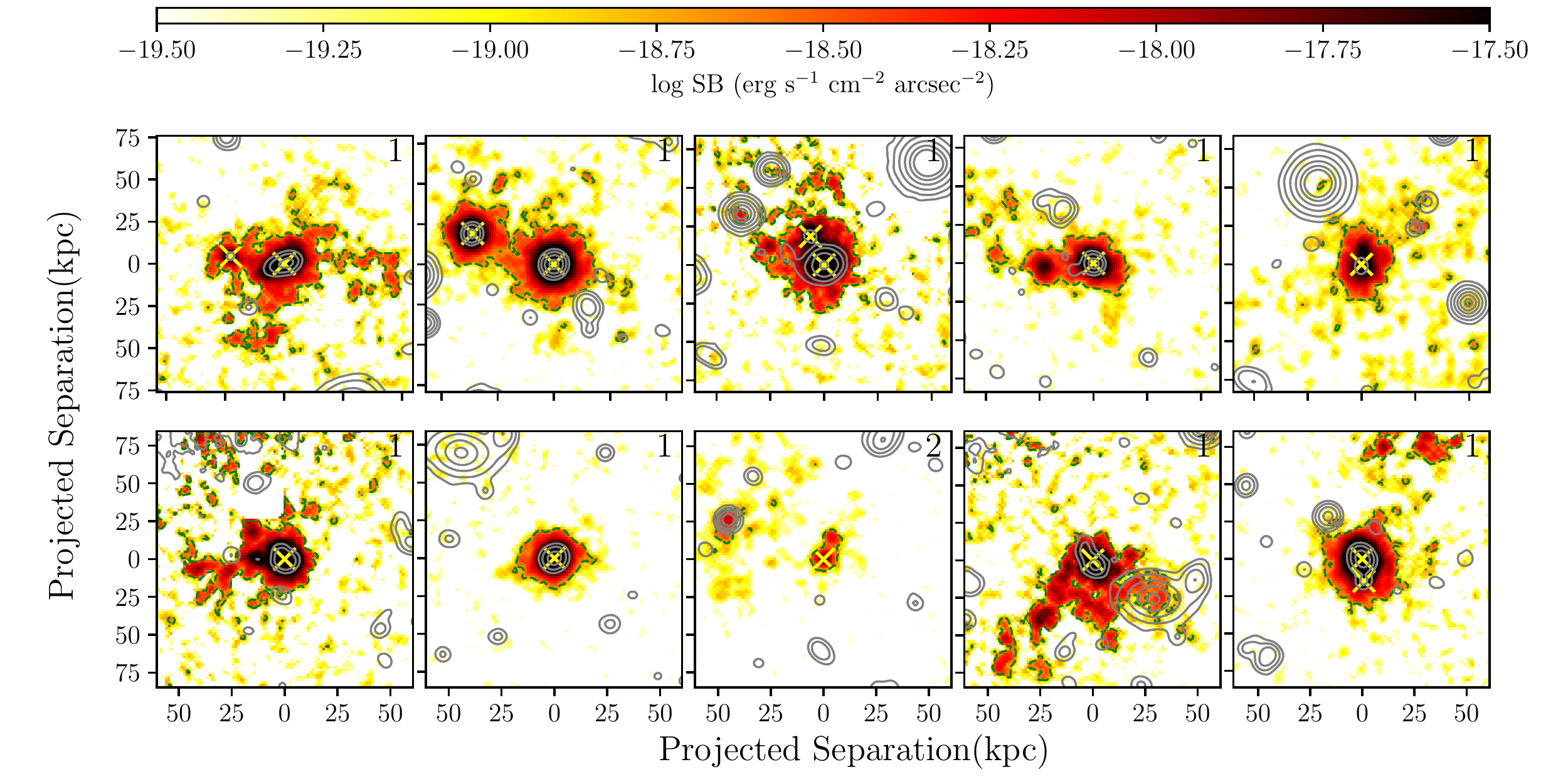}
 \caption{Same as in Fig.~\ref{fig:magg_oii_individual} for the observed SB map of the optimally extracted \oii\ emission around selected group galaxies in MUDF. The emission has been smoothed using a top-hat kernel of width 0.4\,arcsec. The galaxy position is marked by a ``X''. The green contour marks the SB level of $2\times10^{-19}$\,\ergscmarc. The grey contours show the continuum emission at levels of 22, 23, 24, 25, 26 and 27\,mag\,arcsec$^{-2}$. The number in each panel indicates the confidence flag of the detection (1 = higher; 2 = lower).
 }
 \label{fig:mudf_oii_individual}
\end{figure*}
\begin{figure}
 \includegraphics[width=0.5\textwidth]{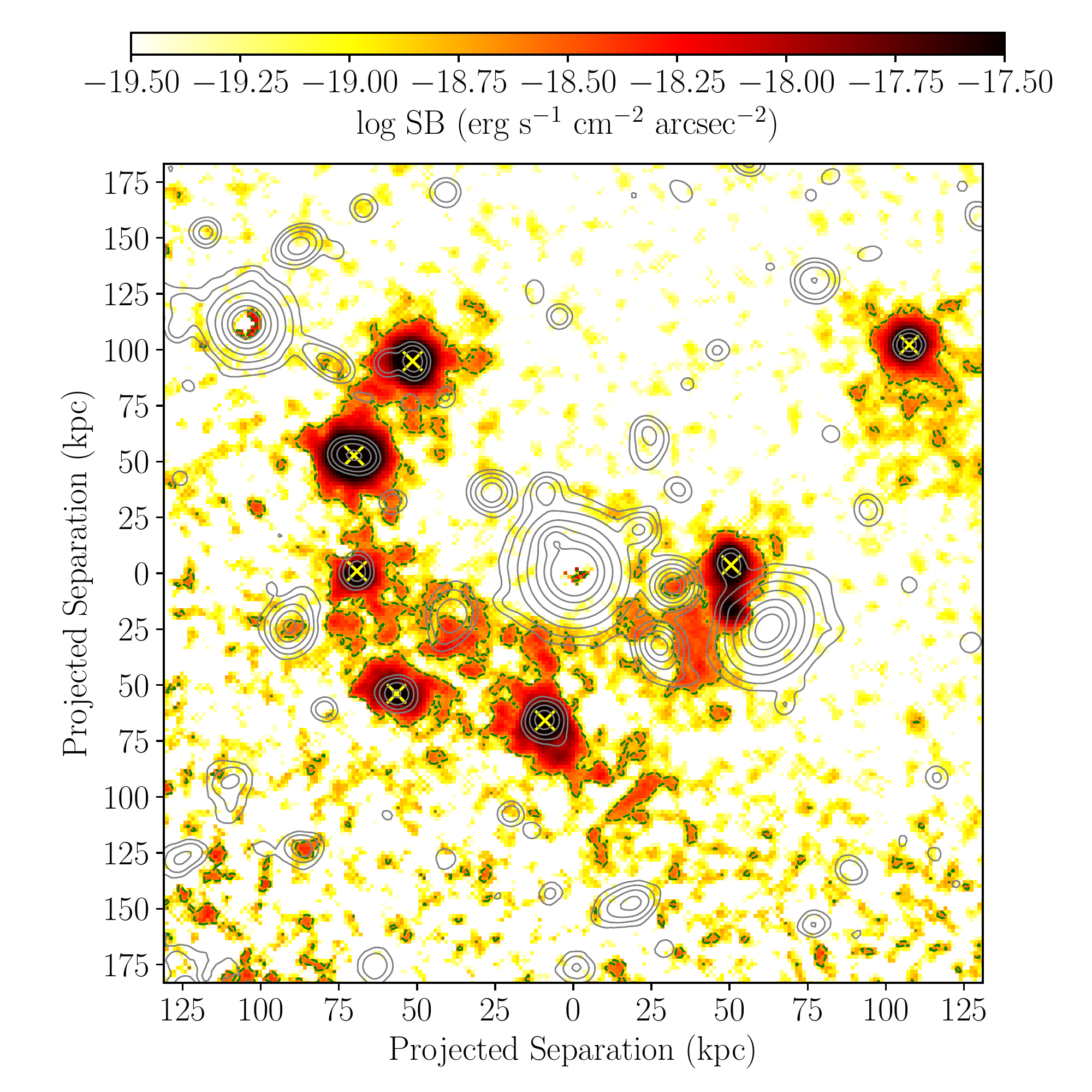}
 \caption{The observed SB map of the optimally extracted \oii\ emission in a galaxy group at $z\approx1$ in MUDF. The emission has been smoothed using a top-hat kernel of width 0.4\,arcsec. The positions of the group galaxies are marked by ``X''. The green contour marks the SB level of $2\times10^{-19}$\,\ergscmarc. The grey contours show the continuum emission at levels of 22, 23, 24, 25, 26 and 27\,mag\,arcsec$^{-2}$. 
 }
 \label{fig:mudf_oii_group}
\end{figure}

We search for extended \mgii\ and \oii\ line emission in 12 galaxy groups in MUDF that comprise of three or more galaxies at $0.7 \le z \le 1.5$. The number of galaxies in these groups range from three to twenty. The MUSE exposure time varies across these galaxies (see Fig.~\ref{fig:mstar_zgal}), with the average $3\sigma$ noise in the pseudo-NB images of width $\pm500$\,\kms\ around the group redshifts being $\approx2\times10^{-18}$\,\ergscmarc. Extended \mgii\ line emission is detected from only one galaxy in this sample ($\approx$1 per cent). Based on the broad emission lines of \mgii\ and \ha\ detected in the optical and NIR spectra, respectively (see top panels of Fig.~\ref{fig:mudf_quasar}), this source is again classified as a quasar. 

This quasar is part of a group of 14 galaxies at $z\approx1.3$. The bottom panels of Fig.~\ref{fig:mudf_quasar} show the SB maps of the \mgii\ and \oii\ emission from this source that is optimally extracted using {\sc CubEx} as described in Section~\ref{sec:results_individual_magg}. The total luminosity of the \mgii\ and \oii\ emission are $\approx5\times10^{42}$\,\ergs\ and $\approx3\times10^{42}$\,\ergs, respectively. The \mgii\ emission extends beyond the region of the continuum emission, and traces structures that extend up to $\approx$30--40\,kpc from the centre. The \oii\ emission, on the other hand, is more widespread than the \mgii\ emission, extending particularly towards the direction a nearby galaxy (\mstar\ $\approx2\times10^{10}$\,\msun) belonging to the same group (marked by a green ``X''). Three additional sources, located between the quasar and this galaxy, are detected in the HST F140W image. These are too small and faint (F140W $\approx$24--25\,mag) to be detected separately in the MUSE continuum image, but \oii\ doublet lines at the redshift of this group are detected in the MUSE spectra of these sources. The \oii\ emission, extending across $\approx$85\,kpc, appears to form a bridge across these sources, and could be arising as a result of tidal interactions or intergalactic transfer between them. 

We detect extended \oii\ emission around $\approx$40 per cent of the group galaxies in this sample, up from $\approx$32 per cent in the shallower MAGG data. Some examples of such extended \oii\ emission around the galaxies are shown in Fig.~\ref{fig:mudf_oii_individual}. The confidence flag given to these detections follows the definition in Section~\ref{sec:results_individual_magg}. As in the case of MAGG group galaxies, the \oii\ emission around these galaxies extends in filamentary-like structures up to few 10s of kpc from the galaxy centres. In $\approx$47 per cent of the cases, \oii\ emission is detected around two or more nearby galaxies such that the emission forms a bridge or common structure around these galaxies.

In one system, we detected \oii\ emission from seven nearby galaxies that are part of a larger group of 20 galaxies at $z\approx1.0$ with halo mass, \mhalo\ $\approx10^{13}$\,\msun\footnote{Halo mass is estimated from the stellar mass of the most massive galaxy in the group following the redshift-dependent stellar-to-halo mass relation of \citet{moster2013}.}. The optimally extracted \oii\ emission from this sub-group is shown in Fig.~\ref{fig:mudf_oii_group}. The \oii\ emission from these galaxies form a ring-like structure extending up to $\approx$200\,kpc around one of the two $z=3.2$ quasars in this field. The total \oii\ luminosity of the structure is $\approx2\times10^{42}$\,\ergs. We have masked the quasar and a few bright foreground galaxies located around the group galaxies (marked by ``X'') before searching for extended emission using {\sc CubEx}, such that the emission from the group is not contaminated by continuum subtraction residuals. It is evident that there is extended \oii\ emission arising around the group galaxies and also from structures connecting them. This \oii\ emission is potentially tracing an intragroup medium or displaced halo gas due to gravitational or hydrodynamic interactions within the group. The deep HST F140W image reveals no large, extended stellar tidal streams between these galaxies. Based on the Gini-$M_{20}$ diagram obtained using {\sc statmorph} morphological measurements (see Section~\ref{sec:results_stack_orientation}) and the classification scheme of \citet{lotz2008}, these galaxies would not be classified as mergers. However, a more detailed analysis of the multi-wavelength data is required to understand whether tidal or ram-pressure stripping could be giving rise to the extended emission. Lastly, we note that the clumpy and filamentary nature of the extended \oii\ emission points towards a patchy distribution of metals in the extended CGM or intragroup medium. 

\section{Discussion}
\label{sec:discussion}

\subsection{Comparison with literature observations}
\label{sec:discussion_literature}

In the local Universe, there are several spatially resolved observations of extended gaseous structures around galaxies in denser environments that arise due to tidal or ram-pressure stripping. Structures such as tails or streams have been observed in emission at different wavelengths that trace different gas phases from cold molecular and neutral gas to warm and hot ionized gas \citep[e.g.][]{chung2007,su2017,fossati2019a,moretti2020}. Using MUSE, the GASP \citep[GAs Stripping Phenomena in galaxies with MUSE;][]{poggianti2017} survey has studied in emission the long tails of ionized gas being stripped away from `jellyfish' galaxies by ram pressure. Gas stripping phenomena in jellyfish galaxies have been studied in cosmological hydrodynamical simulations as well \citep{yun2019}. The extended \oii\ emission that we observe around some of the group galaxies in our sample (Section~\ref{sec:results_individual}) could be higher redshift analogs of the local jellyfish galaxies. Observations of elevated covering fraction of \mgii\ absorption in gaseous haloes of galaxies in overdense group environments lend further support to the picture of gas-stripping $z\approx1$ \citep{dutta2020,dutta2021}. 

Currently, beyond the local Universe, there are only a few detections of extended metal emission, traced through rest-frame optical lines, around galaxies. Most of the extended metal nebulae detected so far are associated with overdense environments. \citet{epinat2018} discovered a large, $10^4$\,kpc$^2$, \oii-emitting gas structure associated with a group of 12 galaxies at $z=0.7$ using $\approx$10\,h of MUSE observations. Based on detailed studies of the ionization and kinematic properties of the galaxies and extended gaseous regions, they suggested that the extended gas has been extracted from the galaxies either from tidal interactions or from AGN outflows induced by the interactions. \citet{chen2019} reported the detection of an extended (up to $\approx$100\,kpc) nebula in \oii, \hb, \oiii, \ha\ and \nii\ emission from a group (\mhalo\ $\approx3\times10^{12}$\,\msun) of 14 galaxies at $z=0.3$ using $\approx$2\,h of MUSE observations. Combining analysis of the morphology and kinematics of the gas detected in emission and in absorption against a background quasar, the study suggested that gas stripping in low-mass galaxy groups is effective in releasing metal-enriched gas from star-forming regions. 

Extended \oii\ emission has been also detected in denser cluster environments. \citet{boselli2019} detected extended (up to $\approx$100\,kpc) tails of diffuse gas in \oii\ emission around two massive (\mstar\ $\approx10^{10}$\,\msun) galaxies in a cluster (\mhalo\ $\approx2\times10^{14}$\,\msun) at $z=0.7$ using $\approx$4\,h of MUSE observations. The observational evidence pointed towards the gas being removed from the galaxies during a ram-pressure stripping event. Recently, \citet{leclercq2022} reported the first detection of extended ($\approx$1000\,kpc$^2$) \mgii\ emission from the intragroup medium of a low-mass (\mhalo\ $\approx5\times10^{11}$\,\msun), compact ($\approx$50\,kpc) group comprising of five galaxies at $z=1.3$ using deep ($\approx$60\,h) MUSE observations. The analysis of the extended \mgii, \feii$^*$ and \oii\ emission suggests that both tidal stripping due to galaxy interactions and outflows are enriching the intragroup medium of this system. 

We do not detect extended \mgii\ emission from the intragroup medium in the sample of groups studied in this work, despite reaching SB limits comparable to that in the study of \citet{leclercq2022} in some of the groups detected in the MUDF. We do however detect extended \oii\ emission from one of the richest groups in the MUDF (see Fig.~\ref{fig:mudf_oii_group}). The \oii\ emission is seen around and between seven galaxies that form a ring-like structure extending up to $\approx$200\,kpc. The extent of the structure is comparable to what has been found in other systems discussed above. The \oii\ SB level ($2-3\times10^{-19}$\,\ergscmarc) in the tails or filaments is also within a factor of few to what has been detected in similar structures in the literature  \citep{boselli2019,leclercq2022}. Based on visual inspection, the extended emission appears to form clumps and filamentary structures. Detailed analysis of the kinematics and physical properties of the extended gas and galaxies are required to establish whether the \oii\ emission originates from gas stripping, outflows or combination of both.

Furthermore, there have been detections of extended nebulae ($\approx10-120$\,kpc) in \oii, \hb\ and \oiii\ emission from galaxy groups hosting quasars at $z\approx0.5-1$ \citep{johnson2018,johnson2022,helton2021}. These observations suggest gas stripping from the interstellar medium of interacting galaxies, and cool, filamentary gas accretion as possible origins of the extended gaseous structures. In this work, we detect extended \mgii\ and \oii\ emission around two active galaxies that we classify as quasars\footnote{We have checked that including the quasars in the stacking analysis in Section~\ref{sec:results_stack} does not have any effects on the results.} based on their broad emission lines (see Figs.~\ref{fig:magg_quasar} and \ref{fig:mudf_quasar}). While \mgii\ emission extending out to $\approx30-40$\,kpc from the centre is detected only around the quasars, \oii\ emission is also detected from nearby galaxies that belong to the same group in both cases. Similar to what has been found for some quasars in the above IFU studies \citep[see also][for detection of a gaseous bridge between a quasar and a galaxy using long-slit spectroscopy]{dasilva2011}, the extended \oii\ emission possibly originates from interactions between the quasar and its companions. Alternatively, the extended emission could be tracing outflows from the quasar as found in spatially resolved observations of some low-redshift AGNs \citep[e.g.][]{smethurst2021}.

Next, extended metal line haloes have been detected around individual galaxies in the literature. \citet{rupke2019} discovered a large ($\approx80\times100$\,kpc$^2$) \oii\ nebula around a massive (\mstar\ $\approx10^{11}$\,\msun), starburst (SFR $\approx100-200$\,\msunyr) galaxy at $z\approx0.5$ using KCWI. \mgii\ emission is also detected on smaller scales up to $\approx$20\,kpc around this galaxy. The study suggests that the extended emission is tracing a bipolar multi-phase outflow that is probably driven by bursts of star formation. In addition, \mgii\ emission extending out to $\approx25-37$\,kpc have been detected around two massive (\mstar\ $\approx10^{10}$\,\msun), star-forming (SFR $\approx20-50$\,\msunyr) galaxies at $z\approx 0.7$ using KCWI \citep{burchett2021} and MUSE \citep{zabl2021}. While \citet{burchett2021} found that the extended emission is most consistent with an isotropic outflow, \citet{zabl2021} found that a biconical outflow can explain the observed emission. Further, \citet{zabl2021} suggested that shocks due to the outflow can power the extended \mgii\ and \oii\ emission in that system. Recently, \citet{shaban2022} reported the highest redshift ($z\approx 1.7$) detection of extended ($\approx$30\,kpc) \mgii\ emission from a gravitationally-lensed galaxy using $\approx$3\,h of MUSE observations. The \mgii\ and \feii$^*$ emission from this galaxy are more extended than the stellar continuum and \oii\ emission, and are probably tracing a clumpy and asymmetric outflow. We note that although all the above observations of extended metal emission are consistent with galactic outflows, there could also be contribution from interactions with nearby, less massive companion galaxies, as detected in the system studied by \citet{zabl2021}.

\begin{figure*}
 \includegraphics[width=0.48\textwidth]{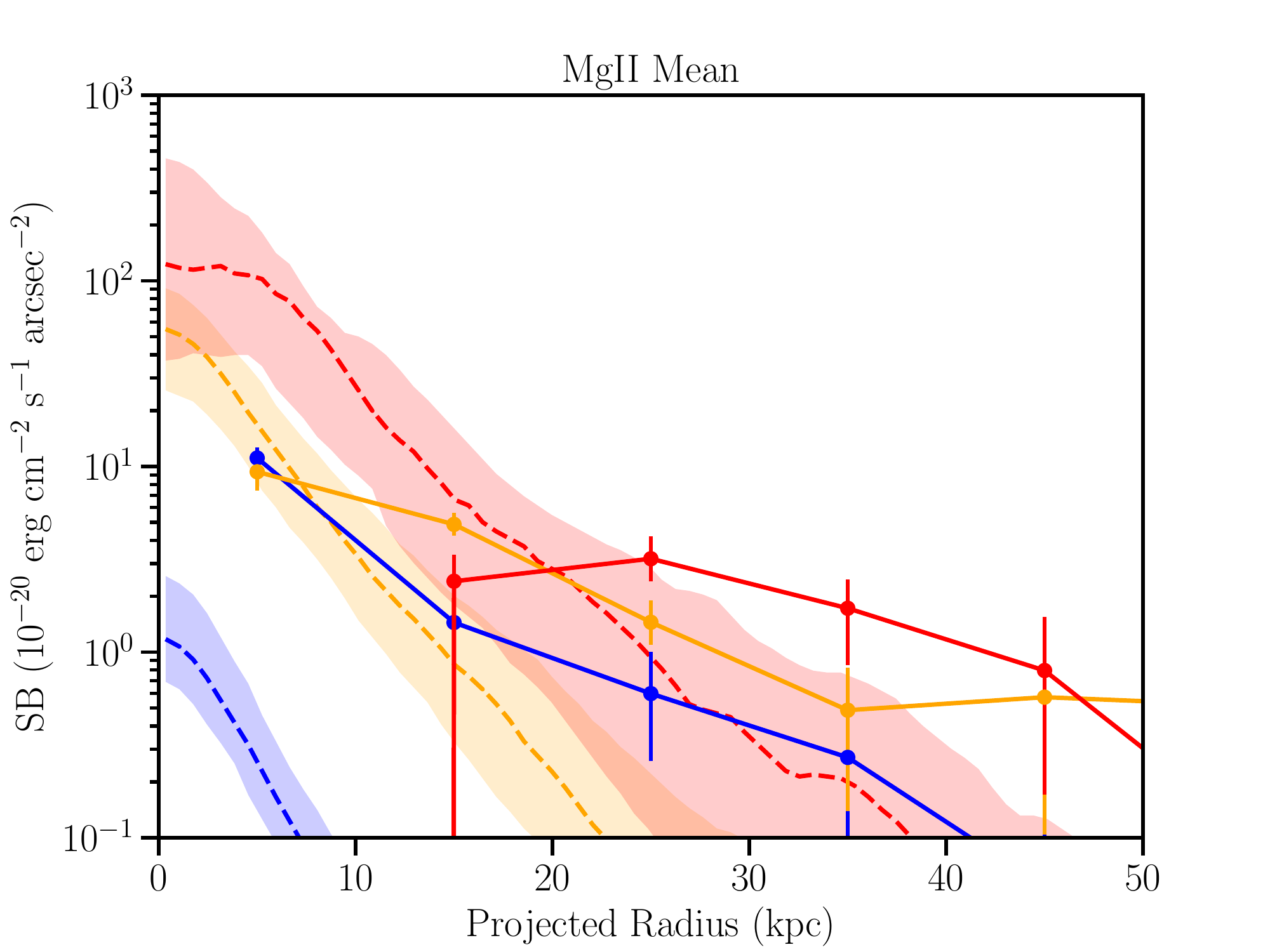}
 \includegraphics[width=0.48\textwidth]{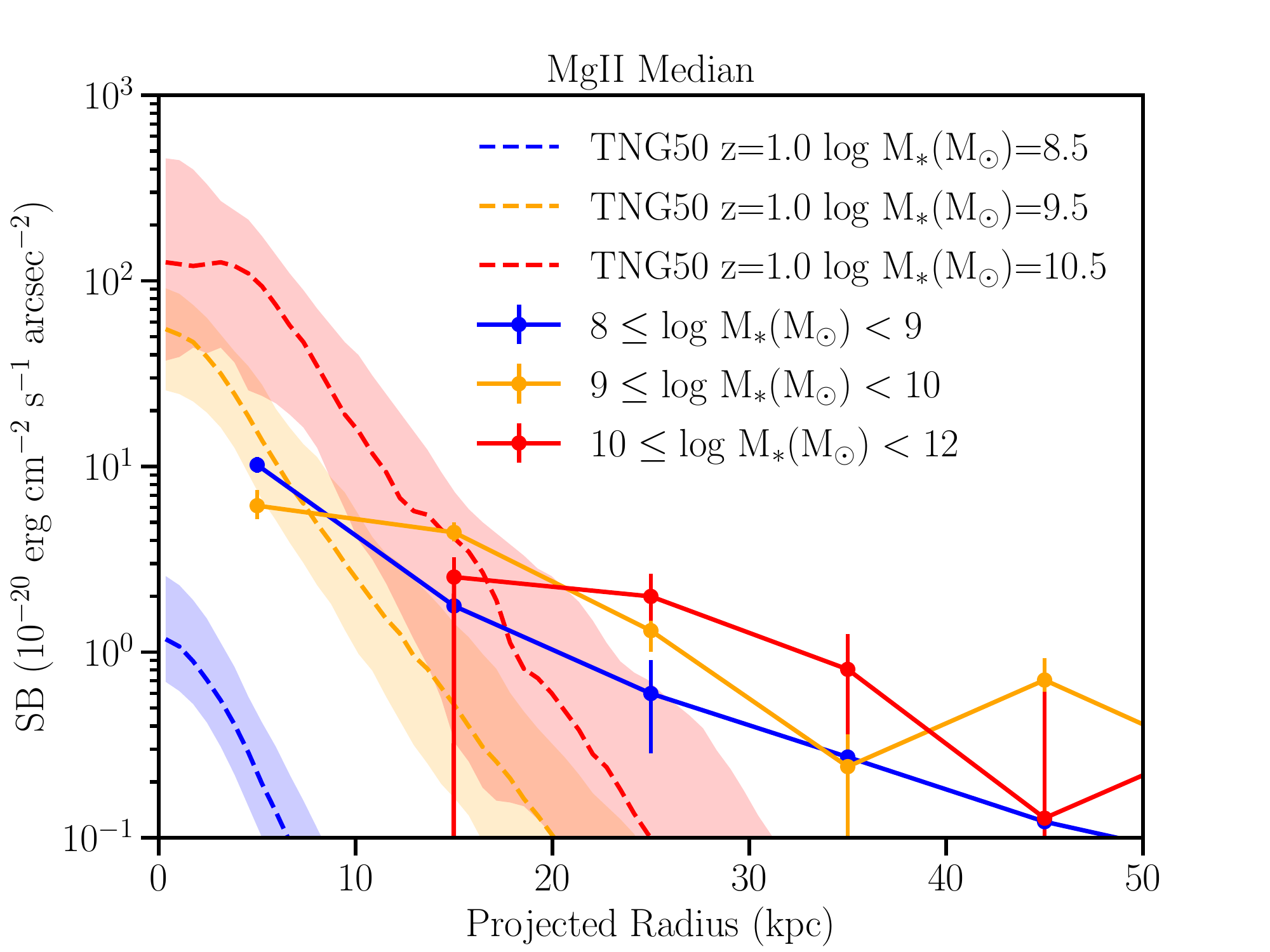}
 \caption{{\it Left:} The mean-stacked azimuthally-averaged radial SB profiles of \mgii\ emission in circular annuli of 10\,kpc from the galaxy centres for different stellar mass bins are shown in solid lines as marked in the legend in the right panel. The error bars represent the $1\sigma$ uncertainties from bootstrapping analysis. The \mgii\ SB profiles from the TNG50 simulation \citep{nelson2021} at comparable redshift and stellar masses are shown in dashed lines. The shaded regions represent the $1\sigma$ halo-to-halo variation.
 {\it Right:} The same as in the left panel for the median-stacked profiles.
 }
 \label{fig:sbprofile_mgii_sim}
\end{figure*}

In this work, we have searched for extended \mgii\ and \oii\ emission around a selection of galaxies in group environments in MAGG and MUDF using {\sc CubEx}. While we detect extended \mgii\ emission only around two active galaxies as discussed above, we do detect extended ($\approx10-50$\,kpc) \oii\ emission around some of the normal group galaxies (see Fig.~\ref{fig:magg_oii_individual} and Fig.~\ref{fig:mudf_oii_individual}). However, the prevalence of such extended emission is moderate ($\approx30-40$\%) at the SB limits we typically reach ($\approx10^{-18}-10^{-19}$\,\ergscmarc). The galaxies from which extended metal line emission have been detected in the literature, as discussed above, are predominantly starburst and/or merging systems, and are thus not characteristic of the general star-forming population. The average emission from a blind galaxy sample should be much dimmer than that obtained from targeting known bright sources. Indeed, the average metal emission SB obtained from stacking all the galaxies in our sample (Fig.~\ref{fig:sbprofile_all}) is fainter by one to two orders of magnitudes (after correcting for cosmological dimming) compared to the detections in the literature \citep[e.g.][]{rupke2019,burchett2021,zabl2021}. The sample used for stacking in this work is based on blind galaxy surveys, with a possible bias against passive galaxies since we select galaxies that have secure redshifts over $0.7 \le z \le 1.5$ based mostly on \oii\ emission. Nevertheless, the sample used for stacking should be representative of the star-forming galaxy population, and the average metal SB profiles obtained here are expected to reflect the typical metal haloes around star-forming galaxies at these redshifts.  

\subsection{Comparison with simulations}
\label{sec:discussion_simulation}

On the theoretical front, there have been several works on predictions of emission from gas around galaxies, focusing on different X-ray, UV, optical and infrared lines, using cosmological simulations \citep[e.g.][]{bertone2012,frank2012,voort2013,corlies2016,sravan2016,lokhorst2019}, zoom-in simulations \citep[e.g.][]{augustin2019,peroux2019,corlies2020}, radiation-hydrodynamical simulations \citep[e.g.][]{katz2019,mitchell2021,byrohl2021}, and analytical models \citep[e.g.][]{faerman2020,piacitelli2022}. Recently, \citet{nelson2021} explored the \mgii\ emission from haloes around a statistical sample of thousands of galaxies with stellar mass in the range $\log(M_*/M_\odot) = 7.5-11.0$ over the redshift range $0.3<z<2$, using the TNG50 cosmological magnetohydrodynamical simulation \citep{nelson2019a,nelson2019b,pillepich2019} from the IllustrisTNG project. This study developed a relatively simple model for \mgii\ emission by post-processing the TNG50 simulation. In brief, the line emissivity was first computed using {\sc cloudy} \citep{ferland2017}. This emissivity, along with the Mg gas-phase abundance from the TNG simulation, was used to estimate the line luminosity, after taking into account the effect of depletion onto dust grains using the observed dependence of the dust-to-metal ratio on metallicity \citep{peroux2020a}. The medium was assumed to be optically thin, i.e. all radiative transfer effects including resonant scattering were neglected, as were interactions between \mgii\ photons and dust, including absorption. We compare here the results from our stacking analysis to the predictions by this study. We note that the \mgii\ emission results from the simulation are based on the combination of both the \mgii\ doublet lines, same as in this work, and take into account convolution by a Gaussian PSF with a full-width at half-maximum of 0.7 arcsec, comparable to that of the MUSE data. 

In Fig.~\ref{fig:sbprofile_mgii_sim}, we show the \mgii\ radial SB profiles from both the mean and median stacks of galaxies in three different stellar mass bins over $0.7 \le z \le 1.5$ (see Section~\ref{sec:results_stack_mass}). For comparison, we show the mean- and median-stacked radial SB profiles of \mgii\ emission in three different stellar mass bins (with median stellar masses similar to that of the observed stacks) at $z=1.0$ from the TNG50 simulation. This figure displays an overall good agreement between the signal detected through stacking and that predicted from the simulation, albeit with some differences. Firstly, we note that the strong increase of the central SB with stellar mass found in the simulation is not reflected in the observed central SB. For the highest stellar mass range, \mstar\ $=10^{10-12}$\,\msun, \mgii\ is observed in absorption in the central 10 kpc region. The observed and simulated central SB values are comparable for stellar mass \mstar\ $\approx10^{9-10}$\,\msun. For lower stellar masses, the simulated SB values are much lower than the observed ones. Secondly, we note that the observed SB profiles are generally flatter than the simulated profiles across stellar masses. The discrepancies between the observed and simulated SB profiles likely arise due to the simplifying assumptions that have gone into the \mgii\ emission model. These are described in detail in section 2.3 of \citet{nelson2021} and we discuss some of them here briefly.

Firstly, due to the inability of the TNG50 simulation to resolve physical mechanisms at small scales ($\lesssim$100\,pc) in the interstellar medium (ISM), emission from highly ionized \hii\ regions around young stars is neglected. The \mgii\ emission from the ISM could be non-negligible, particularly in lower mass galaxies with larger young stellar population, and could propagate and scatter into the CGM. This could explain the brighter and more extended \mgii\ SB profile observed around lower mass (\mstar\ $<10^9$\,\msun) galaxies compared to that in the simulation. Secondly, the model assumes that the \mgii\ emission is optically thin and neglects the impact of resonant scattering. Resonant scattering is expected to redistribute the emergent \mgii\ emission from smaller to larger radii and thereby to flatten the radial SB profiles \citep{prochaska2011b,byrohl2021,mitchell2021}. This could explain why the observed SB profiles are flatter than the simulated ones. Additionally, the model does not take into account the stellar continuum at the \mgii\ wavelength, which if scattered could lead to flattening of the SB profiles, and dust, which could be an additional source of opacity for photons. Indeed, in the case of \lya\ emission, it is found that dust significantly suppresses the flux and that the dust attenuation of the central SB strongly scales with stellar mass \citep[e.g., figure A5 of][]{byrohl2021}. This along with the lack of \mgii\ absorption from the ISM in the models could explain the difference in the central SB between observations and simulations for the most massive galaxies (\mstar\ $\ge10^{10}$\,\msun). 

Overall, the above comparison indicates significant uncertainties in predicting \mgii\ observations, and motivates the need for better modeling of the ISM physics and radiative transfer effects, also including a more rigorous dust treatment. Future work will extend the Monte Carlo radiative transfer technique of \citet{byrohl2021} to the problem of \mgii\ resonant scattering, including empirically calibrated dust models (Byrohl et al. in prep). Simultaneously, all cosmological hydrodynamical simulations have limited spatial resolution, which could impact the size and abundance of cool gas clouds that are resolved in the halo \citep{suresh2019,peeples2019}. We also note that the underlying TNG galaxy formation model yields specific predictions for the abundance, physical properties, and thus emissivity of halo gas \citep{truong2020,pillepich2021,ramesh2022}, and other galaxy formation models will produce different results \citep{fielding2020,ayromlou2022,sorini2022}. Thus, the comparison presented here offers a first analysis, but not a complete one, in comparing metal emission around galaxies in simulations and observations.

From the observational perspective, we note that, as mentioned previously, there maybe a bias against passive galaxies in the sample used for stacking in this work, and the sample is more representative of the star-forming galaxy population. The simulation results, on the other hand, are based on all galaxies and are not restricted to only star-forming galaxies. The highest mass bins in the simulation are dominated by more passive galaxies, which combined with the specific SFR showing a positive correlation with the \mgii\ halo extent in the simulation \citep[see figure 8 of][]{nelson2021}, could lead to a smaller \mgii\ halo extent than in the observations. Next, the observed stacks are obtained in a line-of-sight velocity window and could be affected by projection effects, whereas the simulation results are based on gas that is gravitationally bound to the central halo. In the future, after the \mgii\ scattering problem is addressed with the methodology of \citet{byrohl2021}, we would be able to obtain spatially resolved spectra from the TNG50 simulation, and more faithfully reproduce the observations using the same velocity integration window. Finally, \mgii, being a resonant line, can be detected in emission and in absorption against the background stellar continuum, with part of the absorption infilled by redshifted emission \citep[i.e., P-Cygni-like profiles, see e.g.][]{prochaska2011b,erb2012,martin2012}. We have not attempted to decompose the \mgii\ emission and absorption from the ISM in this work, which would mainly affect the central SB, but this could be explored in the future. All of the above factors, along with the modeling uncertainties, contribute to the differences we find between the observed and simulated \mgii\ SB profiles.  

Although the observed and simulated \mgii\ SB profiles in different stellar mass bins do not match quantitatively, there are some trends that are qualitatively similar in the observational and simulation results. The size or extent at a fixed SB level of the \mgii\ haloes increases with stellar mass in both the observations and simulation. The extent of the \mgii\ emission also correlates with the environment in both cases. The observed \mgii\ SB profile is more extended around group galaxies, whereas the half-light radii of the simulated \mgii\ haloes increases with the local galaxy overdensity due to contributions from nearby satellites.

\subsection{Physical implications of stacked line emission}
\label{sec:discussion_mgii_oii}

\begin{figure}
 \includegraphics[width=0.48\textwidth]{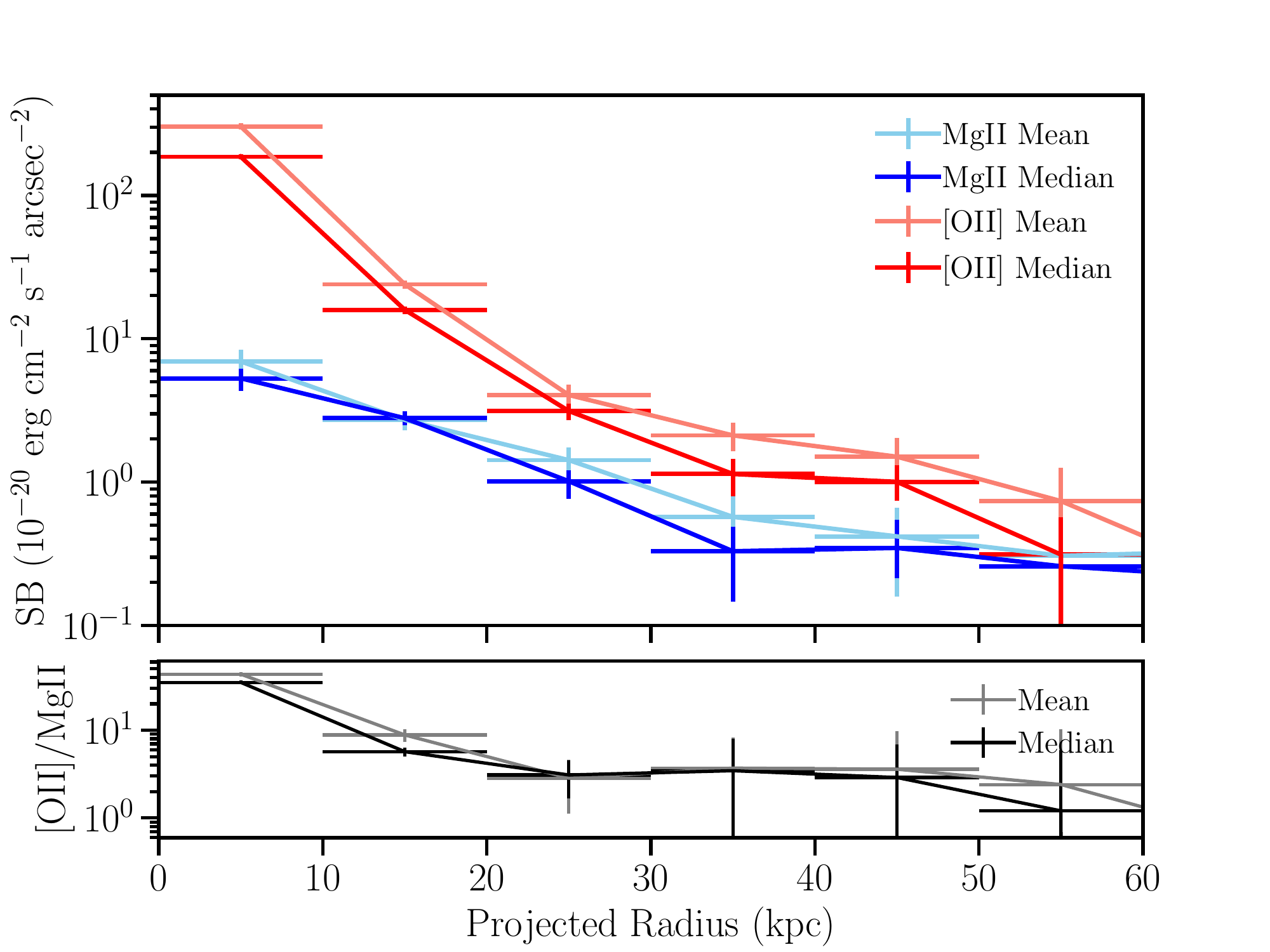}
 \caption{The top panel compares the \mgii\ and the \oii\ SB profiles. The mean- and median-stacked azimuthally-averaged radial SB profiles of \mgii\ emission in circular annuli of 10\,kpc from the galaxy centres for the full sample are shown in light blue and dark blue, respectively. Similarly, the mean- and median-stacked SB profiles of \oii\ emission are shown in pink and red, respectively. The error bars represent the $1\sigma$ uncertainties from bootstrapping analysis. The bottom panel shows the ratio of the SB of \oii\ to \mgii, in grey for the mean profiles and in black for the median profiles. 
 }
 \label{fig:sbprofile_mgii_oii}
\end{figure}

There are several physical mechanisms, that could explain the extended \mgii\ and \oii\ emission detected around the galaxies through the stacked analysis. One possible scenario is that the ionizing photons (ionization potentials of \mgii\ and \oii\ are 15 and 13.6\,eV, respectively) escape from the stellar discs of the galaxies, likely via outflows as indicated by the P-Cygni-like \mgii\ profile in the central region, and ionize the cool gas in the haloes \citep[e.g.][]{chisholm2020}. Another possibility is that the ultraviolet background and/or low levels of star formation in the extended disc or disc-halo interface acts as a local source of ionizing radiation. The presence of more extended \oii\ emission along the disc of the galaxy (Section~\ref{sec:results_stack_orientation}) could support this scenario, since it would be less likely for ionising photons produced in the inner part of the galaxy to escape along the disc. Lastly, gravitational interactions or galactic outflows could lead to shocks, which in turn could produce the required ionizing photons \citep[e.g.][]{heckman1990,pedrini2022}. It is likely that a combination of the different mechanisms contributes to the average extended emission. Observations of additional emission lines (e.g. \oiii, \ha, \nii) are required to distinguish between the different mechanisms.

Constraints on the physical mechanisms and gas conditions can also be obtained by comparing the average \mgii- and \oii-emitting halo gas. The \oii\ emission is brighter than the \mgii\ emission, as also found in observations of individual galaxies in the literature \citep[e.g.][]{rubin2011,feltre2018,leclercq2022}. Based on observations of star-forming galaxies at $z=0.7-2.4$ and predictions from photoionization models of \citet{gutkin2016}, \citet{feltre2018} found that the \oii\ to \mgii\ emission line ratio increases with metallicity and varies between $\approx$3 and $\approx$40. As can be seen from Fig.~\ref{fig:sbprofile_mgii_oii}, the \oii\ to \mgii\ SB ratio from the stack of the full sample is a factor of $\approx$40 in the central 10\,kpc, decreases with increasing distance from the centre and remains constant around $\approx$3 over 20--50\,kpc, before tentatively declining further out. This constant ratio could indicate that the \mgii\ emission in the inner region of the halo does not originate solely from scattering of ionizing photons produced in the stellar disc, but could have a non-negligible in-situ component through collisional ionization similar to the \oii\ emission. The resulting \mgii\ photons can subsequently undergo scattering in the outer halo, leading to a flatter SB profile. 

Indeed, as noted in Section~\ref{sec:results_stack}, it can be seen from Fig.~\ref{fig:sbprofile_mgii_oii} that the \mgii\ SB profile is shallower than the \oii\ SB profile on average. This suggests that the resonant \mgii\ emission is more likely to be affected by dust and radiative transfer effects compared to the non-resonant, collisionally excited \oii\ emission, leading to the observed flattening. Studies of individual systems with coverage of both the lines in the literature have also found that the \mgii\ SB profiles are flatter than the \oii\ ones \citep{zabl2021,leclercq2022}. Currently, there are no studies that have compared the \mgii\ and \oii\ halo emission around galaxies in cosmological simulations. Such a comparison would place useful constraints on the possible physical mechanisms leading to the extended \mgii\ and \oii\ emission around galaxies.

Next, we can obtain an estimate of the electron density using the ratio of the intensities of the \oii\ doublet lines, $\lambda$3729/$\lambda$3727, which varies between 0.35 in the high density limit and 1.5 in the low density limit \citep{osterbrock1989}. We fit the \oii\ emission line extracted from the median-stacked cube of the full sample (see Fig.~\ref{fig:stack_spectra}) with a double Gaussian profile (with the wavelength difference between the two Gaussian components fixed to that of the \oii\ doublet). We find that the ratio of the intensity of the doublet lines is $1.3\pm0.1$ in the central 10\,kpc region and marginally increases to $1.5\pm0.4$ in the outer annulus region between 30--40\,kpc, suggesting that the density is declining with increasing distance into the halo. Following \citet{sanders2016}, we estimate that the above \oii\ doublet ratios imply that the average electron density at $z\approx1$ is $\lesssim$100\,\cc. This is similar to what has been found for star-forming field galaxies at $z\approx1-1.5$ \citep{harshan2020,davies2021}. 

The intrinsic flux ratio of the \mgii\ emission doublet lines, $\lambda$2796/$\lambda$2803, would be 2 if the excitation is dominated by collisions, and would be $\approx$1 if the excitation is dominated by photon absorption \citep{sigut1995,prochaska2011b}. While the stacked spectrum in the central $\approx$10\,kpc region shows \mgii\ in both emission and absorption, the spectra in the outer region (10--30\,kpc) shows \mgii\ emission and no prominent absorption signatures (see Fig.~\ref{fig:stack_spectra}), suggesting that the \mgii\ gas in the halo is likely to be optically thin. From double Gaussian fits (with the wavelength difference between the two Gaussian components fixed to that of the \mgii\ doublet) to the stacked \mgii\ emission lines, the flux ratio is $1.5\pm0.1$ and $1.7\pm0.2$ in the annular region between 10--20\,kpc and 20--30\,kpc, respectively. The ratio being close to 2 suggests that the extended \mgii\ emission could indeed be produced in part through collisional excitation in situ. The \mgii\ gas in the halo could then absorb and scatter the intrinsic light, decreasing the flux ratio from the intrinsic value of 2. Detailed kinematic modeling of the spatially-resolved emission would help to test such a scenario. 

Next, we discuss and compare the dependence of the metal-enriched halo gas, as probed through emission and absorption lines, on redshift, stellar mass and environment. Both the \mgii\ and \oii\ SB profiles are brighter and more extended at $1 < z \le 1.5$ compared to at $0.7 \le z \le 1.0$. It is interesting to note that the covering fractions of the more diffuse halo gas probed using \mgii\ and \civ\ absorption within the virial radius of galaxies are found to be larger at $z>1$ compared to at $z\le1$, indicating a more extended metal-enriched halo at higher redshifts \citep{lan2020,dutta2021}. This dependence of the halo gas on redshift, found independently using emission and absorption probes, could be arising from the higher SFR, ionizing radiation or cool gas density at higher redshifts. Indeed, after matching the samples in SFR, we find that the metal SB profiles in the two redshift bins become consistent within the uncertainties, indicating that this dependence could be mainly driven by the redshift evolution of SFR.

Furthermore, the extents of the \mgii\ and the \oii\ emission evolve with stellar mass, increasing by a factor $\approx$2 from \mstar\ $=10^{7-8}$\,\msun\ to $10^{10-12}$\,\msun\ at a fixed SB limit. While the central SB of \mgii\ emission does not show any significant dependence on stellar mass, the central SB of \oii\ emission increases by a factor of $\approx$3 going from \mstar\ $=10^{7-8}$\,\msun\ to $10^{10-12}$\,\msun. The increase in emission line extent with stellar mass is consistent with absorption line studies finding that more massive galaxies show on average larger equivalent width and covering fraction of \mgii\ absorbing halo gas \citep[e.g.][]{rubin2018,dutta2021}. The more massive galaxies could be populating their haloes with larger amount of metals, e.g. through outflows.

Another complementary result between emission and absorption line studies is the dependence of the gas distribution on the galaxy environment. The metal line SB profiles, \oii\ to a greater extent than \mgii, are brighter and more extended around group galaxies compared to around single galaxies. Consistently, overdense group environments have been shown to be associated with stronger and more prevalent \mgii\ absorption, on average, compared to single galaxies \citep[e.g.][]{bordoloi2011,nielsen2018,fossati2019b,dutta2020,dutta2021}. These studies propose that this could be due to tidal or ram pressure stripping, as also suggested by detections of extended \oii\ emission around individual galaxies in group environments in this work.

\section{Summary and Conclusions}
\label{sec:summary}

We have presented a study of the extended metal emission around galaxies, traced by the rest-frame optical \mgii\ and \oii\ doublet lines, using MUSE IFU data from two large, blind galaxy surveys, MAGG and MUDF. To probe the average metal line emission around galaxies, we have conducted a stacking analysis of the MUSE 3D data totalling $\approx$6000--7000\,h of $\approx$600 galaxies over the stellar mass range of $10^{6-12}$\,\mstar\ and over the redshift range of $0.7 \le z \le 1.5$. To check whether the metal emission traces the halo, we have compared the average metal line emission with the average stellar continuum emission in control windows. We have investigated the dependence of the average \mgii\ and \oii\ emission around galaxies on redshift, stellar mass, environment and orientation. In addition, we have searched for extended \mgii\ and \oii\ line emission in a sample of group galaxies using {\sc CubEx}. Finally, we have compared our results with those obtained from observations in the literature and the TNG50 simulation. Below we summarize the key results from this study.

\begin{enumerate}
    \item[--] Significant \mgii\ and \oii\ line emission is detected out to $\approx$30\,kpc and $\approx$40\,kpc, respectively, from the galaxies on average in the stacked NB images and SB radial profiles. This metal emission is radially more extended than the average stellar continuum emission, which extends up to $\approx$15\,kpc, and therefore it traces the metal-enriched halo gas. The \oii\ emission is more centrally peaked, while the \mgii\ emission profile is shallower. The \oii/\mgii\ SB ratio is $\approx$40 in the central region and decreases to $\approx$3 beyond 20\,kpc. The total luminosity of the average \mgii\ and the \oii\ emission at $z\approx1$ from the halo region ($\approx$15--30\,kpc) is $\approx6\times10^{39}$\,\ergs\ and $\approx2\times10^{40}$\,\ergs, respectively.

    \item[--] The \mgii\ and \oii\ emission around galaxies becomes brighter and more extended at higher redshifts. In particular, the \mgii\ and \oii\ SB profiles are intrinsically brighter by a factor $\approx2-3$ and more radially extended at the observed SB limit by a factor of $\approx1.3$ at $1.0 < z \le 1.5$ than at $0.7 < z \le 1.0$. Our control analysis indicates that this redshift evolution in the metal emission could be related to galaxies having higher SFR at $z>1$, which could be leading to a higher local radiation field and gas density.

    \item[--] The extent of the \mgii\ and \oii\ emission around galaxies shows an increasing trend with the stellar mass of the galaxies. The SB profiles become more radially extended by a factor of $\approx$2 on going from the lowest stellar mass bin ($10^{7-8}$\,\msun) to the highest stellar mass bin ($10^{10-12}$\,\msun). The \oii\ emission in the central region also becomes brighter by a factor of $\approx$3, while the central \mgii\ emission does not show any strong dependence on stellar mass. The dependence of the metal emission on stellar mass could be due to the larger metal enrichment in the haloes of more massive galaxies because of stronger feedback.

    \item[--] The \mgii\ SB profile around a sample of group galaxies, identified through a FoF technique, is brighter by a factor of $\approx$1.4 on average within 30\,kpc, while the \oii\ SB profile is more extended by a factor of $\approx$1.6 compared to that around a sample of isolated galaxies that is matched in stellar mass and redshift. This enhancement could be due to interactions in the group environment, as also indicated by observations of extended \oii\ emission around individual galaxies in groups.
    
    \item[--] For a sub-sample of $\lesssim $60 galaxies at inclination $\ge$30$^\circ$ in the MUDF with deep HST imaging and morphological parameters, the \oii\ SB profile is brighter over $\approx$30--40\,kpc and more extended by a factor of $\approx$1.3 along the major axis of the disc compared to that along the minor axis. No significant dependence on orientation is found for the \mgii\ emission. However, these results need to be verified using a larger galaxy sample with morphological measurements.

    \item[--] Extended ($\approx$10--50\,kpc) \oii\ emission is detected around $\approx$32 per cent and $\approx$40 per cent of galaxies in group environments (comprising of three or more galaxies) in MAGG and MUDF surveys, respectively. Further, extended \oii\ emission connecting seven galaxies in a $z=1.0$ group is detected across $\approx$200\,kpc in the MUDF survey. Extended ($\approx$30--40\,kpc) \mgii\ emission is detected only around two active galaxies (quasars) in the sample ($\approx$1 per cent). Environmental processes (tidal or ram-pressure stripping) or outflows could be responsible for the extended metal line emission detected around the galaxies in overdense group environments.

    \item[--] There are differences in the stacked \mgii\ radial SB profiles obtained from our study and those obtained using TNG50 simulations. Specifically, unlike the simulated profiles, the central SB of the observed profiles do not show a strong increasing trend with stellar mass, and the observed profiles are flatter. Two key modeling assumptions -- neglecting resonant scattering as well as the impact of dust -- likely drive these differences. However, there is an overall agreement in trends between the observational and simulation results given the many complex physical processes at play.
\end{enumerate}

This work demonstrates that we are now able to probe the CGM around normal star-forming galaxies directly in emission using the power of deep and large IFU surveys such as MAGG and MUDF. We have presented here the first characterization of the average metal-emitting halo gas around galaxies at $0.7 \le z \le 1.5$. The stacking results found in this work complement studies of the more diffuse metal-enriched halo gas probed in absorption around $z<2$ galaxies \citep{dutta2020,dutta2021}. A combination of these two different approaches in the future, for a general galaxy population and also for individual systems, will lead to richer constraints on galaxy formation models.

Based on the stacking analysis in this work, the average \oii\ doublet ratio in the halo ($\approx$10--40\,kpc annular region) suggests average electron densities $\lesssim$100\,\cc, while the \mgii\ doublet ratio and the \oii/\mgii\ ratio in the halo point towards in situ collisional ionization, with subsequent scattering of the \mgii\ photons, both playing an important role in the origin of the extended emission. Detailed modeling of the spatially-resolved morpho-kinematics of the extended metal emission is required to place more robust constraints on the potential physical mechanisms and gas conditions. Large galaxy surveys with 4MOST \citep{dejong2019}, DESI \citep{DESI2016}, HET \citep{gebhardt2021}, MOONS \citep{cirasuolo2014}, PFS \citep{tamura2016}, and WEAVE \citep{jin2022}, and deep observations of individual systems with ERIS \citep{davies2018}, JWST \citep{gardner2006}, KCWI \citep{morrissey2018}, and MUSE \citep{bacon2010}, and with future instruments such as ELT/HARMONI \citep{thatte2021}, MIRMOS \citep{konidaris2020}, and VLT/BlueMUSE \citep{richard2019}, will allow us to obtain multiple diagnostics and interpret the physical mechanisms driving the extended emission.


\section*{Acknowledgements}

We thank the reviewer for their helpful comments.
This project has received funding from the European Research Council (ERC) under the European Union’s Horizon 2020 research and innovation programme (grant agreement No 757535) and by Fondazione Cariplo (grant No 2018-2329).
DN acknowledges funding from the Deutsche Forschungsgemeinschaft (DFG) through an Emmy Noether Research Group(grant number NE 2441/1-1).
SC gratefully acknowledges support from the Swiss National Science Foundation grant PP00P2 190092 and from the ERC under the European Union's Horizon 2020 research and innovation programme grant agreement number 864361.
PD acknowledges support from the NWO grant 016.VIDI.189.162 (``ODIN") and from the European Commission's and University of Groningen's CO-FUND Rosalind Franklin program.
This work is based on observations collected at the European Organisation for Astronomical Research in the Southern Hemisphere under ESO programme IDs 197.A-0384, 065.O-0299, 067.A-0022, 068.A-0461, 068.A-0492, 068.A-0600, 068.B-0115, 069.A-0613, 071.A-0067, 071.A-0114, 073.A-0071, 073.A-0653, 073.B-0787, 074.A-0306, 075.A-0464, 077.A-0166, 080.A-0482, 083.A-0042, 091.A-0833, 092.A-0011, 093.A-0575, 094.A-0280, 094.A-0131, 094.A-0585, 095.A-0200, 096.A-0937, 097.A-0089, 099.A-0159, 166.A-0106, 189.A-0424, 094.B-0304, 096.A-0222, 096.A-0303, 103.A-0389, and 1100.A-0528. 
Based on observations with the NASA/ESA Hubble Space Telescope obtained, from the Data Archive at the Space Telescope Science Institute, which is operated by the Association of Universities for Research in Astronomy, Incorporated, under NASA contract NAS 5-26555. Support for Program numbers 15637, and 15968 were provided through grants from the STScI under NASA contract NAS 5-26555.
This work used the DiRAC Data Centric system at Durham University, operated by the Institute for Computational Cosmology on behalf of the STFC DiRAC HPC Facility (\url{www.dirac.ac.uk}). This equipment was funded by BIS National E-infrastructure capital grant ST/K00042X/1, STFC capital grants ST/H008519/1 and ST/K00087X/1, STFC DiRAC Operations grant ST/K003267/1 and Durham University. DiRAC is part of the National E-Infrastructure. 
This research has made use of the following {\sc python} packages: {\sc numpy} \citep{2020NumPy-Array}, {\sc scipy} \citep{2020SciPy-NMeth}, {\sc matplotlib} \citep{Matplotlib2007}, {\sc astropy} \citep{Astropy2013}.


\section*{Data Availability}

The MUSE data used in this work are available from the European Southern Observatory archive (\url{https://archive.eso.org}). The MUSE ESO P3 level data of the MAGG survey are available at \url{https://archive.eso.org/scienceportal/home?data_collection=197.A-0384&publ_date=2022-11-14}. 
The HST data used in this work are available from the MAST archive (\url{https://mast.stsci.edu}).
The IllustrisTNG simulations, including TNG50, are publicly available and accessible at \url{www.tng-project.org/data} \citep{nelson2019a}.



\bibliographystyle{mnras}
\bibliography{mybib} 






\bsp	
\label{lastpage}
\end{document}